\documentclass[a4paper,11pt]{article}
\pdfoutput=1 

\usepackage{jheppub} 

\usepackage[T1]{fontenc} 

\usepackage{amssymb}
\usepackage{amsmath}
\allowdisplaybreaks

\usepackage{graphicx}
\usepackage{subfig}

\usepackage{xltabular}
\usepackage{changepage}
\newcolumntype{C}{>{$}c<{$}}
\usepackage{booktabs}

\usepackage{multirow}

\usepackage{diagbox}

\usepackage{float}

\title{\boldmath 
Unmixing sub-leading Regge trajectories
of 
$\mathcal{N} = 4$ Super-Yang-Mills 
}

\author{Julius Julius${}^{\delta}$}
\author{and Nika Sergeevna Sokolova${}^{\ell}$}

\affiliation{${}^{\delta}$Harish-Chandra Research Institute, Homi Bhabha National Institute, Chhatnag Road, Jhunsi, Allahabad 211019, India}
\affiliation{${}^{\ell}$Department of Mathematics, King's College London, Strand WC2R 2LS, United Kingdom}

\emailAdd{julius@hri.res.in}
\emailAdd{nika.sokolova@kcl.ac.uk}

\abstract{
We study the conformal field theory data 
(CFT-data) 
of planar 4D $\mathcal{N} = 4$ Super-Yang-Mills theory 
in the strong 't Hooft coupling limit.
This regime explores the physics of massive short strings in the flat-space limit of the dual AdS.
We focus on the CFT-data of the massive short strings exchanged in the operator product expansion (OPE) of the four-point function dual to the Virasoro-Shapiro amplitude.  
This CFT-data arranges itself into Regge trajectories in the flat-space limit.
Using inputs from recent advances in the computation of the AdS Virasoro-Shapiro amplitude, integrability, and a stipulation based on analyticity of the CFT-data in spin, 
we are able to fix all the CFT-data 
on the four unique sub-leading Regge trajectories,
at leading non-trivial order, as a function of the string-mass level.
One of our predictions is that one of the four unique sub-leading Regge trajectories decouples from the OPE in the flat-space limit.
This hints at an emergent selection rule in the flat-space limit, similar to our previous results in \href{https://arxiv.org/abs/2310.06041}{arXiv:2310.06041}.
Our procedure 
should be applicable in a variety of similar setups like for the AdS Veneziano amplitude or in ABJM.
}

\begin{document} 
\maketitle
\flushbottom

\section{Introduction}\label{sec:intro}
There is more than meets the eye when one looks at conformal field theory data (CFT-data) of four-dimensional planar $\mathcal{N} = 4$ Super-Yang-Mills theory (SYM) in the strong 't Hooft coupling limit.
In this regime, we are exploring the dual string theory in the flat-space limit of AdS.
The single trace operators of $\mathcal{N} = 4$ SYM are dual to massive short strings, and 
the physics of these states/operators in the flat-space/strong coupling regime display some properties which point at a rich underlying structure. 

More concretely, the CFT-data of these states can be accessed by studying four-point functions of $\mathcal{N} = 4$ SYM. 
These massive short strings or ``stringy'' operators are exchanged when one performs an operator product expansion (OPE) of two external operators in the four-point function.
Thus the stringy operators can be characterised by their scaling dimension, and their OPE coefficients, which capture the three point coupling of a stringy operator to two external operators. 

In this paper, we study the CFT-data of stringy operators exchanged in the four point-function dual to the tree-level four graviton amplitude in AdS, dubbed the AdS Virasoro-Shapiro amplitude, in the flat-space limit.
In this limit, the exchanged operators arrange themselves along Regge trajectories. 
The string mass level and the Lorentz spin of states on any given Regge trajectory are related in a particular way.
We will obtain expressions for the leading non-trivial order CFT-data on the four unique sub-leading Regge trajectories as a function of the string mass level, which parameterises this data on a Regge trajectory. 

We use three inputs: constraints on the CFT-data from the AdS Virasoro Shapiro amplitude~\cite{Alday:2022uxp,Alday:2022xwz,Alday:2023mvu}, partial knowledge of spectral information from integrabililty, and the stipulation that the OPE coefficients of states on sub-leading Regge trajectories have zeros at certain integral spins so as to be consistent with analyticity~\cite{Caron-Huot:2017vep,Alday:2017vkk,Homrich:2022mmd,Henriksson:2023cnh} and representation theory~\cite{Alday:2023flc}. 
Additionally, we impose two assumptions. The first one is regarding the structure of the spectral data on a Regge trajectory, and the second one demands polynomiality of certain combinations of the OPE-coefficients and scaling dimensions.
We are able to characterise the leading non-trivial order CFT-data on all four unique sub-leading Regge trajectories in the flat-space limit, up to a small number of free parameters. At this stage, we get seven possible cases with each case in general having different free parameters.

To fix the remaining freedom, we impose an ad hoc assumption on the structure of the OPE coefficients in a small string mass level expansion.
The number of possible cases now reduces to three, and all the freedom is completely fixed.
For each case, we check its predictions for the spectral data against independent integrability-based results.
Doing so, 
we are able to identify the case whose predictions for the spectral data are closest to the independent integrability-based  data, and therefore we claim that this is our \emph{unique} result.

We continue with analysing the predictions for the CFT-data on sub-leading Regge trajectories given by our unique result. One of its predictions is that one entire Regge trajectory of states decouples from the OPE in the flat-space limit.
This shows more evidence of a hidden emergent symmetry in the flat-space limit, akin to similar evidence obtained by us previously in a related context~\cite{Julius:2023hre}.
These predictions can also be used to construct further constraints on the leading non-trivial order CFT-data that can potentially function as a useful input into the program of~\cite{Alday:2022uxp,Alday:2022xwz,Alday:2023jdk,Alday:2023mvu} towards systematic computation of AdS curvature corrections to the Virasoro-Shapiro amplitude.

Our procedure can potentially be applied in similar setups, especially those where the integrability-based description of the spectrum is either not available, or not developed to same extent as in $\mathcal{N} = 4$ SYM, like the AdS Venziano amplitude~\cite{Alday:2024yax,Alday:2024ksp}
or for ABJM theory.

Our paper is structured as follows. We describe the setup in  Section~\ref{sec:setup}. In Section~\ref{sec:prelim}, we present our procedure, our preliminary results which are based of a small number of free parameters, and checks thereof. 
Then, we fix the remaining freedom in Section~\ref{sec:final} using an ad hoc assumption, to get our final result. We present a check for this result and study its properties. We conclude with a discussion in Section~\ref{sec:disc}.
In Appendix~\ref{adx:prelimResult}, we show some explicit expressions regarding our preliminary results. 
Finally in Appendix~\ref{adx:qscRegge}, we describe more details of our integrability-based checks.

\section{Setup}\label{sec:setup}
In this paper, we focus on planar $\mathcal{N} = 4$ SYM. 
The only free parameter of this theory is the 't Hooft coupling $\lambda$, which is obtained by taking the Yang-Mills coupling $g_\mathtt{YM}$ to zero and the rank of the gauge group $\mathrm{SU}(N)$ to infinity in such a way that the combination $\lambda\equiv g_{\mathtt{YM}}^2\,N$ is held fixed.
As such, we focus solely on single trace operators. 
The symmetry group of this theory is $\mathrm{PSU}(2,2|4)$, whose bosonic subgroup is $\mathrm{SO}(4,2)\times \mathrm{SO}(6)$. The $\mathrm{SO}(4,2)$ is the conformal group in four dimensions, and $\mathrm{SO}(6)$ represents the six-dimensional flavour/$R$-symmetry.
Operators in this theory are labelled according to the quantum numbers of $\mathrm{SO}(4,2)\times \mathrm{SO}(6)$. These are six numbers:
\begin{align}
    [\Delta(\lambda)\;;\;\ell_1\;\ell_2\;;\;q_1\;p\;q_2]
    \;.
\end{align}
Here, the former three, namely the scaling dimension $\Delta(\lambda)$ and Lorentz spins $\ell_1,\ell_2$ label the $\mathrm{SO}(4,2)$ and the latter three are the Dynkin labels of the $\mathrm{SO}(6)$.
The scaling dimension in general is a non-trivial (real-valued) function of the 't Hooft coupling $\lambda$, whereas the other quantum numbers are integers. 

Consider the following four-point function: 
\begin{align}\label{eqn:def4pt}
    \langle \mathcal{O}_2(x_1,y_1)\,\mathcal{O}_2(x_2,y_2)\,\mathcal{O}_2(x_3,y_3)\,\mathcal{O}_2(x_4,y_4)\rangle
    \;.
\end{align}
Here $x_i$ are the coordinates in spacetime where the operators are inserted whereas $y_i$ refer to coordinates in the flavour-/$R$-space.
The operator $\mathcal{O}_k$ is a Lorentz scalar, that transforms in the rank-$k$ symmetric traceless representation of the $\mathrm{SO}(6)$ $R$-symmetry group. Its scaling dimension $\Delta = 2$ is protected from quantum corrections, \textit{i.e.} it doesn't depend on $\lambda$. 

To further analyse this four-point function, we can study the operators that are exchanged as part of an operator product expansion (OPE). 
Such operators include single- and double-trace operators whose scaling dimensions are both protected and unprotected. Our focus will only be on single-trace operators with unprotected scaling dimension. The quantum numbers of such exchanged operators are
\begin{align}\label{eqn:qnums1}
    [\Delta(\lambda)\;;\;\ell_1\;\ell_2\;;\;q_1\;p\;q_2]
    =
    [\Delta(\lambda)\;;\;\ell\;\ell\;;\;0\;0\;0]\;,
    \quad\ell\text{-even}
    \;.
\end{align}
Henceforth, we will refer to $\ell$ as the spin-label of an exchanged operator.

\subsection{AdS Virasoro-Shapiro Amplitude}
The four-point function~\eqref{eqn:def4pt} is dual to the four-point graviton amplitude in Type IIB string theory on AdS${_5\times S^5}$ \cite{Maldacena:1997re}.
In particular, since we are considering the planar theory, the dual observable is the tree-level four-graviton amplitude. 

The series of works~\cite{Alday:2022uxp,Alday:2022xwz,Alday:2023jdk,Alday:2023mvu}, focused on the computation of the tree-level four-graviton amplitude in AdS, by 
developing a systematic method of obtaining AdS curvature corrections to the flat-space four-graviton amplitude known as the Virasoro-Shapiro amplitude. 
For this reason, the authors of these works dubbed the tree-level four-graviton amplitude in AdS as the ``AdS Virasoro-Shapiro amplitude''.

The AdS/CFT dictionary gives us that
\begin{align}
    \frac{\alpha^\prime}{R^2} = \frac{1}{\sqrt{\lambda}}
    \;,
\end{align}
where $R$ is the AdS radius and $\alpha^\prime$ is the string tension.
Thus, the flat-space limit, obtained by taking $R^2\to\infty$ is akin to taking the strong 't Hooft coupling limit: $\sqrt{\lambda}\to\infty$.
In this limit,
the operators with quantum numbers~\eqref{eqn:qnums1}, are dual to 
massive short strings, and they are sometimes called ``stringy'' operators. 
At strong coupling, their scaling dimension goes as~\cite{Gubser:1998bc,Gromov:2023hzc}
\begin{align}\label{eqn:DeltaStrongExp}
    \Delta \simeq 2\,\sqrt{\delta}\,\lambda^{1/4} - 2 + \frac{d_1}{\sqrt{\delta}}\frac{1}{\lambda^{1/4}} + \frac{d_2}{\delta^{3/2}}\frac{1}{\lambda^{3/4}}    
    \;.
\end{align}
Here $\delta$ is a positive integer that labels the string mass level of the dual massive short string. The coefficients $d_1,d_2,\dots,$ are higher order terms in the strong coupling expansion of $\Delta$.

Another observable which captures information about the scaling dimension (as well as the other quantum numbers) is the eigenvalue of the quadratic Casimir of $\mathrm{PSU}(2,2|4)$, denoted by $J^2$, given by 
\begin{align}\label{eqn:Casimir}
    \begin{split}        
        J^2 &= \frac{1}{2}(\Delta+2)^2 - 2 + \frac{1}{4} \ell_1 (\ell_1 + 2) + \frac{1}{4} \ell_2 (\ell_2 + 2) \\&- \frac{1}{4} q_1 (q_1 + 2)- \frac{1}{4} q_2 (q_2 + 2) - \frac{1}{8}(2p + q_1 + q_2)^2 - (2p + q_1 + q_2)\;.
    \end{split}
\end{align}
At strong coupling, it scales as
\begin{align}\label{eqn:StrongExpCasimir}
    J^2 \simeq 2\,\delta\,\sqrt\lambda + j_1 + \frac{j_2}{\delta\,\sqrt\lambda}\;,
\end{align}
with $j_1,j_2,\dots,$ being higher order terms in the strong coupling expansion.
As seen in~\cite{Gromov:2023hzc,Julius:2023hre}, many-a-time, the eigenvalue of the quadratic Casimir is more useful in spectral data analysis than the raw spectral data itself.

\begin{figure}[H]
    \centering
    \includegraphics[width=0.8\textwidth]{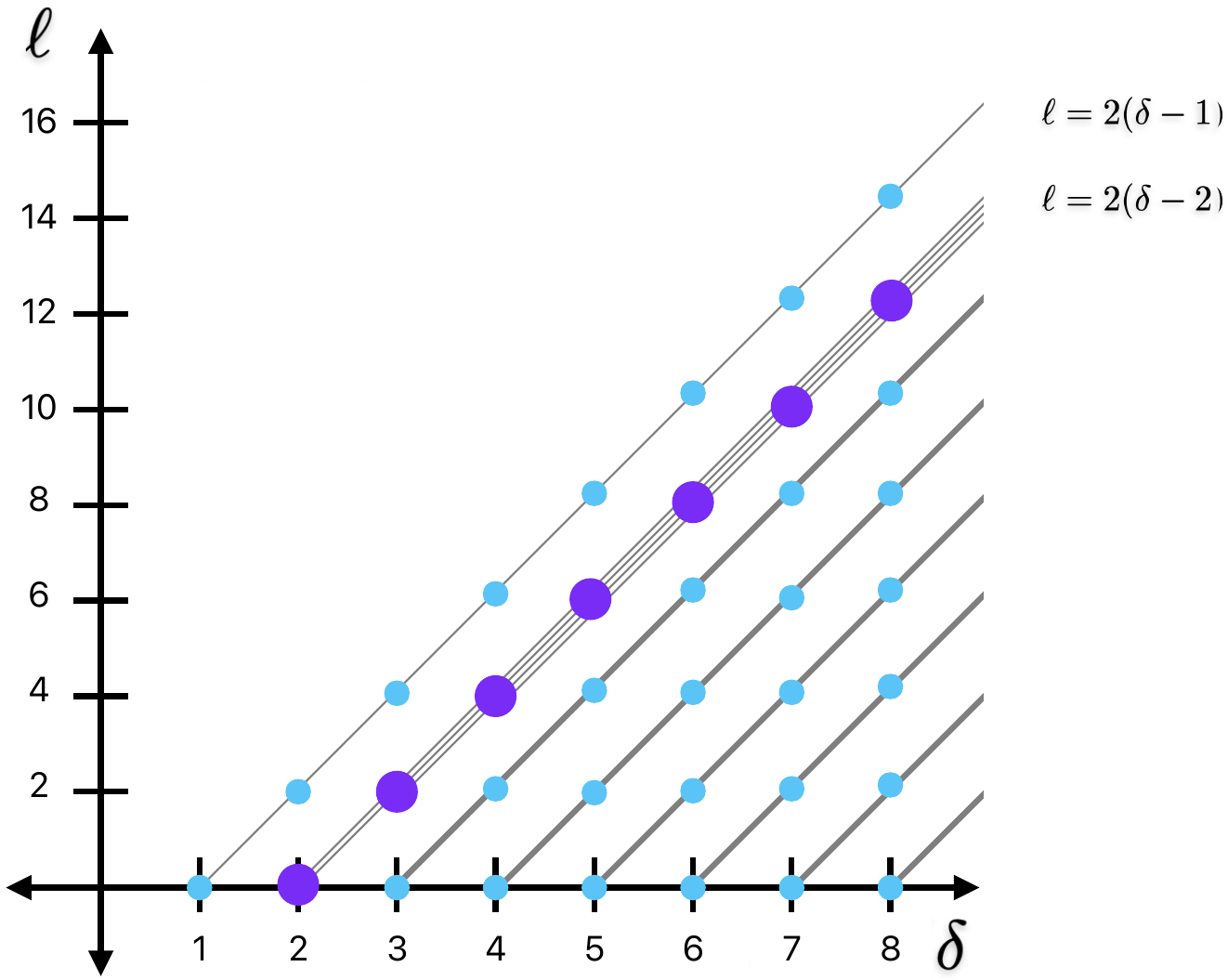}
    \caption{Each point on the plot corresponds to states with string mass $\delta$ and spin $\ell$. The spectrum can be organised into Regge trajectories with the spin-label of states on a particular Regge trajectory given by $\ell = 2(\delta-n)$, where $n$ is the Regge trajectory number. We highlight the sub-leading Regge trajectories, \textit{i.e.} those whose states' spin-label is $\ell = 2(\delta-2)$. As discussed in Section~\ref{sec:CFTRegge} in the main text, whilst the multiplicity of the sub-leading Regge trajectories given by the counting of~\cite{Alday:2023flc} is six, due to certain exact degeneracies among states coming from a parity symmetry of the underlying integrable system, the multiplicity of unique sub-leading Regge trajectories is four.}
    \label{fig:ChewFrautschi}
\end{figure}

In the flat-space limit, the scaling dimension $\Delta(\lambda)$ in~\eqref{eqn:qnums1} can be replaced by the coefficients of the expansion~\eqref{eqn:DeltaStrongExp} or equivalently~\eqref{eqn:StrongExpCasimir}, \textit{i.e.}
\begin{align}\label{eqn:qnums2}
    [\Delta(\lambda)\;;\;\ell\;\ell\;;\;0\;0\;0] \underset{\lambda\to\infty}{\equiv
    }
    [\delta\;d_1\;d_2\;\dots
    \;;\;\ell\;\ell\;;\;0\;0\;0]
    \equiv
    [\delta\;j_1\;j_2\;\dots
    \;;\;\ell\;\ell\;;\;0\;0\;0]
    \;.
\end{align}
For a given choice of the string mass level $\delta$, the allowed values of the spin-label $\ell$ 
are $\ell \in \{ 0,2,\dots,2(\delta -1) \}$. This is neatly summarised in the Chew-Frautschi plot in Figure~\ref{fig:ChewFrautschi}. 

In general, there will be multiple operators for a given choice of $\delta$ and $\ell$. Indeed, according to the labelling~\eqref{eqn:qnums2}, these operators are distinguished by the respective higher order terms in the strong coupling expansion of their scaling dimension.
In other words, every vertex on the plot in Figure~\ref{fig:ChewFrautschi}, in general, represents more than one state. 

Enumerating the degeneracies involves a computation in representation theory that was carried out in~\cite{Alday:2023flc}. We summarise their result in the table below. We have
{
\begin{xltabular}[c]{\textwidth}{c|C|C|C|C|C|C|C}
\diagbox{$\delta$}{$\ell$} &
 0 & 2 & 4 & 6 & 8 & 10 & 12
 \\
 \midrule\midrule
 1 & 1\\
\hline
 2 & 2 & 1\\
 \hline
 3  & 6 & 4 & 1 \\
 \hline
 4 & 22 & 24 & 6 & 1 \\
 \hline
 5 & 99 & 157 & 40 & 6 & 1 \\
 \hline
 6 & 547 & 1104 & 331 & 52 & 6 & 1 \\
 \hline
 7 & 3112 & 7365 & 2570 & 461 & 58 & 6 & 1 \\
  \caption{Degeneracy of scalar ``stringy'' states with string mass level $\delta$, spin $\ell$ and $[0\ 0\ 0]$ $R$-charge  is computed in the flat-space limit \cite{Alday:2023flc}. }
  \label{tab:degTable1}
\end{xltabular}
}
The authors of~\cite{Alday:2023flc} made the following conclusions from the outcome of their enumeration. 
Firstly, they observed that the states with $\ell = 2 (\delta - 1)$ are non-degenerate, \textit{i.e.} there is only one operator for every vertex on the top-most diagonal of the Chew-Frautschi plot in Figure~\ref{fig:ChewFrautschi}.

States with $\ell = 2(\delta - 1)$ are said to 
have Regge trajectory number $n = 1$. 
They are also said to be on the ``leading'' Regge trajectory.
Thus the above passage can be summarised by the following statement: states on the leading Regge trajectory are non-degenerate. 
Similarly, states with $\ell = 2(\delta -n)$ are said to 
have Regge trajectory number $n$. They are also said to be on an $n^{\text{th}}$ Regge trajectory.

The second conclusion by the authors of~\cite{Alday:2023flc} was that for states with Regge trajectory number $2$ (those on a ``sub-leading'' Regge trajectory), \textit{i.e.} those with $\ell = 2(\delta -2)$, 
the multiplicity of states is six except for the first few operators (those with the lowest values of $\delta$). 
Thus, generically, there are six sub-leading Regge trajectories.
This is also illustrated in Figure~\ref{fig:ChewFrautschi}.

In addition to their spectral data, the operators, denoted as $\mathcal{O}_{\Delta}$ and labelled by~\eqref{eqn:qnums2}, which are exchanged under an OPE of $\mathcal{O}_2 \times \mathcal{O}_2$ of the four-point function~\eqref{eqn:def4pt} are associated with another piece of dynamical information.
This is the squared three point structure constant/OPE coefficient $\langle\mathcal{O}_2\,\mathcal{O}_2\,\mathcal{O}_\Delta\,\rangle^2$. At strong coupling, this quantity can be parameterised as~\cite{Alday:2022uxp}
\begin{align}
\label{eqn:OPEmain}
    \mathcal{C}^2_{\Delta} \equiv \langle\mathcal{O}_2\,\mathcal{O}_2\,\mathcal{O}_\Delta\,\rangle^2 \simeq \frac{\pi^3}{2^{12}} \frac{2^{-2\Delta+2\ell}(\Delta-\ell)^6}{\sin^{2}{\left(\frac{\pi (\Delta-\ell)}{2}\right)}}\frac{r_{n-1}(\delta)}{2^{2\ell}(\ell+1)}
    \bigg[
    f_0 + \frac{f_1}{\lambda^{1/4}} + \frac{f_2}{\lambda^{1/2}} 
    \bigg]
    \;,
\end{align}
where the subscript $n$ is the Regge trajectory number and $r_m$ is defined as~\cite{Alday:2022uxp}
\begin{align}\label{eqn:rndef}
    r_m(\delta) \equiv \frac{4^{2-2 \delta} \delta^{2 \delta-2 m-1}(2 \delta-2 m-1)}{\Gamma(\delta) \Gamma\left(\delta-\left\lfloor\frac{m}{2}\right\rfloor\right)}
    \;.
\end{align}
Define the combination $$\mathcal{C}^2_\Delta \bigg[\frac{\pi^3}{2^{12}} \frac{2^{-2\Delta+2\ell}(\Delta-\ell)^6}{\sin^{2}{\left(\frac{\pi (\Delta-\ell)}{2}\right)}}\frac{r_{n-1}(\delta)}{2^{2\ell}(\ell+1)}\bigg]^{-1}
=
f_0 + \frac{f_1}{\lambda^{1/4}} + \frac{f_2}{\lambda^{1/2}} + 
\dots
$$ as the reduced OPE coefficient. 
Thus,
the two pieces of dynamical information associated with 
stringy operators exchanged in the OPE of $\mathcal{O}_2\times\mathcal{O}_2$
at strong coupling are
the set of strong coupling expansion coefficients of the scaling dimensions/quadratic Casimir eigenvalues, and the reduced OPE coefficients:
\begin{align}
    \{
    f_0,f_1,\dots;d_1,d_2,\dots
    \}
    \equiv
    \{
    f_0,f_1,\dots;j_1,j_2,\dots
    \}
    \;.
    \label{eqn:CFTdata}
\end{align}

\paragraph{Constraints on CFT-data.} As mentioned earlier, the program of~\cite{Alday:2022uxp,Alday:2022xwz,Alday:2023jdk,Alday:2023mvu} involves the systematic computation of AdS curvature or $1/R^2$-corrections to the flat-space Virasoro-Shapiro Amplitude. 
A byproduct of this program are constraints on the CFT-data of stringy operators exchanged in the OPE of the dual CFT four-point function~\eqref{eqn:def4pt}. 
These
constraints involve sums over degeneracy, \textit{i.e.} over all operators with a given value of $\delta$ and $\ell$. 
At every order in $1/R^2$, various combinations of the CFT-data are constrained. They are summarised in the table below
{
\begin{xltabular}[c]{\textwidth}{C|C|C}
 \mathcal{O}(1/R^0)\text{~\cite{Alday:2022uxp} }& \mathcal{O}(1/R^2)\text{~\cite{Alday:2022xwz}} & \mathcal{O}(1/R^4)\text{~\cite{Alday:2023mvu}}
 \\
 \midrule\midrule
 \langle f_0 \rangle 
 & \langle f_0\,d_1 \rangle \text{ or } \langle f_0\,j_1 \rangle & \langle f_0\,d_1^2 \rangle \text{ or } \langle f_0\,j_1^2 \rangle
 \\
\hline
  & \langle f_2 \rangle & \langle f_0\,d_2 + f_2\,d_1 \rangle \text{ or } \langle f_0\,j_2 + f_2\,j_1 \rangle 
  \\
 \hline
  &  & \langle f_4 \rangle 
  \\
  \caption{Combinations of the CFT-data \eqref{eqn:CFTdata} constrained at given orders of  the curvature corrections of the Virasoro-Shapiro amplitude. }
  \label{tab:constrTab}
\end{xltabular}
}
At the order $\mathcal{O}(1/R^{2n})$, the combinations of the CFT-data constrained are expected to be $\{\langle f_0\,d_1^{n} \rangle, \dots, \langle f_{2n} \rangle \}$ or $\{\langle f_0\,j_1^{n} \rangle, \dots, \langle f_{2n} \rangle \}$.

Furthermore, if we sort the CFT-data into Regge trajectories, then the constraints can be written in an explicit functional form. We illustrate this on the first and second Regge trajectories below. We have, from~\cite{Alday:2022uxp}, that
\begin{align}
    \langle f_0 \rangle_{\ell = 2(\delta -1)} &= \frac{1}{\delta}\;, \label{eqn:f0t1} \\
    \langle f_0 \rangle_{\ell = 2(\delta - 2)} &= \frac{1}{3}\left(2 \delta^2+3 \delta-8\right) \;, \label{eqn:f0t2} 
\end{align}
Then, from~\cite{Alday:2022xwz}, we have
\begin{align}\label{eqn:f0j1t1}
	\langle f_0\,j_1\rangle_{\ell = 2 (\delta - 1)} &= 5\,\delta - 3
	 \;, \\
\begin{split}\label{eqn:f0j1t2}
	\langle f_0\,j_1\rangle_{\ell = 2 (\delta - 2)} &= \frac{1}{9}\, 
	\big(30\, \delta ^4+7\, \delta ^3 
    -147\, \delta ^2+212\, \delta -120\big)
	\;.
\end{split}
\end{align}
Finally, from~\cite{Alday:2023mvu}, we get
\begin{align}\label{eqn:f0j1sqt1}
	\langle f_0\,j_1^2\rangle_{\ell = 2 (\delta - 1)} &= 
     \,\delta\,\big(5\,\delta - 3\big)^2
     \;, \\
     \label{eqn:f0j1sqt2}
         \langle f_0\,j_1^2\rangle_{\ell = 2 (\delta - 2)} &= \frac{1}{27} \big(
         450 \delta ^6-465 \delta ^5-1888 \delta ^4 
         +6663 \delta ^3-9248 \delta ^2+6180 \delta -1800 \big)
         \;.
\end{align}
In all the above expressions, the notation $\langle \dots \rangle$ means that there is a sum over degeneracy of states with given $\delta$ and $\ell$.

\subsection{CFT-data and Regge trajectories}\label{sec:CFTRegge}
It is a natural question to ask if one can construct the individual Regge trajectories at a particular value of $n$. 
Doing so would mean to start at a state which lives on a particular vertex on the Chew-Frautchi plot in Figure~\ref{fig:ChewFrautschi}, and analytically continue its CFT-data as a function of $\delta$. Since $\ell = 2(\delta - n)$ on the $n^\text{th}$ Regge trajectory, this will enable us to move on a given diagonal connecting various states living on it. 
The correct analytic continuation would give us a function of $\delta$ that evaluates to the CFT-data of various states on the Regge trajectory at integer values.
These states would then be said to be on the same $n^{\text{th}}$ Regge trajectory.

In this paper, we will attempt to do precisely this for the six sub-leading Regge trajectories (Regge trajectory number $n=2$). In particular, we will try to obtain expressions for the sub-leading quadratic Casimir eigenvalue $j_1$ and the leading reduced OPE coefficient $f_0$ for these six states.

\paragraph{Structure of the spectral data.}
Spectral data on the leading and sub-leading Regge trajectories is readily available for the initial few values of the spin-label $\ell$~\cite{Gromov:2009tv,Gromov:2009zb,Roiban:2009aa,Tseytlin:2009fw,Frolov:2010wt,Passerini:2010xc,Gromov:2011de,Roiban:2011fe,Vallilo:2011fj,Basso:2011rs,Gromov:2011bz,Frolov:2012zv,Beccaria:2012xm,Gromov:2014bva,Gromov:2015wca,Hegedus:2016eop,Gromov:2023hzc}. 
We summarise it below:
\begin{xltabular}[c]{\textwidth}{C|C|C|C|C|C|C|C}
\texttt{State ID} & \delta &
 \ell & n & d_1 & j_1 & f_0 & \text{Exactly degenerate to}
 \\
 \midrule\midrule
\text{}_ 2 \text{[0 0 1 1 1 1 0 0]}_ 1 & {1} & {0} & {1} & {2} & {2} & {1\text{~\cite{Costa:2012cb}}}  &  \\
 \midrule
 \text{}_ 4 \text{[0 2 1 1 1 1 2 0]}_ 1 & {2} & {2} & {1} & {6} & {14} & \frac{1}{2} \text{~\cite{Alday:2022uxp}} &  \\
 \midrule
 \text{}_ 6 \text{[0 4 1 1 1 1 4 0]}_ 1 & {3} & {4} & {1} & {13} & {36}  & \frac{1}{3}\text{~\cite{Alday:2022uxp}} 
 & \text{} \\
 \midrule\midrule
 \text{}_ 4 \text{[0 0 2 2 2 2 0 0]}_ 1 & 2 & 0 & 2 & 2 & 2  & 0\text{~\cite{Gromov:2023hzc,Alday:2023mvu}}  & \text{} \\
 \text{}_ 4 \text{[0 0 2 2 2 2 0 0]}_ 2 & 2 & 0 & 2 & 8 & 14  & 2\text{~\cite{Gromov:2023hzc,Alday:2023mvu}} & \text{} \\
 \midrule
 \text{}_ 6 \text{[0 2 2 2 2 2 2 0]}_ 2 & 3 & 2 & 2 & 8 & 18 & {0}\text{~\cite{Julius:2023hre}} & \text{} \\
 \text{}_ 6 \text{[0 2 2 2 2 2 2 0]}_ 3 & 3 & 2 & 2 & 17 & 36 & {3}\text{~\cite{Julius:2023hre}}  & \text{} \\
 \text{}_ 6 \text{[0 2 2 2 2 2 2 0]}_ 4 & 3 & 2 & 2 & 13 & 28 & {\frac{5}{3}}\text{~\cite{Julius:2023hre}} & \text{}_ 6 \text{[0 2 2 2 2 2 2 0]}_ 5 \\
 \text{}_ 6 \text{[0 2 2 2 2 2 2 0]}_ 5 & 3 & 2 & 2 & 13 & 28 & {\frac{5}{3}}\text{~\cite{Julius:2023hre}} & \text{}_ 6 \text{[0 2 2 2 2 2 2 0]}_4 \\
 \midrule
 {}_{8}[0\;4\;2\;2\;2\;2\;4\;0]_{1} & 4 & 4 & 2 & {\color{blue}  17} & {\color{blue} 44} & \\
 {}_{8}[0\;4\;2\;2\;2\;2\;4\;0]_{4} & 4 & 4 & 2 & {\color{blue}  29} & {\color{blue} 68} & \\
{}_{8}[0\;4\;2\;2\;2\;2\;4\;0]_{13} & 4 & 4 & 2 &&&& {}_{8}[0\;4\;2\;2\;2\;2\;4\;0]_{14} \\
{}_{8}[0\;4\;2\;2\;2\;2\;4\;0]_{14} & 4 & 4 & 2 &&&& {}_{8}[0\;4\;2\;2\;2\;2\;4\;0]_{13} \\
{}_{8}[0\;4\;2\;2\;2\;2\;4\;0]_{15} & 4 & 4 & 2 &&&& {}_{8}[0\;4\;2\;2\;2\;2\;4\;0]_{16} \\
{}_{8}[0\;4\;2\;2\;2\;2\;4\;0]_{16} & 4 & 4 & 2 &&&& {}_{8}[0\;4\;2\;2\;2\;2\;4\;0]_{15} \\
 \caption{
    \label{tab:EvenReggeStates} 
Perturbative CFT-data for the states on leading and sub-leading even-spin Regge trajectories.
For every state, we display its \texttt{State ID}, which is a unique identifier of a given state, introduced in~\cite{Gromov:2023hzc} and reviewed in Appendix~\ref{adx:qscRegge}. We also display the string mass level $\delta$, spin $\ell$, Regge trajectory number, sub-sub-leading dimension $d_1$ from~\cite{Gromov:2023hzc}, sub-leading Casimir eigenvalue $j_1$ from~\cite{Gromov:2023hzc}, and operators which are exactly degenerate (see~\cite{Gromov:2023hzc}) to the state. Finally we present the strong coupling expansion coefficients of the OPE coefficient of the state. We have added references to the available results in the literature. 
New results obtained by the method described in Appendix~\ref{adx:qscRegge} are coloured blue.
}
\end{xltabular}
Consider first the leading Regge trajectory, 
\textit{i.e.} states with $\ell = 2(\delta - 1)$.
A prediction for $j_1(\delta)$ can be extracted from~\cite{Basso:2011rs}. We get
\begin{align}\label{eqn:j1Reg1}
    j_{1}(\delta) = 5\,\delta^2 - 3\,\delta\;.
\end{align}
Now let us look at the six sub-leading Regge trajectories. 
There are only two states with $\ell = 0$. Thus, presumably two out of the six sub-leading Regge trajectores begin here. 
At $\ell = 2$, there are four states. Two out of these four states should be on the Regge trajectories identified at $\ell = 0$, and the other two should begin here. 
Finally, the last two  Regge trajectories should begin at $\ell = 4$.

Our strategy to identify states on Regge trajectories is to consider the integrability-based quantum spectral curve description of a particular candidate state \cite{Gromov:2014caa, Gromov:2015wca, Gromov:2023hzc}, and analytically continue its spin-label $\ell$ to recover the spectral data of other predicted states on the candidate Regge trajectory \cite{Alfimov:2018cms}. We present the details in Appendix~\ref{adx:qscRegge}.

Before we continue, there is an important detail in the spectral information to point out. 
This is the fact that due to some parity symmetries of the underlying integrable system (see~\cite{Gromov:2023hzc} for the details), some states in the spectrum are \emph{exactly degenerate}, \textit{i.e.} their scaling dimension is exactly the same non-perturbatively, and to all orders in perturbation theory. Examples of such states are those with \texttt{State ID} ${}_ 6 {[0 2 2 2 2 2 2 0]}_ {4/5}$.
All exactly degenerate states discussed in this paper are of ``Type II'' in the notations of~\cite{Gromov:2023hzc}, whereas the non-degenerate states are of ``Type I'' (see~\cite{Gromov:2023hzc} for the details).
The act of varying the spin-label $\ell$ along a Regge trajectory should not change anything about the underlying integrability description~\cite{Gromov:2017blm,Alfimov:2018cms}. Thus we can expect that all states on a Regge trajectory have the same integrability properties. In particular, this means that if two states are exactly degenerate, then all the states on their respective Regge trajectories must be exactly degenerate, \textit{i.e.} the entire Regge trajectories are exactly degenerate. 

Since the states ${}_ 6 {[0 2 2 2 2 2 2 0]}_ {4/5}$ are exactly degenerate, 
the third/fourth sub-leading Regge trajectories are exactly degenerate. 
Similarly, we also find that the fifth/sixth sub-leading Regge trajectories are also exactly degenerate. 
Since there is no way to distinguish exactly degenerate states, their CFT-data should also be indistinguishable.
Thus the effective multiplicity of the sub-leading Regge trajectories gets modified, and these are summarised 
in the table below:
\begin{xltabular}[c]{\textwidth}{C|C|C|C}
 \delta & \ell = 2(\delta - 2) & \text{Multiplicity} &
 \text{Unique} 
 \\
 \midrule\midrule
 2  & 0 & 2 & 2 \\
 3  & 2 & 4 & 3 \\
 4  & 4 & 6 & 4 \\
 5  & 6 & 6 & 4 \\
 6  & 8 & 6 & 4 \\
 7  & 10 & 6 & 4  \\
 \caption{
    \label{tab:deg} 
    Multiplicity of states on the sub-leading Regge trajectories. As found by~\cite{Alday:2023flc}, the multiplicity of the sub-leading Regge trajectories saturates at six. However, due to exact degeneracies of the underlying integrable system, the effective multiplicity is four, as this is the number of unique sub-leading Regge trajectories.
    }
\end{xltabular}
Thus there are only \emph{four} unique sub-leading Regge trajectories. 
Denote the sub-leading coefficient of the quadratic Casimir eigenvalue of states on the first, second, third/forth and fifth/sixth sub-leading Regge trajectories as $j_{1;1}$, $j_{1;2}$, $j_{1;3}$ and $j_{1;4}$ respectively. 
To parameterise these quantities as a function of $\delta$, let us consider the analysis done in~\cite{Beccaria:2012xm}. In that paper, the 2D marginality condition on the world-sheet was used to argue that the strong coupling expansion of the square of the scaling dimension of a state with string mass level $\delta$ and $R$-charge $p$ goes as
\begin{align}
    \Delta^2 \simeq 4\,\delta\,\sqrt{\lambda} + p^2 + a\,\delta^2 + b\,\delta
    \;,
\end{align}
up to shifts of $p$ and $\Delta$ corresponding to the position of the state in the supermultiplet. Here $a$ and $b$ are undetermined coefficients. 
Therefore the most general form of $\delta$-dependence that we \emph{assume} for the $j_1$ 
is
\begin{align}\label{eqn:j1Anz}
    j_{1;m} = a_m\,\delta^2 + b_m\,\delta + c_m
    \;.
\end{align}
For the first two sub-leading Regge trajectories, using the method described in Appendix~\ref{adx:qscRegge}, we obtain
\begin{align}
    \label{eqn:j11pred}
    j_{1;1} &= 5\,\delta^2 - 3\,\delta\;, \\
    \label{eqn:j12pred}
    j_{1;2} &= 5\,\delta^2 - 9\,\delta\;.
\end{align}
We are unable to fix the $\delta$-dependence of $j_{1;3}$ and $j_{1;4}$ with the current precision of our integrability-based data.

Finally, we denote the leading order reduced OPE coefficient on the four unique sub-leading Regge trajectories as $f_{0;m}$, $m = 1,\dots,4$.
Due to the exact degeneracy, the states corresponding to the third/fourth and fifth/sixth sub-leading Regge trajectories will always come with a factor of two. This means for example that a constraint of the form $\langle f_0\,j_1^n \rangle$ is expanded as
\begin{align}
    \langle f_0\,j_1^n \rangle_{\ell = 2\,(\delta - 2)}
    = f_{0;1}\,j_{1;1}^n + f_{0;2}\,j_{1;2}^n
    + 2\,f_{0;3}\,j_{1;3}^n + 2\,f_{0;4}\,j_{1;4}^n
    \;.
\end{align}
This completes the setup of our problem. 
We want to obtain expressions for $j_{1;3}$, $j_{1;4}$, $f_{0;1}$, $f_{0;2}$, $f_{0;3}$ and $f_{0;4}$ as a function of $\delta$.
Doing this would allow us to completely characterise the leading non-trivial order CFT-data on all six (four unique) sub-leading Regge trajectories. 
In order to do so, we solve the constraint~\eqref{eqn:f0t2} on $\langle f_0 \rangle$ from~\cite{Alday:2022uxp}, the constraint~\eqref{eqn:f0j1t2} on $\langle f_0\,j_1 \rangle$ from~\cite{Alday:2022xwz} and the constraint~\eqref{eqn:f0j1sqt2}) on $\langle f_0\,j_1^2 \rangle$ from~\cite{Alday:2023mvu}.
We describe our solution in the next section. 

\section{Solution}
\label{sec:prelim}

In this section, we combine various methods and inputs, namely analysis at large string mass level $\delta$, the structure of higher order constraints, a stipulation based on the analyticity of CFT-data in spin, as well as already known results to obtain expressions for $j_{1;3}$, $j_{1;4}$, $f_{0;1}$, $f_{0;2}$, $f_{0;3}$ and $f_{0;4}$ as a function of $\delta$, given in terms of as few free parameters as possible.   
It will also be crucial in our analysis, to impose that the leading order reduced OPE coefficients $f_{0;m}$ are non-negative, as this follows from the fact that the squared OPE coefficient $\mathcal{C}_\Delta^2$~\eqref{eqn:OPEmain} of a unitary theory must be non-negative.

\subsection{Analysis at large string mass level}
\label{sec:largeDeltaAnalysis}

Let us begin by considering the limit of large string mass level, \textit{i.e.} large-$\delta$.
Notice that under a large order in $\delta$ expansion, the leading-$\delta$ term of any $f_{0;m}$ cannot be higher than $\mathcal{O}(\delta^2)$. 
To see this consider the following argument.
The highest power of $\delta$ in the constraint \eqref{eqn:f0t2}: $$\langle f_0\rangle_{\ell = 2(\delta-2)} = f_{0;1} + f_{0;2} + 2\,f_{0;3} + 2\,f_{0;4} \sim \mathcal{O}(\delta^2)\;,$$ is $\delta^2$.
Wlog, suppose that one of the leading OPE coefficients, say $f_{0;m}$ goes as $\delta^3$ at large-$\delta$. 
This would mean that in order to still satisfy the constraint \eqref{eqn:f0t2}, there should be leading OPE coefficients, say $f_{0;m^\prime}$, $f_{0;m^{\prime\prime}}$ \textit{etc.}, which also go as $\delta^3$ at large-$\delta$, with the respective $\delta^3$ terms of these constrained in such a way that their sum adds up to exactly the negative of the $\delta^3$ coefficient of $f_{0;m}$.
It is easy to see that this implies that the $\delta^3$ term of least one of the leading OPE coefficients, say $f_{0;m^\prime}$, will be negative.
This contradicts the fact that all leading OPE coefficients must be non-negative as there will be a critical value of $\delta$, say $\delta^{*}$, so that for $\delta>\delta^{*}$, $f_{0;m^\prime} < 0$.
Therefore, the highest power possible in a large-$\delta$ expansion of the leading OPE coefficients must be $\delta^2$.
Thus, let us impose the following ansatz for the large-$\delta$ expansion of $f_{0;m}$. We have
\begin{equation}
\label{eqn:f0LargeDeltaExpn}
    f_{0;m} \underset{\delta\to\infty}{=} B_{2;m} \,\delta^2 + B_{1;m} \,\delta + B_{0;m} + \mathcal{O}\left(\frac{1}{\delta}\right) \;.
\end{equation}
We also know the $\delta$-dependence of $j_{1;1}$ \eqref{eqn:j11pred} and $j_{1;2}$ \eqref{eqn:j12pred}.
For $j_{1;3}$ and $j_{1;4}$, we use the ansatz \eqref{eqn:j1Anz}.
Plugging this and the large-$\delta$ ansatz for $f_{0;2}$ from~\eqref{eqn:f0LargeDeltaExpn}, into the constraints \eqref{eqn:f0t2}, \eqref{eqn:f0j1t2}, \eqref{eqn:f0j1sqt2}, we can solve for $f_{0;1}$, $f_{0;3}$ and $f_{0;4}$ in the large-$\delta$ limit. The solution will therefore be in terms of the parameters
\begin{align}
    \{ \{a_3, b_3, c_3 \}, \{a_4, b_4, c_4 \}, \{B_{2;2}, B_{1;2}, B_{0;2}, \dots \} \}
    \;,
\end{align}
where the ellipsis in the last set of parameters represents higher order terms in the ansatz~\eqref{eqn:f0LargeDeltaExpn}.

Our solution at the leading order at large-$\delta$ takes the form
\begin{align}
\begin{split}
f_{0;1} &\simeq -\frac{1}{3} \left(3 B_{2;2}-2\right) \delta ^2\;,\\
f_{0;2} &\simeq B_{2;2}\, \delta ^2\;,\\
f_{0;3} &\simeq -\frac{\left(27 B_{2;2}-10\right) \left(a_4-5\right) \delta }{9 \left(a_3-5\right) \left(a_3-a_4\right)}\;,\\
f_{0;4} &\simeq \frac{\left(27 B_{2;2}-10\right) \left(a_3-5\right) \delta }{9  \left(a_4-5\right) \left(a_3-a_4\right)}\;.
\end{split}
\end{align}
\paragraph{Case 0:} Let us first consider the case that $a_3 \neq a_4$, $a_3 \neq 5$ and $a_4  \neq 5$. 
Imposing non-negativity, we get some ranges of allowed values of the parameters, along with the stipulation that $B_{2;2} = \frac{10}{27}$. Plugging the latter in, we get
\begin{align}
\begin{split}
f_{0;1} &\simeq \frac{8}{27} \delta^2\;, \\
f_{0;2} &\simeq \frac{10}{27} \delta^2\;, \\
f_{0;3} &\simeq -\frac{81 a_4 B_{1;2}-405 B_{1;2}+80}{27 \left(a_3-5\right) \left(a_3-a_4\right)} \delta\;,\\
f_{0;4} &\simeq \frac{81 a_3 B_{1;2}-405 B_{1;2}+80}{27 \left(a_4-5\right) \left(a_3-a_4\right)} \delta\;.
\end{split}
\end{align}
However, the non-negativity conditions cannot be solved for any values of $a_3$, $a_4$ and $B_{1;2}$. 
Thus, the solution is inconsistent.
This means that we need to relax the requirements of $a_3 \neq a_4$, $a_3 \neq 5$ and $a_4 \neq 5$.
There are then four cases of how this can be done: 
$a_3 = a_4 = 5$,
$a_3 = 5 \neq a_4 $, 
$a_3 \neq 5 = a_4$
and 
$a_3 = a_4 \neq 5$.
Let us go case by case and see where they lead us.

\paragraph{Case 1: $a_3 = a_4 = 5$.}
The leading terms at large-$\delta$ are
\begin{align}
\begin{split}
f_{0;1}&\simeq-\frac{\left(243 B_{2;2} b_3+27 B_{2;2} b_3 b_4+243 B_{2;2} b_4+2187 B_{2;2}-114 b_3-18 b_3 b_4-114 b_4-722\right)}{27 \left(b_3+3\right) \left(b_4+3\right)}\delta ^2\;,\\
f_{0;2}&\simeq B_{2;2}\, \delta ^2\;,\\
f_{0;3}&\simeq-\frac{ \left(81 B_{2;2} b_4+729 B_{2;2}-30 b_4-190\right)}{27 \left(b_3+3\right) \left(b_3-b_4\right)}\delta ^2\;,\\
f_{0;4}&\simeq\frac{\left(81 B_{2;2} b_3+729 B_{2;2}-30 b_3-190\right)}{27 \left(b_4+3\right)\left(b_3-b_4\right) } \delta ^2.
\end{split}
\end{align}
Let us first consider $b_3 \neq -3$, $b_4 \neq -3$ and $b_3\neq b_4$. Then, we get
\begin{align}
\begin{split}
f_{0;1} &\simeq\frac{2 \left(3 b_3+19\right) \left(3 b_4+19\right) \delta ^2}{27 \left(b_3+3\right)\left(b_4+3\right)}\;,\\
f_{0;2} &\simeq B_{1;2}\,\delta\;,\\
f_{0;3} &\simeq \frac{10 \left(3 b_4+19\right) \delta ^2}{27 \left(b_3+3\right) \left(b_3-b_4\right)}\;,\\
f_{0;4} &\simeq -\frac{10 \left(3 b_3+19\right) \delta ^2}{27 \left(b_4+3\right)\left(b_3-b_4\right)}\;. 
\end{split}
\end{align}
In this sub-case, non-negativity sets either $b_3 = -\frac{19}{3}\neq b_4$, which gives
\begin{align}
\begin{split}
f_{0;1} &\simeq\frac{\left(-3 b_4 c_3+12 b_4 B_{1;2}+108 B_{1;2}+15 b_4-19 c_3+35\right) \delta}{15 (b_4 + 3)},\\
f_{0;2} &\simeq B_{1;2}\,\delta\;,\\
f_{0;3} &\simeq \frac{\delta^2}{3}\;,\\
f_{0;4} &\simeq -\frac{2\left(36 B_{1;2}-5 c_3-5\right) \delta}{3 \left(b_4+3\right) \left(3 b_4+19\right)}
\end{split}
\end{align}
or $b_3 \neq -\frac{19}{3} = b_4$, which gives
\begin{align}
\begin{split}
f_{0;1} &\simeq\frac{\left(-3 b_3 c_4+12 b_3 B_{1;2}+108 B_{1;2}+15 b_3-19 c_4+35\right) \delta}{15 (b_3 +3)},\\
f_{0;2} &\simeq B_{1;2}\,\delta\;,\\
f_{0;3} &\simeq -\frac{2\left(36 B_{1;2}-5 c_4-5\right) \delta}{3 \left(b_3+3\right) \left(3 b_3+19\right)}\;,\\
f_{0;4} &\simeq \frac{\delta^2}{3}\;.\\
\end{split}
\end{align}
The next two sub-cases are $b_3 = -3 \neq b_4$ and $b_3 \neq -3 = b_4$. For the former sub-case, from non-negativity we get
$b_4 = -\frac{19}{3}$, $c_3 \neq 0$, $c_4 = \frac{36 B_{1;2} -5}{5}$
with
\begin{align}
\begin{split}
    f_{0;1} &\simeq -\frac{\left(16 B_{1;2} c_3-324 B_{1;2}^2+45 B_{1;2}-120 B_{0;2}-30 c_3-300\right)}{25 c_3}\delta \;,\\
    f_{0;2} &\simeq B_{1;2}\, \delta \;,\\
    f_{0;3} &\simeq -\frac{3\left(108 B_{1;2}^2-15 B_{1;2}+40 B_{0;2}+100\right)}{50 c_3}\delta \;,\\
    f_{0;4} &\simeq \frac{\delta ^2}{3}\;.
\end{split}
\end{align}
For the latter sub-case, we get from non-negativity that $b_3 = -\frac{19}{3}$, $c_3 = \frac{36 B_{1;2}-5}{5}$, $c_4 \neq 0$ giving
\begin{align}
\begin{split}
	f_{0;1} &\simeq -\frac{\left(16 B_{1;2} c_4-324 B_{1;2}^2+45 B_{1;2}-120 B_{0;2}-30 c_4-300\right)}{25 c_4}\delta  \;,\\
    f_{0;2} &\simeq B_{1;2}\, \delta \;,\\
    f_{0;3} &\simeq \frac{\delta ^2}{3}\;,\\
    f_{0;4} &\simeq -\frac{3 \left(108 B_{1;2}^2-15 B_{1;2}+40 B_{0;2}+100\right)}{50 c_4}\delta\;.
\end{split}
\end{align}
Finally, let us set $b_3 = b_4$. The only consistent solution is $b_3 = b_4 = -\frac{19}{3}$, $c_3 \neq c_4$ with the following leading order behaviour: 
\begin{align}
\begin{split}
	f_{0;1} &\simeq (\frac{6}{5} - \frac{16}{25}B_{1;2}) \delta \;,\\
	f_{0;2} &\simeq B_{1;2}\, \delta\;,\\
	f_{0;3} &\simeq \frac{\left(36 B_{1;2}-5 c_4-5\right)}{15 \left(c_3-c_4\right)} \delta ^2\;,\\
	f_{0;4} &\simeq -\frac{ \left(36 B_{1;2}-5 c_3-5\right)}{15 \left(c_3-c_4\right)} \delta ^2\;.
\end{split}
\end{align}

\paragraph{Case 2: $a_3 = 5\neq a_4$.}
In this case the leading order terms are 
\begin{align}
\begin{split}
f_{0;1} &\simeq -\frac{\left(9 B_{2;2} b_3+81 B_{2;2}-6 b_3-38\right)}{9 \left(b_3+3\right)}\delta ^2 \;,\\
f_{0;2} &\simeq B_{2;2}\, \delta ^2\;,\\
f_{0;3} &\simeq \frac{\left(27 B_{2;2}-10\right) }{9 \left(b_3+3\right)}\delta ^2\;,\\
f_{0;4} &\simeq  -\frac{81 B_{2;2} b_3+729 B_{2;2}-30 b_3-190}{27 \left(a_4-5\right)^2}\;.
\end{split}
\end{align}
Let us first consider $b_3 \neq -3$. Non-negativity gives us that $B_{2;2} = 0$ and $b_3 = -\frac{19}{3}$,. Therefore, we have
\begin{align}
\begin{split}
f_{0;1} &\simeq \left( 1 + \frac{4 B_{1;2}-c_3}{5} \right)\delta \;,\\
f_{0;2} &\simeq B_{1;2}\, \delta\;,\\
f_{0;3} &\simeq \frac{1}{3} \delta^2\;,\\
f_{0;4} &\simeq -\frac{2 \left(36 B_{1;2}-5 c_3-5\right)}{9 \left(a_4-5\right)^2 \delta}\;. 
\end{split}
\end{align}
Now let us consider $b_3 = -3$. We get
\begin{align}
  f_{0;1} \simeq  -\frac{2 \left(27 B_{2;2}-10\right) \delta ^3}{9 c_3}\;,\quad 
  f_{0;3} \simeq  \frac{\left(27 B_{2;2}-10\right) \delta ^3}{9 c_3}\;,
\end{align}
and then we get $B_{2;2} =\frac{10}{27}$ from non-negativity. 
Continuing the analysis leads to an us not being able to solve the non-negativity condition for the obtained set of parameters. 

\paragraph{Case 3: $a_3 \neq 5 = a_4$.}
This case is analogous to \textbf{Case 2}. For the leading terms, we get
\begin{align}
\begin{split}
f_{0;1} &\simeq -\frac{ \left(9 B_{2;2} b_4+81 B_{2;2}-6 b_4-38\right)}{9 \left(b_4+3\right)}\delta ^2\;,\\
f_{0;2} &\simeq B_{2;2}\, \delta ^2\;,\\
f_{0;3} &\simeq  -\frac{81 B_{2;2} b_4+729 B_{2;2}-30 b_4-190}{27 \left(a_3-5\right)^2},\\
f_{0;4} &\simeq \frac{\left(27 B_{2;2}-10\right)}{9 \left(b_4+3\right)}\delta ^2\;.
\end{split}
\end{align}
and for $b_4 \neq -3$ non-negativity gives $B_{2;2} = 0$, $b_4 = -\frac{19}{3}$, which implies

\begin{align}
\begin{split}
f_{0;1} &\simeq \left( 1 + \frac{4 B_{1;2}-c_4}{5} \right)\delta \;,\\
f_{0;2} &\simeq B_{1;2}\, \delta\;,\\
f_{0;3} &\simeq -\frac{2 \left(36 B_{1;2}-5 c_4-5\right)}{9 \left(a_3-5\right)^2 \delta}\;,\\
f_{0;4} &\simeq \frac{1}{3} \delta^2\;,
\end{split}
\end{align}
being the only consistent solution.

\paragraph{Case 4: $a_3 = a_4 \neq 5$.} 
The leading terms are as follows. We have
\begin{align}
\begin{split}
f_{0;1} &\simeq -\frac{1}{3} \left(3 B_{2;2}-2\right) \delta ^2\;,\\
f_{0;2} &\simeq B_{2;2} \delta ^2\;,\\
f_{0;3} &\simeq-\frac{\left(27 B_{2;2}-10\right) \delta ^2}{9 \left(b_3-b_4\right)}\;,\\
f_{0;4} &\simeq \frac{\left(27 B_{2;2}-10\right) \delta ^2}{9 \left(b_3-b_4\right)}\;.
\end{split}
\end{align}
We go both paths  $b_3 \neq b_4$ and $b_3 = b_4$, and we cannot satisfy the non-negativity condition for any parameters. Therefore, we discard this case.

\paragraph{Summary:}
Putting everything together, we the allowed values of the parameters are summarised in the table below. We have
\setlength\LTleft{-16mm}
\setlength\LTright\fill
\begin{xltabular}{\textwidth}{C|C|C|C|C|C|C|C|C|C|C}
\text{Case} &
 a_3 & b_3 & c_3 &
 a_4 & b_4 & c_4 &
 \mathcal{O}(f_{0;1}) &
 \mathcal{O}(f_{0;2}) &
 \mathcal{O}(f_{0;3}) &
 \mathcal{O}(f_{0;4}) 
 \\
 \midrule\midrule
 \rule{0pt}{3.5ex} 1.1 & 5 & -\frac{19}{3} & & 5 & \neq \{-\frac{19}{3},-3\}& & \delta & \delta & \delta^2 & \delta\\[1ex]
\hline
 \rule{0pt}{3.5ex} 1.2 & 5 & \neq \{-\frac{19}{3},-3\} & & 5 & -\frac{19}{3}& & \delta & \delta & \delta & \delta^2  \\[1ex]
 \hline
\rule{0pt}{3.5ex} 1.3 & 5 & -3 & \neq 0 & 5 & -\frac{19}{3}& \frac{36 B_{1;2} - 5}{5}& \delta & \delta & \delta & \delta^2 \\[1ex]
\hline
\rule{0pt}{3.5ex} 1.4 & 5 & -\frac{19}{3} & \frac{36 B_{1;2} - 5}{5} & 5 & -3 & \neq 0 & \delta & \delta & \delta^2 & \delta \\[1ex]
\hline
\rule{0pt}{3.5ex} 1.5 & 5 & -\frac{19}{3} & \neq c_4 & 5 & -\frac{19}{3}& \neq c_3 & \delta & \delta & \delta^2 & \delta^2 \\[1ex]
 \hline
 \rule{0pt}{3.5ex} 2 & 5 & -\frac{19}{3} & & \neq 5 & & &\delta & \delta & \delta^2 & \frac{1}{\delta} \\[1ex]
 \hline
 \rule{0pt}{3.5ex} 3 & \neq 5 & & & 5 & -\frac{19}{3}& &\delta & \delta & \frac{1}{\delta} & {\delta^2}  \\[1ex]
  \caption{
  Outcome of large-$\delta$ analysis. We tabulate the various cases that are consistent with the non-negativity of the leading order term in a large-$\delta$ expansion.
  }
  \label{tab:AsymptoticSummary}
\end{xltabular}
In the sequel, we will input more information to nail down the CFT-data by obtaining predictions for the reduced OPE coefficients $f_{0;m}$, as well as the parameters that characterise $j_{1;3}: \{a_3, b_3, c_3 \},$ and $j_{1;4}: \{a_4, b_4, c_4 \}$. In doing so, will always proceed case by case and consider each time, the seven cases described above.
\subsection{Another constraint on the CFT-data}
In the previous section, we solved three constraints on the CFT-data. These were respectively of the form: $\langle f_0 \rangle$  \eqref{eqn:f0t2}, $\langle f_0 \, j_1\rangle$ \eqref{eqn:f0j1t2}, and $\langle f_0 \, j_1^2\rangle$ \eqref{eqn:f0j1sqt2}. Since we have four reduced OPE coefficients $f_{0;m}$, $m = 1,\dots,4$, it would really help us if we had another constraint, so that we would have a system of four equations (constraints on CFT-data) on four variables (reduced OPE coefficients). As discussed in section \ref{sec:setup}, a constraint of the form $\langle f_0 \, j_1^n\rangle$ is generated by the order-$1/R^{2n}$ correction to the flat-space Virasoro Shapiro amplitude, when one applies the methods of \cite{Alday:2023mvu}.
The current state of the art is at order-$1/R^4$. However, let us imagine that we had access to further curvature corrections and consequently further constraints on the CFT-data of the form $\langle f_0 \, j_1^n\rangle$. In particular, let us parameterise the constraints that stem from the next curvature correction as
\begin{align}
    \label{eqn:f0j1p3t2}
    \langle f_0 \, j_1^3\rangle = &\sum_{n = 0}^{8} \alpha_n\,\delta^n\;.
\end{align}
Here, $\alpha_n$, 
are so far undetermined coefficients. The polynomial form of the above expression has been \emph{assumed} in analogy with the structure of hitherto known constraints. 
The degree of the polynomial follows from the fact that $f_{0;m}$ can be at most $\mathcal{O}(\delta^2)$ and $j_{1;m}$ are assumed to have degree two in~\eqref{eqn:j1Anz}.

We will include equation~\eqref{eqn:f0j1p3t2} to our existing three constraints \eqref{eqn:f0t2}, \eqref{eqn:f0j1t2}, \eqref{eqn:f0j1sqt2}. Then, we can solve for all four $f_{0;m}$, in terms of $j_{1;m}$, $m = 1,\dots,4$ and $\alpha_n$, $n = 1,8$. Doing so, we get
\begin{align}
    f_{0;m} &= \frac{F_m}{\prod_{\substack{n \neq m\\n =  1}}^{4} (j_{1;m} - j_{1;n})} \;,
    \label{eqn:f0withnewconst}
\end{align}
with
\begin{align}
    \begin{split}
    \label{eqn:F1def}
        F_1 &=
        \Bigg[
        \left(-\frac{80}{9} a_3 b_3+\frac{16}{9} a_3 a_4 b_3+\alpha _8-\frac{250}{3}\right) \delta ^8
        \\&+\left(6 a_3 b_3 a_4-\frac{80 a_4}{9}-\frac{506 a_3 b_3}{27}+\frac{16}{9} a_3 b_3 b_4+\alpha _7+\frac{2125}{9}\right) \delta ^7
        \\&+\left(-\frac{4}{9} a_3 b_3 a_4-\frac{506 a_4}{27}+18 a_3 b_3+6 a_3 b_3 b_4-\frac{80 b_4}{9}+\frac{16}{9} a_3 b_3 c_4+\alpha _6+\frac{5255}{27}\right) \delta ^6
        \\&+\left(-\frac{40}{3} a_3 b_3 a_4+\frac{16 c_3 a_4}{9}+18 a_4+\frac{1724 a_3 b_3}{27}-\frac{4}{9} a_3 b_3 b_4-\frac{506 b_4}{27}-\frac{80 c_3}{9} \right.\\&+\left.6 a_3 b_3 c_4-\frac{80 c_4}{9}+\alpha _5-\frac{16769}{9}\right) \delta ^5+\left(6 c_3 a_4+\frac{1724 a_4}{27}-\frac{980 a_3 b_3}{9}-\frac{40}{3} a_3 b_3 b_4\right.\\&+\left.18 b_4+\frac{16 b_4 c_3}{9}-\frac{506 c_3}{27}-\frac{4}{9} a_3 b_3 c_4-\frac{506 c_4}{27}+\alpha _4+\frac{106207}{27}\right) \delta ^4
        \\&+\left(-\frac{4}{9} c_3 a_4-\frac{980 a_4}{9}+\frac{200 a_3 b_3}{3}+\frac{1724 b_4}{27}+6 b_4 c_3+18 c_3-\frac{40}{3} a_3 b_3 c_4+\frac{16 c_3 c_4}{9}\right.\\&+\left.18 c_4+\alpha _3-\frac{38044}{9}\right) \delta ^3+\left(-\frac{40}{3} c_3 a_4+\frac{200 a_4}{3}-\frac{980 b_4}{9}-\frac{4 b_4 c_3}{9}+\frac{1724 c_3}{27}\right.\\&+\left.6 c_3 c_4+\frac{1724 c_4}{27}+\alpha _2+\frac{7180}{3}\right) \delta ^2+\left(-\frac{40}{3} c_3 b_4+\frac{200 b_4}{3}-\frac{980 c_3}{9}-\frac{4 c_3 c_4}{9}\right.\\&-\left.\frac{980 c_4}{9}+\alpha _1-600\right) \delta +\frac{200 c_3}{3}-\frac{40 c_3 c_4}{3}+\frac{200 c_4}{3}+\alpha _0
        \Bigg]
        \;,
    \end{split}
\end{align}
\begin{align*}
        F_2 &= \Bigg[
        \left(\frac{100 a_3 b_3}{9}-\frac{20}{9} a_3 a_4 b_3+\alpha _8-\frac{250}{3}\right) \delta ^8
        \\ 
        &+\left(\frac{100 a_4}{9}-\frac{380 a_3 b_3}{27}-\frac{20}{9} a_3 b_3 b_4+\alpha _7+\frac{1225}{9}\right) \delta ^7 
        \\&+\left(\frac{140}{9} a_3 b_3 a_4-\frac{380 a_4}{27}-80 a_3 b_3+\frac{100 b_4}{9}-\frac{20}{9} a_3 b_3 c_4+\alpha _6+\frac{8045}{27}\right) \delta ^6\\&+\left(-\frac{40}{3} a_3 b_3 a_4-\frac{20 c_3 a_4}{9}-80 a_4+\frac{5540 a_3 b_3}{27}+\frac{140}{9} a_3 b_3 b_4 \right.\\&-\left. \frac{380 b_4}{27}+\frac{100 c_3}{9}+\frac{100 c_4}{9}\alpha _5-\frac{4331}{3}\right) \delta ^5
        \tag{\stepcounter{equation}\theequation}
        \\&+\left(\frac{5540 a_4}{27}-\frac{1700 a_3 b_3}{9}-\frac{40}{3} a_3 b_3 b_4-80 b_4-\frac{20 b_4 c_3}{9}-\frac{380 c_3}{27}\right.\\&+\left.\frac{140}{9} a_3 b_3 c_4-\frac{380 c_4}{27}+\alpha _4+\frac{66229}{27}\right) \delta ^4\\&+\left(\frac{140 c_3 a_4}{9}-\frac{1700 a_4}{9}+\frac{200 a_3 b_3}{3}+\frac{5540 b_4}{27}-80 c_3-\frac{40}{3} a_3 b_3 c_4-\frac{20 c_3 c_4}{9} \right.\\&-\left. 80 c_4+\alpha _3-2172\right) \delta ^3+\left(-\frac{40}{3} c_3 a_4+\frac{200 a_4}{3}-\frac{1700 b_4}{9}+\frac{140 b_4 c_3}{9}\right.\\&+\left.\frac{5540 c_3}{27}+\frac{5540 c_4}{27}+\alpha _2+1020\right) \delta ^2+\left(-\frac{40}{3} c_3 b_4+\frac{200 b_4}{3}-\frac{1700 c_3}{9}+\frac{140 c_3 c_4}{9}\right.\\&-\left.\frac{1700 c_4}{9}+\alpha _1-200\right) \delta +\frac{200 c_3}{3}-\frac{40 c_3 c_4}{3}+\frac{200 c_4}{3}+\alpha _0
        \Bigg]\;,
\end{align*}
\begin{align}
    \label{eqn:F3def}
    \begin{split}
        F_3 &= \Bigg[
        \left(\frac{\alpha _8}{2}-\frac{125}{3}\right) \delta ^8+\left(\frac{\alpha _7}{2}+\frac{575}{6}\right) \delta ^7+\left(\frac{80 a_4}{27}+\frac{\alpha _6}{2}+\frac{3445}{54}\right) \delta ^6\\&+\left(-\frac{10 a_4}{9}+\frac{80 b_4}{27}+\frac{\alpha _5}{2}-\frac{1717}{2}\right) \delta ^5+\left(-\frac{20 a_4}{27}-\frac{10 b_4}{9}+\frac{80 c_4}{27}+\frac{\alpha _4}{2}+\frac{113369}{54}\right) \delta ^4\\&+\left(-\frac{310 a_4}{9}-\frac{20 b_4}{27}-\frac{10 c_4}{9}+\frac{\alpha _3}{2}-\frac{22334}{9}\right) \delta ^3
        \\&+\left(\frac{100 a_4}{3}-\frac{310 b_4}{9}-\frac{20 c_4}{27}+\frac{\alpha _2}{2}+\frac{4580}{3}\right) \delta ^2\\&+\left(\frac{100 b_4}{3}-\frac{310 c_4}{9}+\frac{\alpha _1}{2}-400\right) \delta +\frac{100 c_4}{3}+\frac{\alpha _0}{2}
        \Bigg]\;.
    \end{split}
\end{align}
\begin{align*}
        F_4 &= \Bigg[
        \left(\frac{\alpha _8}{2}-\frac{125}{3}\right) \delta ^8+\left(\frac{80 a_3 b_3}{27}+\frac{\alpha _7}{2}+\frac{575}{6}\right) \delta ^7+\left(-\frac{10}{9} a_3 b_3+\frac{\alpha _6}{2}+\frac{3445}{54}\right) \delta ^6\\&+\left(-\frac{20}{27} a_3 b_3+\frac{\alpha _5}{2}-\frac{1717}{2}\right) \delta ^5+\left(-\frac{310}{9} a_3 b_3+\frac{80 c_3}{27}+\frac{\alpha _4}{2}+\frac{113369}{54}\right) \delta ^4\\&+\left(\frac{100 a_3 b_3}{3}-\frac{10 c_3}{9}+\frac{\alpha _3}{2}-\frac{22334}{9}\right) \delta ^3+\left(-\frac{20 c_3}{27}+\frac{\alpha _2}{2}+\frac{4580}{3}\right) \delta ^2\\&+\left(-\frac{310 c_3}{9}+\frac{\alpha _1}{2}-400\right) \delta +\frac{100 c_3}{3}+\frac{\alpha _0}{2}
        \Bigg]\;. \tag{\stepcounter{equation}\theequation}
\end{align*}
In the above expressions for $F_{m}$, we have substituted the known expressions for $j_{1;1}$ \eqref{eqn:j11pred} and $j_{1;2}$ \eqref{eqn:j12pred}, as well as the ansatz \eqref{eqn:j1Anz} for $j_{1;3}$ and $j_{1;4}$. Therefore, the set of parameters that characterises the $F_m$ are
\begin{align}
    \{ \{a_3, b_3, c_3 \}, \{a_4, b_4, c_4 \}, \{\alpha_1,\dots,\alpha_8 \} \}
    \;,
\end{align}
Depending on which case we consider, some of the parameters $\{ a_3, b_3, c_3, a_4, b_4, c_4 \}$ may be fixed by the analysis of the previous section.

We can compare the large-$\delta$ expansions of the above expressions \eqref{eqn:f0withnewconst} to the expected behaviour of the $f_{0;m}$ in this regime from the previous section (\textit{cf.}~Table~\ref{tab:AsymptoticSummary}). This fixes some parameters. The results for each case are summarised in the table below. We have
\begin{xltabular}[c]{\textwidth}{C|C|C|C|C|C}
\text{Case} &
 \alpha_8 & \alpha_7 & \alpha_6 &
 \alpha_5 & \alpha_4 
 \\
 \midrule\midrule
\rule{0pt}{3.5ex}  1.1 & \frac{250}{3} & -\frac{575}{3} & -\frac{1415}{9} & \frac{143017}{81}\\[1ex]
\hline
\rule{0pt}{3.5ex} 1.2 &  \frac{250}{3} & -\frac{575}{3} & -\frac{1415}{9} & \frac{143017}{81}\\[1ex]
 \hline
\rule{0pt}{3.5ex} 1.3 & \frac{250}{3} & -\frac{575}{3} & -\frac{1415}{9} & \frac{143017}{81} & - \frac{113549}{27} - \frac{160}{27} c_4\\[1ex]
\hline
\rule{0pt}{3.5ex}1.4 & \frac{250}{3} & -\frac{575}{3} & -\frac{1415}{9} & \frac{143017}{81} & - \frac{113549}{27} - \frac{160}{27} c_4\\[1ex]
\hline
\rule{0pt}{3.5ex}1.5 & \frac{250}{3} & -\frac{575}{3} & -\frac{1415}{9} & \frac{143017}{81}\\[1ex]
 \hline
\rule{0pt}{3.5ex} 2 & \frac{250}{3} & -\frac{575}{3} & -\frac{1415}{9}\\[1ex]
 \hline
\rule{0pt}{3.5ex} 3 & \frac{250}{3} & -\frac{575}{3} & -\frac{1415}{9}\\[1ex]
  \caption{Coefficients in the ansatz of the $\langle f_0 j_1^3 \rangle$ \eqref{eqn:f0j1p3t2} which are fixed by comparing the known large-$\delta$ limit of all cases \eqref{tab:AsymptoticSummary}.}
  \label{tab:AsymptoticsFixAlpha}
\end{xltabular}
Different numbers of $\alpha_n$ are fixed in the various different cases, however, they are all fixed to the same numbers for all cases. 
In order to fix further parameters, in the next section, we will input some more information.

\subsection{Input of existing results and missing zeros}
\begin{figure}[ht!]
    \centering
    \includegraphics[width=0.8\textwidth]{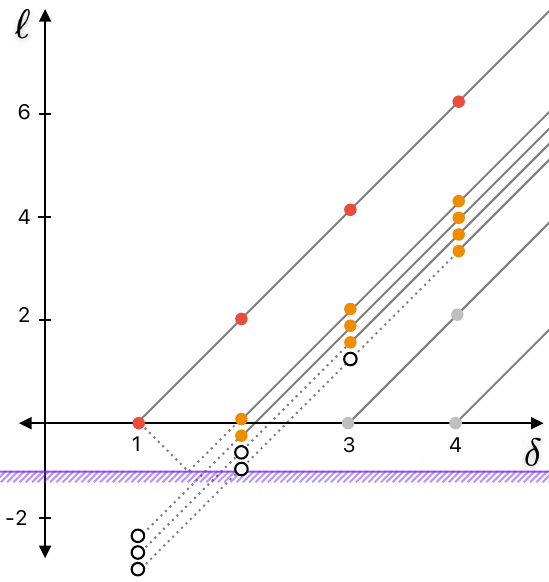}
    \caption{
    Chew-Frautschi plot with analytically continued leading and sub-leading Regge trajectories till $\ell = -2$.
    Physical states on the leading Regge trajectory are denoted in red and those on the sub-leading trajectories are denoted in orange. 
    While undertaking analytic continuation on the third/fourth and fifth/sixth sub-leading Regge trajectories, we come across certain unphysical states at $\delta = 2$ and $\delta = 3$, denoted by the black circular holes. 
    Analytically continuing all the way to $\ell = -2$ on the sub-leading Regge trajectories can be visualised as reflecting them on a line situated at $\ell = -1$. 
    Then, one of the sub-leading Regge trajectories coincides with the state which has $(\delta,\ell) = (1,0)$ on the leading Regge trajectory.
    Hence, 
    apart from the reduced OPE coefficient of the coincident state, 
    the other three reduced OPE coefficients have zeros at this point.
    }
    \label{fig:MissingZeros}
\end{figure}

\paragraph{Existing results.}
Explicit expressions for the reduced OPE coefficients \eqref{eqn:OPEmain} were worked out for the cases of $\delta = 2$~\cite{Gromov:2023hzc,Alday:2023mvu} and $\delta = 3$~\cite{Julius:2023hre}. We summarise them here. 
For $\delta = 2$, we have~\cite{Gromov:2023hzc,Alday:2023mvu}
\begin{align}\label{eqn:del2input}
    f_{0;1}(\delta = 2) = 2\;,\quad f_{0;2}(\delta = 2) = 0 
    \;.
\end{align}
For $\delta = 3$, we have~\cite{Julius:2023hre}
\begin{align}\label{eqn:del3input}
    f_{0;1}(\delta = 3) = 3\;,\quad f_{0;2}(\delta = 3) = 0 \;,\quad f_{0;3}(\delta = 3) = \frac{5}{3}\;,\quad j_{1;3}(\delta = 3) = 28\;.
\end{align}
\paragraph{Missing zeros.}
As summarised in the Setup, for $\delta = 2$ (or for $\delta = 3$), there are only two (or three) reduced OPE coefficients~\cite{Alday:2023flc}, as opposed to four reduced OPE coefficients in the case of $\delta \geq 4$. However, as illustrated in Figure~\ref{fig:MissingZeros}, the spectral data can be analytically continued to any value of $\delta$. In particular, this means that we can obtain a value for $j_{1;3}$ at $\delta = 2$, and for $j_{1;4}$ for $\delta = 2,3$, by analytically continuing in $\delta$, using the method explained in Appendix~\ref{adx:qscRegge}. 
At these points, no physical operator is present in the spectrum corresponding to the obtained values of $j_{1;3}$ and $j_{1;4}$. A consistent way to reconcile these apparently spurious states was given by the authors \cite{Homrich:2022mmd, Henriksson:2023cnh}, whose proposal states that the OPE coefficients of these spurious states must vanish. Incorporating this proposal we get that
\begin{align}\label{eqn:missingZeros}
    f_{0;3}(\delta = 2) = f_{0;4}(\delta = 2) = f_{0;4}(\delta = 3) = 0\;.
\end{align}
\paragraph{Reduction to leading trajectory.}
We can analytically continue the sub-leading trajectories all the way to $\delta=1$ such that $\ell=-2$ in this case. 
This is possible to do in $\mathcal{N} = 4$ SYM due to supersymmetry, as pointed out in~\cite{Alday:2017vkk} (see also~\cite{Caron-Huot:2017vep}).
Under the transformation $\ell \to -2-\ell$, the 4D conformal block $g^{4\text{D}}_{\Delta,\ell}$, transforms as $g^{4\text{D}}_{\Delta,\ell} = - g^{4\text{D}}_{\Delta,-2-\ell}$, whereas the pre-factor 
$$
\frac{\mathcal{C}^2_\Delta}{r_{n-1}}{\bigg[
f_0 + \frac{f_1}{\lambda^{1/4}} + \frac{f_2}{\lambda^{1/2}} + \dots 
\bigg]}^{-1}
$$
from~\eqref{eqn:OPEmain} also goes to itself, up to a minus sign. 
Thus, their product, which contributes to the OPE, is invariant.
Additionally, the states on the sub-leading Regge trajectories, which have $(\delta,\ell) = (1,-2)$ get mapped to $(\delta,\ell) = (1,0)$,
which is precisely a vertex on the leading Regge trajectory \textit{cf.} Figure~\ref{fig:MissingZeros}.
Moreover, evaluating constraints on $\langle r_{1}\,f_0\rangle$\footnote{By this we mean that we take, for instance the usual constraint~\eqref{eqn:f0t2}, but multiply both LHS and RHS by $r_{1}$ defined in~\eqref{eqn:rndef}.}, $\langle r_1\,f_0\,j_1\rangle$ and $\langle r_1\,f_0\,j_1^2\rangle$ at $\delta = 1$ for states on sub-leading Regge trajectories, we see that they reduce to the respective constraints $\langle r_0\,f_0\rangle$ $\langle r_0\,f_0\,j_1\rangle$ and $\langle r_0\,f_0\,j_1^2\rangle$ on the leading Regge trajectory, at $\delta = 1$! 
Furthermore $j_1$ on the leading Regge trajectory~\eqref{eqn:j1Reg1} has the same functional form as $j_{1;1}$ on a sub-leading Regge trajectory~\eqref{eqn:j11pred}. 
This means that 
by analytic continuing this particular sub-leading Regge trajectory, we can actually access the operator on the leading trajectory with $(\delta,\ell) = (1,0)$!
This is a very interesting phenomenon and deserves further study. 
It is important to mention at this point that the authors of~\cite{Brizio:2024nso} independently found the same phenomenon, both for $\mathcal{N} = 4$ SYM and ABJM theory. 
Theirs is a non-perturbative construction based on exploiting the integrable structure of these theories. 
This phenomenon can be visualised as a sort of reflection of the sub-leading Regge trajectories about a line at $\ell = -1$, as illustrated in Figure~\ref{fig:MissingZeros}.
\footnote{We thank Tobias Hansen for a detailed and insightful discussion about the preceding passage.}

The upshot from the point of view of decoding $f_{0;m}$ on the sub-leading Regge trajectories is that since the latter set of constraints from the discussion above have the solution that $r_0(\delta = 1)\,f_0(\delta = 1) = 1$ on the leading Regge trajectory,
this implies that the solution of the former system of constraints should yield $r_{1}(\delta = 1)\,f_{0;1}(\delta = 1) = 1$ and set the other three reduced OPE coefficients to zero. Therefore, we have
\begin{align}\label{eqn:reduceToLeading}
    r_{1}(\delta = 1)\,f_{0;1}(\delta = 1) = 1\;,\quad
    f_{0;2}(\delta = 1) = f_{0;3}(\delta = 1) = f_{0;4}(\delta = 1) = 0\;.
\end{align}

\subsection{Preliminary result}
Plugging all the inputs from equations~\eqref{eqn:del2input} --~\eqref{eqn:reduceToLeading} into our expressions for $f_{0;m}$, we can fix more parameters. We get
\begin{xltabular}[c]{\textwidth}{C|C|C|C}
\text{Case} &
 c_3 & \alpha_5 & \alpha_4 
 \\
 \midrule\midrule
\rule{0pt}{3.5ex} 1.1 & 2 & \text{fixed in Table~\ref{tab:AsymptoticsFixAlpha}} &  -\frac{11 \alpha_0}{36}-\frac{\alpha_1}{6}-\frac{36283}{9}  \\[1ex]
\hline
\rule{0pt}{3.5ex} 1.2 & -3 b_3 - 17 & \text{fixed in Table~\ref{tab:AsymptoticsFixAlpha}} &  -\frac{11 \alpha_0}{36}-\frac{\alpha_1}{6}-\frac{36283}{9}  \\[1ex]
 \hline
\rule{0pt}{3.5ex}1.3 & -8 & \text{fixed in Table~\ref{tab:AsymptoticsFixAlpha}} & \text{fixed in Table~\ref{tab:AsymptoticsFixAlpha}} \\[1ex]
\hline
\rule{0pt}{3.5ex}1.4 &  2 & \text{fixed in Table~\ref{tab:AsymptoticsFixAlpha}} & \text{fixed in Table~\ref{tab:AsymptoticsFixAlpha}}\\[1ex]
\hline
\rule{0pt}{3.5ex}1.5 & 2 & \text{fixed in Table~\ref{tab:AsymptoticsFixAlpha}} & -\frac{11 \alpha_0}{36}-\frac{\alpha_1}{6}-\frac{36283}{9} \\[1ex]
 \hline
\rule{0pt}{3.5ex} 2 & 2 & -\frac{85 \alpha_0}{216}-\frac{11 \alpha_1}{36}-\frac{\alpha_2}{6}+\frac{120857}{81} & \frac{37 \alpha_0}{18}+\frac{5 \alpha_1}{3}+\alpha_2-\frac{64529}{27} \\[1ex]
 \hline
\rule{0pt}{3.5ex} 3 & - 9 a_3 - 3 b_3 + 28 & -\frac{85 \alpha_0}{216}-\frac{11 \alpha_1}{36}-\frac{\alpha_2}{6}+\frac{120857}{81} & \frac{37 \alpha_0}{18}+\frac{5 \alpha_1}{3}+\alpha_2-\frac{64529}{27} \\[1ex]
  \caption{Parameters fixed for different cases after imposing the existing data at some $\delta$ points \eqref{eqn:del2input}, \eqref{eqn:del3input}, implementing the ``missing zeros'' constraints \eqref{eqn:missingZeros} and studying the reduction to the leading trajectory \eqref{eqn:reduceToLeading}.}
  \label{tab:FixMoreAlphaFromInput1}
\end{xltabular}
{
\begin{xltabular}[c]{\textwidth}{C|C|C|C}
\text{Case} &
 \alpha_3 & \alpha_2 & \alpha_1
 \\
 \midrule\midrule
\rule{0pt}{3.5ex} 1.1  & \frac{5 \alpha_0}{3}+\alpha_1+\frac{337352}{81} & -\frac{85\alpha_0}{36}-\frac{11\alpha_1}{6}  -\frac{44320}{27} \\[1ex]
\hline
\rule{0pt}{3.5ex} 1.2 & \frac{5 \alpha_0}{3}+\alpha_1+\frac{337352}{81} & -\frac{85\alpha_0}{36}-\frac{11\alpha_1}{6}  -\frac{44320}{27} \\[1ex]
 \hline
\rule{0pt}{3.5ex}1.3  & \frac{320 c_4}{9}-\frac{\alpha_0}{6}+\frac{421952}{81} & -\frac{1760 c_4}{27}+\alpha_0 -\frac{96020}{27} &  \frac{320 c_4}{9}-\frac{11 \alpha_0}{6}+\frac{9400}{9}\\[1ex]
\hline
\rule{0pt}{3.5ex}1.4  & \frac{427712}{81}-\frac{\alpha_0}{6} & \alpha_0-\frac{11060}{3} &  \frac{10040}{9}-\frac{11 \alpha_0}{6}\\[1ex]
\hline
\rule{0pt}{3.5ex}1.5  & \frac{5 \alpha_0}{3}+\alpha_1+\frac{337352}{81} & -\frac{85 \alpha_0}{36}-\frac{11 \alpha_1}{6}-\frac{44320}{27}\\[1ex]
 \hline
\rule{0pt}{3.5ex} 2  & -\frac{575 \alpha_0}{216}-\frac{85 \alpha_1}{36}-\frac{11 \alpha_2}{6}+\frac{93592}{81}\\[1ex]
 \hline
\rule{0pt}{3.5ex} 3  & -\frac{575 \alpha_0}{216}-\frac{85 \alpha_1}{36}-\frac{11 \alpha_2}{6}+\frac{93592}{81}\\[1ex]
  \caption{Parameters fixed for different cases after imposing the existing data at some $\delta$ points \eqref{eqn:del2input}, \eqref{eqn:del3input}, implementing the ``missing zeros'' constraints \eqref{eqn:missingZeros} and studying the reduction to the leading trajectory \eqref{eqn:reduceToLeading}.}
  \label{tab:FixMoreAlphaFromInput2}
\end{xltabular}
}
We can put everything together to obtain expressions for $f_{0;m}$. For each of the cases, the $f_{0;m}$ are fixed in terms of a different number of free parameters. This is our preliminary result. 
We present the explicit formulas for $f_{0;m}$ in appendix~\ref{adx:prelimResult}. In the table below, we summarise which parameters are free in which case. We have
{
\begin{xltabular}[c]{\textwidth}{C|C}
\text{Case} &
 \text{Free parameters}
 \\
 \midrule\midrule
 1.1  & \left\{b_4,c_4,\alpha _0,\alpha _1\right\}\\
\hline
 1.2 & \left\{b_3,c_4,\alpha _0,\alpha _1\right\}\\
 \hline
1.3  & \left\{c_4,\alpha _0\right\} \\
\hline
1.4  & \left\{c_4,\alpha _0\right\}\\
\hline
1.5  & \left\{c_4,\alpha _0,\alpha _1\right\}\\
 \hline
 2  & \left\{a_4,b_4,c_4,\alpha _0,\alpha _1,\alpha _2\right\} \\
 \hline
 3  & \left\{a_3,c_3,c_4,\alpha _0,\alpha _1,\alpha _2\right\} \\
  \caption{Free parameters which remain unfixed for all different cases.}
  \label{tab:freeParamsPrelim}
\end{xltabular}
}
\subsection{Properties and Checks}
Analysing the preliminary results in Appendix~\ref{adx:prelimResult}, we see that four out of seven cases 
give the same prediction for $j_{1;3}$, which is
\begin{align}\label{eqn:j13pred}
    j_{1;3} = 5\,\delta^2 - \frac{19}{3}\,\delta + 2\;.
\end{align}
Out of the remaining three cases, $j_{1;3}$ is fully fixed only in Case 1.3. The prediction can be seen in Table~\ref{tab:PrelimResult} in Appendix~\ref{adx:prelimResult}.
In order to check these predictions, we need to produce spectral data on this particular Regge trajectory. We describe how we did this in Appendix~\ref{adx:qscRegge}. On this trajectory, apart from the hitherto known data point of $j_{1;3}(\delta = 3) = 28$, we produced spectral data for states with $\delta = 4,5$. 
We display the comparison below. We have
{
\begin{xltabular}[c]{\textwidth}{C|C|C|C}
\delta &
 \text{From~\eqref{eqn:j13pred}} & \text{From Case } 1.3 \text{ (\textit{cf.} Table~\ref{tab:PrelimResult})} &  \text{From Fit (\textit{cf.} Table~\ref{tab:j13j14fitdata})} 
 \\
 \midrule\midrule
 \rule{0pt}{3.5ex}4 & \frac{170}{3} = 56.6666\dots & 60 &  56.445 \\[1ex]
\hline
 \rule{0pt}{3.5ex}5 & \frac{286}{3} = 95.3333\dots & 102 &  94.99 \\[1ex]
  \caption{Comparison of various prediction for $j_{1;3}$ with data obtained by integrability.}
  \label{tab:checkj13pred}
\end{xltabular}
}
Immediately, we see that Case 1.3 can be \emph{ruled out} as the predictions for $j_{1;3}$ in that case are very far away from the integrability-based prediction.
In the four cases which predicted $j_{1;3}$ to be given by~\eqref{eqn:j13pred} however, our prediction seems to be consistent with the independently obtained integrability-based data. However, due to the lack of precision on the integrability side, it is hard to establish this with a high degree of certainty. 
Nevertheless, the remaining cases, namely Case 1.2 and Case 3, which are yet unfixed fully at this stage, do provide some leeway in case that improved integrability-based predictions can definitively rule out these four cases.

\paragraph{Constructing more constraints.}
Equipped with the explicit expressions for $f_{0;m}$ and $j_{1;m}$, we can now construct more constraints of the form $\langle f_0\,j_1^n \rangle$, with $n\geq 3$. These constraints take the form of polynomials in $\delta$, with degree $2\,n + 2$.
For all the different values of $n$ that we checked, we see that
for a given $n$, the coefficients of the first three highest powers of $\delta$: $\{\delta^{2\,n+2},\delta^{2\,n + 1},\delta^{2\,n}\}$, are the same across the seven different cases.
We checked this for $n = 3,\dots,6$, and we display these coefficients below
{
\begin{xltabular}[c]{\textwidth}{C|C}
\text{Constraint} &
 \text{Terms with highest powers of $\delta$}
 \\
 \midrule\midrule
\rule{0pt}{3.5ex} \langle f_0\,j_1^3\rangle  & \frac{250 \delta ^8}{3}-\frac{575 \delta ^7}{3}-\frac{1415 \delta ^6}{9} + \dots \\[1ex]
\hline
\rule{0pt}{3.5ex} \langle f_0\,j_1^4\rangle & \frac{1250 \delta ^{10}}{3}-\frac{13375 \delta ^9}{9}+\frac{7600 \delta ^8}{9}  +\dots\\[1ex]
 \hline
 \rule{0pt}{3.5ex}\langle f_0\,j_1^5\rangle  & \frac{6250 \delta ^{12}}{3}-\frac{90625 \delta ^{11}}{9}+\frac{424375 \delta ^{10}}{27} + \dots  \\[1ex]
\hline
\rule{0pt}{3.5ex}\langle f_0\,j_1^6\rangle  & \frac{31250 \delta ^{14}}{3}-\frac{190625 \delta ^{13}}{3}+\frac{1375000 \delta ^{12}}{9} + \dots\\[1ex]
  \caption{We display the first three highest powers of $\delta$ of the constraints of the type $\langle f_0 j_1^n \rangle$ with $n=3,\dots,6$. These powers are the same for all cases \eqref{tab:AsymptoticsFixAlpha}.}
  \label{tab:constraintsHighestPowers}
\end{xltabular}
}
Notice that the fact that the first few highest powers of $\delta$ are fixed in these constraints indicates that there are some subtle cancellations among terms involving the free parameters going on, rendering final expression free parameter independent. 
These expressions and indeed any expression of the form $\langle f_0\,j_1^n \rangle$, generatable this way, are a non-trivial result of our procedure. As discussed in the Setup and the previous section, constraints of the form $\langle f_0\,j_1^n \rangle$ should be generated when one obtains the $1/R^{2n}$ curvature to the flat-space Virasoro-Shapiro amplitude. Therefore, a non-trivial check of our result would be to compute the next curvature correction to the amplitude, \textit{i.e.} at order-$1/R^6$ using the methods of~\cite{Alday:2023mvu}, extract a constraint of the form $\langle f_0\,j_1^3\rangle$, and compare its highest-power-in-$\delta$ terms with our result above. 
In principle one could also attempt to do the reverse, and use constraints generated using our procedure as an input into the program of~\cite{Alday:2023mvu}, to try to capture higher curvature corrections.

\section{More assumptions}
\label{sec:final}

The results described in the previous section and displayed in Appendix~\ref{adx:prelimResult} are the furthest that we can go to nail down the leading order reduced OPE coefficients $f_{0;m}$, by using the current assumptions. In this section, we will analyse expressions for the leading OPE coefficients and constraints of the form $\langle f_0\,j_1^n\rangle$ in a small-$\delta$ expansion. Then looking at their general structure, we will make certain mild assumptions, which will allow us to fix all the remaining freedom. Whilst these assumptions are somewhat ad hoc, the final result that we obtain using them has certain properties, which suggest that it could be correct. We describe everything in detail below. 
\subsection{Analysis at small string mass level}
Akin to what we did in Section~\ref{sec:largeDeltaAnalysis}, we can analyse the reduced OPE coefficients at small-$\delta$. Doing so, we see that in all cases, the reduced OPE coefficients go as
\begin{align}
    \begin{split}\label{eqn:f0smalldelexpanz}
        f_{0;1} &\underset{\delta\to0}{=} \frac{A_{-1;1}}{\delta} + A_{0;1} + A_{1;1}\,\delta + \mathcal{O}\left(\delta^2\right)\;, \\
        f_{0;2} &\underset{\delta\to0}{=} \frac{A_{-1;2}}{\delta} + A_{0;2} + A_{1;2}\,\delta + \mathcal{O}\left(\delta^2\right)\;, \\
        f_{0;3} &\underset{\delta\to0}{=} A_{0;3} + A_{1;3}\,\delta + \mathcal{O}\left(\delta^2\right)\;, \\
        f_{0;4} &\underset{\delta\to0}{=} A_{0;4} + A_{1;4}\,\delta + \mathcal{O}\left(\delta^2\right)\;. \\
    \end{split}
\end{align}
For each case, the $A_{k;m}$ are fixed up to the set of free parameters given in Table~\ref{tab:freeParamsPrelim}. Explicit expressions for them can be read off from the small-$\delta$ expansion of the preliminary result from Table~\ref{tab:PrelimResult}. 

Let us now combine these expressions into the form $\langle f_0\,j_1^n\rangle$, and analyse them order by order in $\delta$.
Up to the $\delta^0$ term we get
\begin{align}\label{eqn:delta0terms}
    \begin{split}
        \langle f_0 \rangle  &= {\color{red} \frac{A_{-1;1} + A_{-1;2}}{\delta} } {\color{blue}+ A_{0;1} + A_{0;2}} + 2\, A_{0;3} + 2\, A_{0;4} + \mathcal{O}(\delta)  \;, \\
        \langle f_0\,j_1 \rangle  &= {\color{blue} -3\,A_{-1;1} - 9\, A_{-1;2}} + 2\,c_3\,A_{0;3} + 2\,c_4\,A_{0;4} + \mathcal{O}(\delta)\;, \\
        \langle f_0\,j_1^n \rangle \bigg|_{n\geq 2} &= 2\,c_3^n\,A_{0;3} + 2\,c_4^n\,A_{0;4} + \mathcal{O}(\delta) \;.
    \end{split}
\end{align}
Firstly, as shown in red in the above expressions, we immediately see that $\langle f_0\rangle$ has a term of order-$1/\delta$. This clearly contradicts \eqref{eqn:f0j1t2}, which states that $\langle f_0\rangle$ is a polynomial of degree two. Therefore, we should have that $A_{-1;1} = -A_{-1;2}$. Indeed, in all the cases, expression for these terms does satisfy this equality. 

Next, let us look at the terms coloured in blue in the above expressions~\eqref{eqn:delta0terms}. We see that, excluding these terms, the rest of the terms in the expressions~\eqref{eqn:delta0terms} fall into nice patterns, for any value of $n$. We can demand that these spurious terms in blue, vanish. Depending on the case, the explicit expressions in terms of the free parameters are given in Table~\ref{tab:freeParamsPrelim}. Demanding that they vanish will imply some non-trivial constraints on these parameters, potentially allowing us to fix some of them. Indeed this is what happens that we are able to fix some freedom in each of the cases. Our results are summarised in the table below. We have
{
\begin{xltabular}[c]{\textwidth}{C|C}
\text{Case} &
 \text{Fixed parameters}
 \\
 \midrule\midrule
 \rule{0pt}{3.5ex}1.1  & c_4 = 5\;,\quad \alpha_0 = -\frac{1000}{3}\\[1ex]
\hline
\rule{0pt}{3.5ex} 1.2.1 & b_3 = -\frac{22}{3}\;,\quad \alpha_0 = -\frac{1000}{3}\\[1ex]
 \hline
\rule{0pt}{3.5ex}1.2.2 & c_4 = 5\;,\quad \alpha_0 = -\frac{1000}{3}\\[1ex]
 \hline
\rule{0pt}{3.5ex}1.4  & c_4 = 5\;,\quad \alpha_0 = -\frac{1000}{3}\\[1ex]
\hline
\rule{0pt}{3.5ex}1.5  & c_4 = 5\;,\quad \alpha_0 = -\frac{1000}{3}\\[1ex]
 \hline
 \rule{0pt}{3.5ex}2  & c_4 = 5\;,\quad \alpha_0 = -\frac{1000}{3} \\[1ex]
 \hline
 \rule{0pt}{3.5ex}3.1  & c_3 = 5\;,\quad \alpha_0 = -\frac{1000}{3} \\[1ex]
 \hline
\rule{0pt}{3.5ex} 3.2  & c_4 = 5\;,\quad \alpha_0 = -\frac{1000}{3} \\[1ex]
  \caption{{Parameters fixed as we impose that the spurious terms in \eqref{eqn:delta0terms} vanish. Two cases, 1.2 and 3 get further stratified, so we consider this extended set of cases. Note that Case 1.3, which was ruled out in the previous section, is not analysed further. }}
  \label{tab:fixdelta0}
\end{xltabular}
}
There are three things to note here. Firstly, we see that some of the cases, namely Case 1.2 and Case 3, admit two solutions, and therefore, these cases get further stratified into two sub-cases each.
Secondly, we see that in all cases, the parameter $\alpha_0$ gets fixed to the same value. 
Finally, the Case 1.4, which only had $\alpha_0$ and $c_4$ as their free parameters, gets completely fixed. 

Let us inspect this case further. We do so by asking what is the range of $\delta$, for which all four $f_{0;m}$ are non-negative. 
We get that
\begin{align}
    \textbf{Case 1.4:}&\qquad f_{0;m} \geq 0 \quad \forall m \quad \Rightarrow \delta = 2 \texttt{ OR } \delta \geq 3\;.
\end{align}
Physical states can potentially occur on the sub-leading Regge trajectories for positive integer values of $\delta\geq 2$.
Therefore, our reduced OPE coefficients should also be non-negative for all integer values of $\delta\geq 2$.
Case 1.4 yields non-negative reduced OPE coefficients in the stipulated range, and therefore is not ruled out. We will continue studying it in the next section.  

We proceed now by considering higher order in $\delta$ contributions to the constraints of the form $\langle f_0\,j_1^n\rangle$. For the $\delta^1$-term, we have
\begin{align}
    \begin{split}\label{eqn:delta1terms}
        \langle f_0\rangle &= \mathrm{o}(\delta) + \bigg[{\color{blue}A_{1;1} + A_{1;2}} + 2\,A_{1;3} + 2\,A_{1;4}\bigg]\delta + \mathcal{O}(\delta^2)\;, \\
        \langle f_0\,j_1\rangle &= \mathrm{o}(\delta) + 
        \bigg[
        {\color{blue} 
        6\,A_{0;1} 
        }
        +2\,b_3\,A_{0;3} + 2\,b_4\,A_{0;4} + 2\,c_3^1\,A_{1;3} + 2\,c_4^1\,A_{1;4}
        \bigg]\delta
        + \mathcal{O}(\delta^2)
        \;, \\
        \langle f_0\,j_1^n\rangle\bigg|_{n\geq 2} 
        &= 
        o(\delta) + \bigg[
        2\,n\,b_3\,c_3^{n-1}\,A_{0;3} + 2\,n\,b_4\,c_4^{n-1}\,A_{0;4} + 2\,c_3^n\,A_{1;3} + 2\,c_4^n\,A_{1;4}\bigg]\delta + \mathcal{O}(\delta^2)
        \;.
    \end{split}
\end{align}
Here again, we can demand that the spurious terms in blue vanish. This allows us to fix more free parameters. We get
{
\begin{xltabular}[c]{\textwidth}{C|C}
\text{Case} &
 \text{Fixed parameters}
 \\
 \midrule\midrule
 \rule{0pt}{3.5ex} 1.1  & b_4 = -\frac{37}{4}\;,\quad \alpha_1 = \frac{5210}{3}\\[1ex]
\hline
 \rule{0pt}{3.5ex} 1.2.1 & c_4 = -\frac{40}{3}\;,\quad \alpha_1 = -\frac{13100}{9}\\[1ex]
 \hline
\rule{0pt}{3.5ex} 1.2.2 & b_3 = -11\;,\quad \alpha_0 = \frac{5980}{3}\\[1ex]
 \hline
\rule{0pt}{3.5ex} 1.5  & \text{No solution}\\[1ex]
 \hline
\rule{0pt}{3.5ex}  2  & b_4 = -\frac{37}{4}\;,\quad \alpha_1 = \frac{5210}{3}\\[1ex]
 \hline
\rule{0pt}{3.5ex}  3.1  & a_3 = \frac{117\,c_4 - 640}{24(c_4 - 5)} \;,\quad \alpha_1 = \frac{55\,c_4}{3} + 1700 \\[1ex]
 \hline
\rule{0pt}{3.5ex} 3.2  & c_3 = 16\;,\quad \alpha_1 = \frac{5980}{3} \\[1ex]
  \caption{Parameters fixed after imposing that the spurious terms in \eqref{eqn:delta1terms} vanish at the $\mathcal{O}(\delta)$ order. }
  \label{tab:fixdelta1}
\end{xltabular}
}
Cases 1.1, 1.2.1 and 1.2.2 get totally fixed at this point. 
For Case 1.5, we get no solution with respect to the given constraints and therefore, we discard it. 
We inspect the other three cases for non-negativity, which yields
\begin{align}
    \textbf{Case 1.1:}&\qquad f_{0;m} \geq 0 \quad \forall m \quad \Rightarrow \delta = 2 \texttt{ OR } \delta \geq 3\;. \\
    \textbf{Case 1.2.1:}&\qquad f_{0;m} \geq 0 \quad \forall m \quad \Rightarrow 2 \leq \delta \leq 3\;, \\
    \textbf{Case 1.2.2:}&\qquad f_{0;m} \geq 0 \quad \forall m \quad \Rightarrow \delta \geq 3\;. 
\end{align}
Again, we see that only Case 1.1 is non-negative for the whole range of physical states, and therefore we rule the other cases out. We will continue studying Case 1.1 in the next section.
Finally, we can consider the $\delta^2$-order contributions to the constraints. We have
\begin{align}\label{eqn:delta2terms}
    \begin{split}
        \langle f_0
        \rangle &= 
        \mathrm{o}(\delta^2) +
        \bigg[{\color{blue} 
        A_{2;1} + A_{2;2}
        } +
        2\,A_{2;3} + 2\,A_{2;4}\bigg]\delta^2
        + \mathcal{O}(\delta^3)
        \;,\\
        \langle f_0\,j_1^1
        \rangle &= 
        \mathrm{o}(\delta^2) + \bigg[
        {\color{blue}
        6\,A_{1;1}
        }
        + 2\,\big[a_3\,A_{0;3} + b_3\,A_{1;3}\big] \\&+ 2\,\big[a_4\,A_{0;4} + b_4\,A_{1;4}\big] +
        2\,c_3\,A_{2;3} + 2\,c_4\,A_{2;4}\bigg]
        + \mathcal{O}(\delta^3)
        \;,\\
        \langle f_0\,j_1^n
        \rangle \bigg|_{n\geq 2} &= \mathrm{o}(\delta^2) + \bigg[
        n\,(n-1)\,b_3^2\,c_3^{n-2}\,A_{0;3} + n\, (n-1)\,b_4^2\,c_4^{n-2}\,A_{0;4} 
        + 2\,n\,c_3^{n-1}\,\big[a_3\,A_{0;3}
        \\
        &+ b_3\,A_{1;3}\big] + 2\,n\,c_3^{n-1}\,\big[a_4\,A_{0;4} + b_4\,A_{1;4}\big] +
        2\,c_3^n\,A_{2;3} + 2\,c_4^n\,A_{2;4} \bigg] + \mathcal{O}(\delta^3)
        \;.
    \end{split}
\end{align}
Demanding again that the spurious terms in blue vanish, we get that
{
\begin{xltabular}[c]{\textwidth}{C|C}
\text{Case} &
 \text{Fixed parameters}
 \\
 \midrule\midrule
 \rule{0pt}{3.5ex} 2  & a_4 = \frac{365}{48}\;,\quad \alpha_2 = -\frac{8285}{2}\\[1ex]
 \hline
 \rule{0pt}{3.5ex} 3.1.1  &  c_4 = -30 - 10\sqrt{\frac{83}{7}}
 \;,\quad \alpha_2 = - \frac{28295}{9} + \frac{3925}{9}\sqrt{\frac{83}{7}} \\[1ex]
 \hline
 \rule{0pt}{3.5ex} 3.1.2  & 
 c_4 = -30 + 10\sqrt{\frac{83}{7}}
 \;,\quad \alpha_2 = - \frac{28295}{9} -\frac{3925}{9}\sqrt{\frac{83}{7}} 
 \\[1ex]
 \hline
 \rule{0pt}{3.5ex} 3.2  & a_3 = \frac{31}{3}\;,\quad \alpha_2 = -\frac{44420}{9} \\[1ex]
  \caption{Parameters fixed after imposing that the spurious terms in \eqref{eqn:delta2terms} vanish at the $\mathcal{O}(\delta^2)$ order. }
  \label{tab:fixdelta2}
\end{xltabular}
}
Again on inspection, we see that only Case 2 is non-negative on the entire physical range:
\begin{align}
     \textbf{Case 2:}&\qquad f_{0;m} \geq 0 \quad \forall m \quad \Rightarrow \delta = 2 \texttt{ OR } \delta \geq 3\;. \\
     \textbf{Case 3.1.1:}&\qquad f_{0;m} \geq 0 \quad \forall m \quad \Rightarrow 2 \leq \delta \leq 3\;, \\
     \textbf{Case 3.1.2:}&\qquad f_{0;m} \geq 0 \quad \forall m \quad \Rightarrow 2 \leq \delta \leq 3
    \texttt{ OR } \delta\geq -\frac{442}{143} + \frac{56}{143}\sqrt{581}
    \;, \\
     \textbf{Case 3.2:}&\qquad f_{0;m} \geq 0 \quad \forall m \quad \Rightarrow 2 \leq \delta \leq 3\;. 
\end{align}
We stop now, since we have now fixed the free parameters for all the cases. 

From the analysis above,  we see that after fixing all the freedom using our somewhat ad hoc prescription that the ``spurious'' terms vanish in the small-$\delta$ expansion of constraints of the form $\langle f_0\,j_{1}^n \rangle$, only \emph{three} cases survive, namely \textbf{Cases 1.1}, \textbf{1.4} and \textbf{2}. In the sequel, we will analyse these cases further, and see if can rule out any further cases. 
\subsection{Final result}
Below, we summarise our final result: the leading non-trivial order at strong coupling CFT-data, on the four unique sub-leading Regge trajectories, of stringy operators exchanged in the four-point function~\eqref{eqn:def4pt} of planar $\mathcal{N} = 4$ Super-Yang-Mills theory
Case by case, we have
{
\begin{xltabular}[c]{\textwidth}{C|C|C|C}
& {\text{Case } 1.1} & {\text{Case } 1.4} & {\text{Case } 2} 
 \\
 \midrule\midrule
 \rule{0pt}{3.5ex}j_{1;3} & 5\,\delta^2 - \frac{19}{3}\,\delta + 2
 & 5\,\delta^2 - \frac{19}{3}\,\delta + 2
 & 5\,\delta^2 - \frac{19}{3}\,\delta + 2
 \\[1ex]
 \hline
 \rule{0pt}{3.5ex}j_{1;4} & 5\,\delta^2 - \frac{37}{4}\,\delta + 5
 & 5\,\delta^2 - 3\,\delta + 5
 & \frac{365}{48}\,\delta^2 - \frac{37}{4}\,\delta + 5
 \\[1ex]
 \hline
\rule{0pt}{3.5ex}f_{0;1} &  \frac{\delta ^2 (23 \delta -25)}{(5 \delta -4) (5 \delta -3)}
& \frac{\delta  (3 \delta +7)-12}{5 \delta -3}
& \frac{\delta ^3 (75 \delta -101)}{(5 \delta -3) (5 \delta  (5 \delta -12)+48)}
\\[1ex]
\hline
\rule{0pt}{3.5ex} f_{0;2} & 0
& \frac{10 (\delta -3) (\delta -2) (\delta -1)}{(4 \delta +3) (6 \delta +5)} & 0 \\[1ex]
 \hline
\rule{0pt}{3.5ex}f_{0;3}  & \frac{5 (\delta -2) (\delta -1) \delta  (35 \delta +33)}{3 (5 \delta -3) (35 \delta -36)} 
& \frac{5 (\delta -2) (\delta -1) \delta  (\delta  (40 \delta +141)+387)}{3 (4 \delta +3) (5 \delta -3) (10 \delta +9)}
& \frac{5 (\delta -2) (\delta -1) \delta  (5 \delta  (25 \delta +47)-132)}{3 (5 \delta -3) (5 \delta  (25 \delta -28)+144)}
\\[1ex]
\hline
\rule{0pt}{3.5ex} f_{0;4}  & \frac{32 (\delta -3) (\delta -2) (\delta -1)}{(5 \delta -4) (35 \delta -36)} 
& \frac{10 (\delta -3) (\delta -2) (\delta -1)}{(6 \delta +5) (10 \delta +9)}
& \frac{1536 (\delta -3) (\delta -2) (\delta -1)}{(5 \delta  (5 \delta -12)+48) (5 \delta  (25 \delta -28)+144)}
\\[1ex]
  \caption{Final results for $j_{1;3}$, $j_{1;4}$, $f_{0;1}$, $f_{0;2}$, $f_{0;3}$, $f_{0;4}$ for three cases which satisfy constraints imposed. Together with \eqref{eqn:j11pred}, \eqref{eqn:j12pred}, we then have CFT-data for all four Regge trajectories as a function of $\delta$ at the first non-trivial order.}
  \label{tab:MainResult}
\end{xltabular}
}
In the sequel, we will study the properties of each of these three cases. 
\subsection{Properties and Checks}
In each case, we have a different prediction for $j_{1;4}$, given in the second row of Table~\ref{tab:MainResult}. Again in order to check the predictions, we need to compare against independently obtained integrability data on this Regge trajectory. As before, we have described the procedure that we used to obtain this data in Appendix~\ref{adx:qscRegge}. We produced spectral data for $\delta = 4,5$ and present the results and comparison below. We have
{
\begin{xltabular}[c]{\textwidth}{C|C|C|C|C}
\delta &
 \text{Case 1.1} &
 \text{Case 1.4} &
 \text{Case 2} & \text{From Fit (\textit{cf.} Table~\ref{tab:j13j14fitdata})} 
 \\
 \midrule\midrule
 \rule{0pt}{3.5ex}4 & 48 & 73 & \frac{269}{3} = 89.6666\dots & 47.2215 \\[1ex]
\hline
 \rule{0pt}{3.5ex}5 & \frac{335}{4} = 83.75 & 115 & \frac{7145}{48} = 148.85416666\dots & 83.331\\[1ex]
  \caption{Comparison of the prediction for $j_{1;4}$ (see Table \ref{tab:MainResult}) with integrability data at the points at $\delta=4$ and $\delta=5$ for all cases of the main result.}
  \label{tab:checkj14pred}
\end{xltabular}
}
Clearly, we see from the above table, that \emph{only} \textbf{Case 1.1} is in the ballpark of the independently obtained numerical data from integrability. Therefore, of the three possible cases discussed in this section, only this case can be considered as a prediction for the leading non-trivial order CFT-data, on the four unique sub-leading Regge trajectories.

Note that in this case, $f_{0;2}$ identically vanishes. This is potentially a powerful prediction, as it tells us that infinitely many leading order OPE coefficients on a particular sub-leading Regge trajectory should vanish. 
This hints at some kind of emergent selection rule in the flat-space limit. We will ponder upon this in the Discussion. 

\begin{figure}[H]
    \centering
    \includegraphics[width=0.8\textwidth]{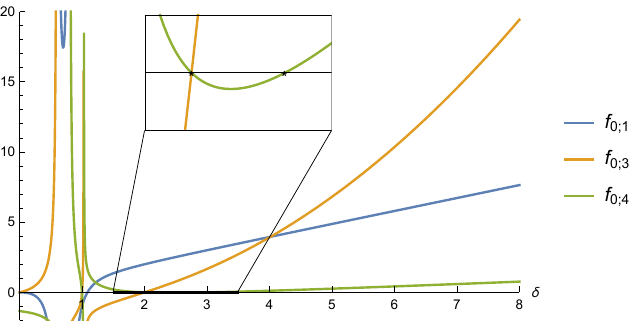}
    \caption{
    The leading order reduced OPE coefficients of Case 1.1, our main result, plotted as a function of $\delta$. 
    The poles in the $f_{0;m}$ are outside the physical range of $\delta$.
    All the reduced OPE coefficients are non-zero on all physical states. In particular $f_{0;3}$ is negative between $\delta = 2$ and $\delta = 3$, and has zeros at $\delta = 2$ and $\delta = 3$, and is positive for $\delta>3$.
    $f_{0;3}$ dominates over the other two at large-$\delta$.
    }
    \label{fig:f0plots}
\end{figure}

Analysing the expressions for $f_{0;m}$ further, we see that while the $f_{0;m}$ do have poles, these occurs at $\delta<2$, \textit{i.e.} outside the physical range of $\delta$. We also plotted the predictions of Case 1.1 for $f_{0;m}$, in Figure~\ref{fig:f0plots}. It is clear from the plot, as well as the expression for $f_{0;3}$ in Table~\ref{tab:MainResult}, that as we get to higher values of $\delta$, that $f_{0;3}$ dominates over the other two non-zero reduced OPE coefficients. It would be interesting to see what, if any physical significance this has.
Finally, as all four unique reduced OPE coefficients on sub-leading Regge trajectories are unmixed, the constraints of the type $\langle f_0 j_1^n \rangle$ can be constructed for any $n$. 
Such constraints could potentially serve as an input into the program of~\cite{Alday:2023mvu}, potentially helping fix some of the freedom when computing the $1/R^{2\,n}$ term of the AdS Virasoro-Shapiro amplitude. 
We display the constraints with $n=3,\dots,6$ in the table below. We have
{
\setlength\LTleft{-28.75mm}
\setlength\LTright\fill
\begin{xltabular}{\textwidth}{C|C}
\text{Constraint} & \text{Explicit Expression}
 
 \\
 \midrule\midrule
\rule{0pt}{3.5ex} \langle f_0\,j_1^3\rangle  & \frac{250 \delta ^8}{3}-\frac{575 \delta ^7}{3}-\frac{1415 \delta ^6}{9}+\frac{143017 \delta ^5}{81}-\frac{113914 \delta ^4}{27}+\frac{433022 \delta ^3}{81}-\frac{12115 \delta ^2}{3}+\frac{5210 \delta }{3}-\frac{1000}{3} \\[1ex]
\hline
\rule{0pt}{3.5ex} \multirow{2}{*}{$\langle f_0\,j_1^4\rangle$} & \frac{1250 \delta ^{10}}{3}-\frac{13375 \delta ^9}{9}+\frac{7600 \delta ^8}{9}+\frac{815990 \delta ^7}{81}-\frac{9379078 \delta ^6}{243}+\frac{23821459 \delta ^5}{324}\\[1ex]
\rule{0pt}{3.5ex} & 
-\frac{43145911 \delta ^4}{486}+\frac{23150965 \delta ^3}{324}-\frac{675325 \delta ^2}{18}+\frac{105460 \delta }{9}-\frac{5000}{3}\\[1ex]
 \hline
 \rule{0pt}{3.5ex}\multirow{2}{*}{$\langle f_0\,j_1^5\rangle$}  & \frac{6250 \delta ^{12}}{3}-\frac{90625 \delta ^{11}}{9}+\frac{424375 \delta ^{10}}{27}+\frac{3669250 \delta ^9}{81}-\frac{72085450 \delta ^8}{243}+\frac{2277121483 \delta ^7}{2916}-\frac{5072058409 \delta ^6}{3888}\\[1ex]
\rule{0pt}{3.5ex} & 
 +\frac{8896931761 \delta ^5}{5832}-\frac{4969488875 \delta ^4}{3888}+\frac{490837195 \delta ^3}{648}-\frac{16372505 \delta ^2}{54}+\frac{665170 \delta }{9}-\frac{25000}{3}  \\[1ex]
\hline
\rule{0pt}{3.5ex} \multirow{2}{*}{$\langle f_0\,j_1^6\rangle$}  & \frac{31250 \delta ^{14}}{3}-\frac{190625 \delta ^{13}}{3}+\frac{1375000 \delta ^{12}}{9}+\frac{10001875 \delta ^{11}}{81}-\frac{162974750 \delta ^{10}}{81}+\frac{6876983455 \delta ^9}{972}-\frac{267207052619 \delta ^8}{17496}\\[1ex]
\rule{0pt}{3.5ex} & 
+\frac{1093732131439 \delta ^7}{46656}-\frac{1875583992391 \delta ^6}{69984}+\frac{358931037655 \delta ^5}{15552}-\frac{38477697665 \delta ^4}{2592}+\frac{2249522735 \delta ^3}{324}-2238430 \delta ^2+\frac{1339280 \delta }{3}-\frac{125000}{3}\\[1ex]
  \caption{We display constraints of the type $\langle f_0 j_1^n \rangle$ with $n=3,\dots,6$.}
  \label{tab:constraintsFinal}
\end{xltabular}
}

\section{Discussion}
\label{sec:disc}

In this paper, we have obtained expressions for the leading non-trivial order CFT-data at strong coupling, of the stringy operators exchanged in the OPE of the four-point function $\langle \mathcal{O}_2\,\mathcal{O}_2\,\mathcal{O}_2\,\mathcal{O}_2 \rangle$, which lie in the four unique sub-leading Regge trajectories. This comprises eight observables:
\begin{align}\label{eqn:CFTdata2}
    \{j_{1;m};f_{0;m}\}\;,\quad m = 1,\dots,4
    \;,
\end{align}
which we obtained as function of the string mass level $\delta$. Here $j_{1;m}$ is the sub-leading coefficient of the eigenvalue of the quadratic Casimir of an exchanged operator and $f_{0;m}$ is the squared leading order reduced OPE coefficient of two external $\mathcal{O}_2$ and one exchanged operator. 

Our only assumptions were minimal ones that $j_{1;m}$ on a Regge trajectory are a polynomials of degree two in the string mass level~\eqref{eqn:j1Anz}, and that the combination $\langle f_0\,j_1^3 \rangle$ is a polynomial of the string mass level. The the former has motivation from string theory perspective mentioned in~\cite{Beccaria:2012xm} and the latter, from looking at the structure of constraints on the CFT-data obtained in~\cite{Alday:2022uxp,Alday:2022xwz,Alday:2023mvu}.
Firstly, we obtained expressions for $j_{1;1}$~\eqref{eqn:j11pred} and $j_{1;2}$~\eqref{eqn:j12pred} using the integrability-based Quantum Spectral Curve method. 
Then, for the other six pieces of information in~\eqref{eqn:CFTdata2}: $\{j_{1;3},j_{1;4},f_{0;1},f_{0;2},f_{0;3},f_{0;4}\}$, we proceeded in the following way:
\begin{enumerate}
    \item We analysed the CFT-data at large-$\delta$, and imposed non-negativity.
    \item We imposed that the reduced OPE coefficients should vanish at certain unphysical points (\textit{i.e.} at certain integral values of $\delta$ where no physical state is exchanged).
    \item We inputted already known results for certain values of $\delta$.
\end{enumerate}
Doing so allowed us to reduce to \emph{seven} cases of possible solutions. Each case was fixed up to a small number of free parameters given in Table~\ref{tab:freeParamsPrelim}. 
The specific free parameters that we have for each case quantify exactly how the yet unfixed integrability-data and the yet unfixed string amplitudes-input enter into expressions for the CFT-data. 
The explicit expressions for each case are given in Table~\ref{tab:PrelimResult}. 
This was our \emph{Preliminary Result}.

At this point, our solutions already had some predictive power.
Namely, irrespective of which of the seven cases we considered, it was possible to extract a prediction for the top three highest-power-of-$\delta$ terms in constraints on the CFT-data of the form $\langle f_0\,j_1^n\rangle$ for any $n\geq 3$, where there are no predictions in the literature. 
As constraints of the form $\langle f_0\,j_1^n\rangle$ are expected to be obtained as a byproduct of the order-$1/R^{2\,n}$ correction to the flat-space Virasoro-Shapiro amplitude using the method explained in~\cite{Alday:2023mvu}, extracting such constraints once one has obtained the $1/R^6$ or higher correction can then serve as a consistency check of our result. 

Five out of the seven cases also straight away gave a prediction for $j_{1;3}$ on the Regge trajectory. Moreover, in four out of five cases, we got the same prediction~\eqref{eqn:j13pred}.
The prediction for the remaining case can be referred to under Case 1.3 of Table~\ref{tab:PrelimResult}.
These predictions were tested against independently produced integrability-based prediction for states on this Regge trajectory given in Table~\ref{tab:j13j14fitdata}.
This immediately rules out the latter prediction, \textit{i.e.} Case 1.3.
Whereas, with respect to the prediction~\eqref{eqn:j13pred}, we see that there is only a tentative match, at best. 
This could be due to the fact that the integrability-based data, obtained by performing numerical fits, does not have very high precision. 

It is very important to stress a particular point about the structure of the CFT-data on the sub-leading Regge trajectories. In implementing Steps 2.~and 3.~above, we analytically continued the sub-leading Regge trajectories to $\delta = 1$. Surprisingly, for one of the \emph{sub-leading} Regge trajectories, this landed us on precisely a state living on the \emph{leading} Regge trajectory. 
Whilst the immediate use of this for us was to fix a bit more of the freedom in the reduced OPE coefficients, this phenomenon points at something deeper in the structure of the CFT-data itself, and how it arranges itself in Regge trajectories. It would be very interesting to explore and unravel this structure further.  
It is important to note that the authors of~\cite{Brizio:2024nso} also found the same phenomenon independently, which they call as ``Regge bridges''.
Their construction is non-perturbative and applies to both $\mathcal{N} = 4$ SYM and ABJM theory. 

To finally fix all the freedom in the reduced OPE coefficients, we proceeded as follows:
\begin{enumerate}
    \item We analysed each of the seven cases at small-$\delta$ and observed the same behaviour in all seven cases. We parameterised this expansion using some coefficients, \textit{cf.} equation~\eqref{eqn:f0smalldelexpanz}. 
    \item We constructed constraints of the form $\langle f_0\,j_1^n \rangle$ for $n\geq 0$ in a small-$\delta$ expansion.
    \item We analysed the structure of these constraints, and used observed patterns to impose certain \emph{ad hoc} relations among the coefficients in equation~\eqref{eqn:f0smalldelexpanz}.
\end{enumerate}
Doing so allowed us to fix all the remaining freedom in all but one case, which we discarded.
The result however, was that not all the cases now returned OPE coefficients that were always non-negative on physical states. 
We discarded these pathological cases as well.
Finally we were left with three cases that formed our \emph{Final Result}; it is tabulated in Table~\ref{tab:MainResult}.

In all the three cases, the expression for $j_{1;3}$ was the same, and it is given by equation~\eqref{eqn:j13pred}.
We tested the Final Result by comparing the predictions for $j_{1;4}$ from the three cases in Table~\ref{tab:MainResult} against independently obtained integrability-based data from Table~\ref{tab:j13j14fitdata} on these Regge trajectories. 
The result was that two out of the three cases were ruled out and only \textbf{Case 1.1} remained.
However, it is important to point out that even in this case, the match between the prediction for $j_{1;4}$ and the integrability-based data is tentative at best. 
Given the fact that the integrability-based data comes from numerical fits which are not particularly precise, there is a need to produce more precise integrability-based data on this trajectory. Perhaps this could be done by extending and specialising the strong coupling-tailored QSC numerical method developed in~\cite{Ekhammar:2024rfj}. 

Let us discuss an important property of \textbf{Case 1.1} of the Final Result.
We see that the OPE coefficient $f_{0;2}$ is predicted to be identically zero. 
This means at in the flat-space limit of AdS, one of the Regge trajectories effectively decouples from the OPE $\mathcal{O}_2 \times \mathcal{O}_2$.
By applying the methods of~\cite{Caron-Huot:2022sdy}, it should be possible to numerically establish that an entire sub-leading Regge trajectory decouples from the OPE as predicted here. 
This decoupling points towards the potential of a selection due to an emergent hidden symmetry in this limit.
This is very similar to the conclusion drawn in~\cite{Julius:2023hre}, albeit in a different setup of a four-point function with two external operators of higher $R$-charge.
It is possible that this hidden emergent symmetry is the ten-dimensional dual conformal symmetry that becomes emergent in the flat-space limit of string amplitudes~\cite{Aprile:2018efk,Caron-Huot:2018kta,Aprile:2020mus,Caron-Huot:2021usw}.
This very interesting direction deserves further investigation as it may shed light on the structure of the CFT-data of planar $\mathcal{N} = 4$ SYM in the flat-space limit of the dual AdS.
Potentially, many sub-sub-leading Regge trajectories also decouple in the flat-space limit, (in analogy with the conclusion drawn in~\cite{Julius:2023hre}), which could hint at a further simplification of the physics in this limit, as well as make calculations more tractable by reducing the number of variables.

It would also be interesting to see if one can focus on specific limits like the high-energy limit~\cite{Alday:2023pzu} or sub-leading Regge limit following~\cite{Alday:2024xpq}, to isolate the CFT-data studied here, and to leverage them to either make comparative predictions to check our results, or to use our results as an input there.
Another approach could be to truncate the sum over string mass levels, as pioneered in~\cite{Saha:2024qpt}. Such a truncation could isolate contributions to the string amplitude from CFT-data which is more easily accessible using integrability techniques.

It is important to highlight some properties of our procedure. 
We analysed the leading order reduced OPE coefficients on a Regge trajectory using a variety of inputs and stipulations. 
As enumeratated above, these comprised analysis at large string mass level, imposition of non-negativity, a requirement that certain OPE coefficients vanish at certain unphysical points, and the input of known results. 
All these steps are completely model independent (except the specific known results and their extent), and therefore our procedure potentially has a wide range of applicability.
In particular, our procedure could potentially be very useful in similar setups, where there is a lack of integrability-based data available currently.
Examples include the study of the AdS Veneziano amplitude~\cite{Alday:2024yax,Alday:2024ksp}, as well as the study of similar setups in other examples of the holographic duality like ABJM.
Furthermore, we only considered short strings in the discussions of this paper, whereas our procedure should be applicable in the case of long strings as well.
Moreover, since the main observables studied in this paper, the $f_0$~\cite{Alday:2023flc,Fardelli:2023fyq} and the $j_1$~\cite{Gromov:2023hzc}, are both $R$-charge independent, as shown in~\cite{Julius:2023hre}, our conclusions should also apply to the case of the $t$-channel squared OPE coefficients in the four-point function with two external operators with higher $R$-charge considered in~\cite{Fardelli:2023fyq,Julius:2023hre}.

It is also intriguing to notice that the analysis of the reduced OPE coefficients in three regions of parameter space, namely large-$\delta$, small-$\delta$, and particular small finite values of $\delta$ is strong enough to fix all the freedom in the leading non-trivial order CFT-data, and fix them as a function of $\delta$ for all values of $\delta$. 
It would be interesting to understand why this ``Pade-like'' analysis is sufficient to fix the reduced OPE coefficients  and partial spectral data completely.

In conclusion, using our procedure, we fixed the leading non-trivial order CFT-data at strong coupling, of the stringy operators exchanged in the OPE of the four-point function $\langle \mathcal{O}_2\,\mathcal{O}_2\,\mathcal{O}_2\,\mathcal{O}_2 \rangle$, which lie in the four unique sub-leading Regge trajectories~\eqref{eqn:CFTdata2}.
Thus, we have successfully unmixed the CFT-data of infinitely many states that live on these trajectories, as well as for infinitely many of their Kaluza-Klein descendents. 
Furthermore, it could be worthwhile to use our procedure to generate inputs for two ongoing research programs.
Firstly, to use its output to generate constraints on the CFT-data that could potentially capture some part of higher curvature corrections to the AdS Virasoro-Shaprio amplitude in the program of~\cite{Alday:2023mvu}, and indeed potentially in other contexts like the AdS Veneziano amplitude~\cite{Alday:2024yax,Alday:2024ksp} as well.
Secondly, our procedure can also be used to generate spectral data itself, which could feed into the program precision computation of the spectral data of planar $\mathcal{N} = 4$ SYM~\cite{Gromov:2023hzc} as well as in other contexts.

\acknowledgments

We thank Luis Fernando Alday, Frank Coronado, James Drummond, Tobias Hansen, Alok Laddha, Joseph Minahan, Ashoke Sen and Arkady Tseytlin
for many insightful questions, suggestions and discussions that led to the inception of the idea behind this paper.
We express our sincere gratitude 
to Alok Laddha and Ashoke Sen for their guidance and 
support as well as 
for inspiring discussions that greatly informed the physical picture and narrative of this paper; 
to Luis Fernando Alday and Tobias Hansen for a very encouraging chat at Eurostrings meets FPUK 2024 at the University of Southampton;
to Tobias Hansen for a careful reading of our manuscript, and insightful and helpful suggestions and clarifications on various points;
and to Nikolay Gromov for numerous suggestions on the manuscript and an insightful discussion thereafter.
We thank the Chennai Mathematical Institute, the International Centre for Theoretical Sciences and the Tata Institute for Fundamental Research, and the staff and students thereof for their kind hospitality and for providing a stimulating environment during critical phases of this work.
The work of NS is supported by the European Research Council (ERC) under the European Union’s Horizon 2020 research and innovation programme (grant agreement No. 865075) EXACTC.
JJ expresses gratitude to the people of India for their continuing support towards the study of basic sciences.

\appendix

\section{Explicit preliminary result}
\label{adx:prelimResult}

Below we tabulate our preliminarty result, which was obtained as a result of the considerations of Section~\ref{sec:prelim}. We have
{
\setlength\LTleft{-26mm}
\setlength\LTright\fill
\begin{xltabular}{\textwidth}{C|C}
\multicolumn{2}{C}{\text{Case } 1.1} 
 \\
 \midrule\midrule
 \rule{0pt}{3.5ex} j_{1;3} & 
 5 \delta ^2-\frac{19 \delta }{3}+2
 \\[1ex]
 \hline
 \rule{0pt}{3.5ex} j_{1;4} & 
 5 \delta ^2+b_4 \delta +c_4
 \\[1ex]
 \hline
\rule{0pt}{3.5ex} \multirow{2}{*}{$F_1$} &  
\left(-\frac{56 b_4}{3}-\frac{11 \alpha _0}{36}-\frac{\alpha _1}{6}+\frac{3508}{27}\right) \delta ^4+\left(12 b_4-\frac{56 c_4}{3}+\frac{5 \alpha _0}{3}+\alpha _1-\frac{9916}{9}\right) \delta ^3
\\[1ex]
\rule{0pt}{3.5ex} &
+\left(-\frac{76 b_4}{3}+12 c_4-\frac{85 \alpha _0}{36}-\frac{11 \alpha _1}{6}+\frac{56768}{27}\right) \delta ^2+\left(40 b_4-\frac{76 c_4}{3}+\alpha _1-1240\right) \delta +40 c_4+\alpha _0+\frac{400}{3}
\\[1ex]
\hline
 \rule{0pt}{3.5ex} \frac{F_2}{(\delta-1)\,(\delta -2)\,(\delta - 3)} & 
 -\frac{20 c_4}{3}-\frac{\alpha _0}{6}+\delta  \left(-\frac{20 b_4}{3}-\frac{11 \alpha _0}{36}-\frac{\alpha _1}{6}+\frac{3400}{27}\right)-\frac{200}{9}
 \\[1ex]
 \hline
\rule{0pt}{3.5ex} \frac{F_3}{(\delta-1)\,(\delta -2)}  &  
\left(\frac{80 b_4}{27}+\frac{1520}{81}\right) \delta ^3+\left(\frac{70 b_4}{9}+\frac{80 c_4}{27}-\frac{11 \alpha _0}{72}-\frac{\alpha _1}{12}+\frac{3680}{27}\right) \delta ^2+\left(\frac{50 b_4}{3}+\frac{70 c_4}{9}+\frac{3 \alpha _0}{8}+\frac{\alpha _1}{4}-200\right) \delta +\frac{50 c_4}{3}+\frac{\alpha _0}{4}
\\[1ex]
\hline
\rule{0pt}{3.5ex} \frac{F_4}{(\delta-1)\,(\delta -2)\,(\delta - 3)}  & 
-\frac{\alpha _0}{12}+\delta  \left(-\frac{11 \alpha _0}{72}-\frac{\alpha _1}{12}+\frac{2510}{27}\right)-\frac{100}{9}
\\[1ex]
\midrule\midrule
 \multicolumn{2}{C}{ \rule{0pt}{6.5ex} \text{Case } 1.2} 
 \\
 \midrule\midrule
 \rule{0pt}{3.5ex} j_{1;3} & 
 5 \delta ^2+b_3 \delta -3 b_3-17
 \\[1ex]
 \hline
 \rule{0pt}{3.5ex} j_{1;4} & 
 5 \delta ^2-\frac{19 \delta }{3}+c_4
 \\[1ex]
 \hline
 \rule{0pt}{3.5ex} \multirow{3}{*}{$F_1$} &  
 \left(\frac{16 c_4 b_3}{9}-\frac{200 b_3}{9}+\frac{304 c_4}{27}-\frac{11 \alpha _0}{36}-\frac{\alpha _1}{6}+\frac{2900}{27}\right) \delta ^4+\left(\frac{2 c_4 b_3}{3}+\frac{200 b_3}{3}-\frac{130 c_4}{9}+\frac{5 \alpha _0}{3}+\alpha _1-\frac{6800}{9}\right) \delta ^3
 \\[1ex]
 \rule{0pt}{3.5ex} &
 +\left(-\frac{166}{9} c_4 b_3-\frac{220 b_3}{9}-\frac{2830 c_4}{27}-\frac{85 \alpha _0}{36}-\frac{11 \alpha _1}{6}+\frac{56920}{27}\right) \delta ^2+\left(-12 c_4 b_3+140 b_3-\frac{304 c_4}{3}+\alpha _1-\frac{1820}{3}\right) \delta 
 \\[1ex]
 \rule{0pt}{3.5ex} & -200 b_3+40 b_3 c_4+\frac{880 c_4}{3}+\alpha _0-\frac{3400}{3} \\[1ex]
\hline
 \rule{0pt}{3.5ex} \frac{F_2}{(\delta-1)\,(\delta -2)\,(\delta - 3)} & 
 -\frac{20}{3} c_4 b_3+\frac{100 b_3}{3}-\frac{440 c_4}{9}-\frac{\alpha _0}{6}+\delta  \left(-\frac{20}{9} c_4 b_3-\frac{20 b_3}{9}-\frac{380 c_4}{27}-\frac{11 \alpha _0}{36}-\frac{\alpha _1}{6}+\frac{4160}{27}\right)+\frac{1700}{9}
 \\[1ex]
 \hline
\rule{0pt}{3.5ex} \frac{F_3}{(\delta-1)\,(\delta -2)}  & 
\left(\frac{80 c_4}{27}-\frac{11 \alpha _0}{72}-\frac{\alpha _1}{12}+\frac{2350}{27}\right) \delta ^2+\left(\frac{70 c_4}{9}+\frac{3 \alpha _0}{8}+\frac{\alpha _1}{4}-\frac{2750}{9}\right) \delta +\frac{50 c_4}{3}+\frac{\alpha _0}{4}
\\[1ex]
\hline
\rule{0pt}{3.5ex} \frac{F_4}{(\delta-1)\,(\delta -2)\,(\delta - 3)}  & 
\left(\frac{80 b_3}{27}+\frac{1520}{81}\right) \delta ^2+\left(\frac{70 b_3}{9}-\frac{11 \alpha _0}{72}-\frac{\alpha _1}{12}+\frac{1280}{9}\right) \delta +\frac{50 b_3}{3}-\frac{\alpha _0}{12}+\frac{850}{9}
\\[1ex]
\midrule\midrule 
\multicolumn{2}{C}{\rule{0pt}{6.5ex}\text{Case } 1.3} 
 \\
 \midrule\midrule
 \rule{0pt}{3.5ex}j_{1;3} & 5 \delta ^2-3 \delta -8 \\[1ex]
 \hline
 \rule{0pt}{3.5ex}j_{1;4} & 5 \delta ^2-\frac{19 \delta }{3}+c_4 \\[1ex]
 \hline
 \rule{0pt}{3.5ex} F_1 & \left(\frac{172 c_4}{9}-\frac{\alpha _0}{6}+\frac{800}{9}\right) \delta ^3+\left(-\frac{344 c_4}{3}+\alpha _0+\frac{800}{3}\right) \delta ^2+\left(-\frac{268 c_4}{9}-\frac{11 \alpha _0}{6}+\frac{160}{9}\right) \delta +\frac{520 c_4}{3}+\alpha _0-\frac{1600}{3} \\[1ex]
\hline
  \rule{0pt}{3.5ex} \frac{F_2}{(\delta-1)\,(\delta -2)\,(\delta - 3)} & \delta  \left(-\frac{40 c_4}{3}-\frac{40}{3}\right)-\frac{260 c_4}{9}-\frac{\alpha _0}{6}+\frac{800}{9} \\[1ex]
 \hline
\rule{0pt}{3.5ex} \frac{F_3}{(\delta-1)\,(\delta -2)}  & \frac{50 c_4}{3}+\delta  \left(\frac{50 c_4}{3}-\frac{\alpha _0}{12}-\frac{400}{9}\right)+\frac{\alpha _0}{4} \\[1ex]
\hline
\rule{0pt}{3.5ex} \frac{F_4}{(\delta-1)\,(\delta -2)\,(\delta - 3)}  & \frac{800 \delta ^2}{81}+\left(\frac{860}{27}-\frac{80 c_4}{27}\right) \delta -\frac{\alpha _0}{12}+\frac{400}{9} \\[1ex]
\midrule\midrule 
\multicolumn{2}{C}{\rule{0pt}{6.5ex}\text{Case } 1.4} 
 \\
 \midrule\midrule
\rule{0pt}{3.5ex} j_{1;3} & 5 \delta ^2-\frac{19 \delta }{3}+2 \\[1ex]
 \hline
\rule{0pt}{3.5ex} j_{1;4} & 5 \delta ^2-3 \delta +c_4 \\[1ex]
 \hline
\rule{0pt}{3.5ex} F_1 & \left(-\frac{56 c_4}{3}-\frac{\alpha _0}{6}-\frac{200}{9}\right) \delta ^3+\left(12 c_4+\alpha _0+\frac{400}{3}\right) \delta ^2+\left(-\frac{76 c_4}{3}-\frac{11 \alpha _0}{6}-\frac{2200}{9}\right) \delta +40 c_4+\alpha _0+\frac{400}{3} \\[1ex]
\hline
\rule{0pt}{3.5ex} \frac{F_2}{(\delta-1)\,(\delta -2)\,(\delta - 3)} & -40 \delta -\frac{20 c_4}{3}-\frac{\alpha _0}{6}-\frac{200}{9} \\[1ex]
 \hline
\rule{0pt}{3.5ex} \frac{F_3}{(\delta-1)\,(\delta -2)}  & \frac{800 \delta ^3}{81}+\left(\frac{80 c_4}{27}+20\right) \delta ^2+\left(\frac{70 c_4}{9}-\frac{\alpha _0}{12}+\frac{260}{9}\right) \delta +\frac{50 c_4}{3}+\frac{\alpha _0}{4} \\[1ex]
\hline
\rule{0pt}{3.5ex} \frac{F_4}{(\delta-1)\,(\delta -2)\,(\delta - 3)}  & -\frac{\alpha _0}{12}-\frac{100}{9} \\[1ex]
\midrule\midrule 
\multicolumn{2}{C}{\rule{0pt}{6.5ex}\text{Case } 1.5} 
 \\
 \midrule\midrule
 \rule{0pt}{3.5ex} j_{1;3} & 5 \delta ^2-\frac{19 \delta }{3}+2 \\[1ex]
 \hline
\rule{0pt}{3.5ex} j_{1;4} & 5 \delta ^2-\frac{19 \delta }{3}+c_4 \\[1ex]
 \hline
\rule{0pt}{3.5ex} \multirow{2}{*}{$F_1$} & \left(-\frac{11 \alpha _0}{36}-\frac{\alpha _1}{6}+\frac{6700}{27}\right) \delta ^4+\left(-\frac{56 c_4}{3}+\frac{5 \alpha _0}{3}+\alpha _1-\frac{10600}{9}\right) \delta ^3+\left(12 c_4-\frac{85 \alpha _0}{36}-\frac{11 \alpha _1}{6}+\frac{61100}{27}\right) \delta ^2\\[1ex]
\rule{0pt}{3.5ex} & 
+\left(-\frac{76 c_4}{3}+\alpha _1-\frac{4480}{3}\right) \delta +40 c_4+\alpha _0+\frac{400}{3} 
\\[1ex]
\hline
\rule{0pt}{3.5ex} \frac{F_2}{(\delta-1)\,(\delta -2)\,(\delta - 3)} & 
-\frac{20 c_4}{3}-\frac{\alpha _0}{6}+\delta  \left(-\frac{11 \alpha _0}{36}-\frac{\alpha _1}{6}+\frac{4540}{27}\right)-\frac{200}{9}
\\[1ex]
 \hline
\rule{0pt}{3.5ex} \frac{F_3}{(\delta-1)\,(\delta -2)}  & 
\left(\frac{80 c_4}{27}-\frac{11 \alpha _0}{72}-\frac{\alpha _1}{12}+\frac{2350}{27}\right) \delta ^2+\left(\frac{70 c_4}{9}+\frac{3 \alpha _0}{8}+\frac{\alpha _1}{4}-\frac{2750}{9}\right) \delta +\frac{50 c_4}{3}+\frac{\alpha _0}{4}
\\[1ex]
\hline
\rule{0pt}{3.5ex} \frac{F_4}{(\delta-1)\,(\delta -2)\,(\delta - 3)}  &
-\frac{\alpha _0}{12}+\delta  \left(-\frac{11 \alpha _0}{72}-\frac{\alpha _1}{12}+\frac{2510}{27}\right)-\frac{100}{9}
\\[1ex]
\midrule\midrule 
\multicolumn{2}{C}{\rule{0pt}{6.5ex}\text{Case } 2} 
 \\
 \midrule\midrule
\rule{0pt}{3.5ex} j_{1;3} & 5 \delta ^2-\frac{19 \delta }{3}+2 \\[1ex]
 \hline
\rule{0pt}{3.5ex} j_{1;4} & a_4 \delta ^2+b_4 \delta +c_4 \\[1ex]
 \hline
\rule{0pt}{3.5ex} \multirow{3}{*}{$F_1$} & 
\left(-\frac{56 a_4}{3}-\frac{85 \alpha _0}{216}-\frac{11 \alpha _1}{36}-\frac{\alpha _2}{6}-\frac{14600}{81}\right) \delta ^5+\left(12 a_4-\frac{56 b_4}{3}+\frac{37 \alpha _0}{18}+\frac{5 \alpha _1}{3}+\alpha _2+\frac{46208}{27}\right) \delta ^4
\\[1ex]
\rule{0pt}{3.5ex}  & 
+\left(-\frac{76 a_4}{3}+12 b_4-\frac{56 c_4}{3}-\frac{575 \alpha _0}{216}-\frac{85 \alpha _1}{36}-\frac{11 \alpha _2}{6}-\frac{322744}{81}\right) \delta ^3+\left(40 a_4-\frac{76 b_4}{3}+12 c_4+\alpha _2+3544\right) \delta ^2
\\[1ex]
\rule{0pt}{3.5ex}  & 
+\left(40 b_4-\frac{76 c_4}{3}+\alpha _1-1240\right) \delta +40 c_4+\alpha _0+\frac{400}{3}
\\[1ex]
\hline
\rule{0pt}{3.5ex} \frac{F_2}{(\delta-1)\,(\delta -2)\,(\delta - 3)} &
\left(-\frac{20 a_4}{3}-\frac{85 \alpha _0}{216}-\frac{11 \alpha _1}{36}-\frac{\alpha _2}{6}-\frac{19460}{81}\right) \delta ^2+\left(-\frac{20 b_4}{3}-\frac{11 \alpha _0}{36}-\frac{\alpha _1}{6}+\frac{3400}{27}\right) \delta -\frac{20 c_4}{3}-\frac{\alpha _0}{6}-\frac{200}{9}
\\[1ex]
 \hline
\rule{0pt}{3.5ex} \multirow{2}{*}{$\frac{F_3}{(\delta-1)\,(\delta -2)}$}  & 
\left(\frac{80 a_4}{27}-\frac{400}{27}\right) \delta ^4+\left(\frac{70 a_4}{9}+\frac{80 b_4}{27}-\frac{85 \alpha _0}{432}-\frac{11 \alpha _1}{72}-\frac{\alpha _2}{12}-\frac{12710}{81}\right) \delta ^3
\\[1ex]
\rule{0pt}{3.5ex}  & 
+\left(\frac{50 a_4}{3}+\frac{70 b_4}{9}+\frac{80 c_4}{27}+\frac{7 \alpha _0}{16}+\frac{3 \alpha _1}{8}+\frac{\alpha _2}{4}+\frac{1390}{3}\right) \delta ^2+\left(\frac{50 b_4}{3}+\frac{70 c_4}{9}+\frac{3 \alpha _0}{8}+\frac{\alpha _1}{4}-200\right) \delta +\frac{50 c_4}{3}+\frac{\alpha _0}{4}
\\[1ex]
\hline
 \rule{0pt}{3.5ex} \frac{F_4}{(\delta-1)\,(\delta -2)\,(\delta - 3)}  & 
 \left(-\frac{85 \alpha _0}{432}-\frac{11 \alpha _1}{72}-\frac{\alpha _2}{12}-\frac{11080}{81}\right) \delta ^2+\left(-\frac{11 \alpha _0}{72}-\frac{\alpha _1}{12}+\frac{2510}{27}\right) \delta -\frac{\alpha _0}{12}-\frac{100}{9}
 \\[1ex]
\midrule\midrule 
\multicolumn{2}{C}{\rule{0pt}{6.5ex}\text{Case } 3} 
 \\
 \midrule\midrule
\rule{0pt}{3.5ex} j_{1;3} & a_3 \delta ^2+\left(-3 a_3-\frac{c_3}{3}+\frac{28}{3}\right) \delta +c_3 \\[1ex]
 \hline
\rule{0pt}{3.5ex} j_{1;4} & 5 \delta ^2-\frac{19 \delta }{3}+c_4 \\[1ex]
 \hline
\rule{0pt}{3.5ex} \multirow{5}{*}{$F_1$} & 
\left(\frac{16 c_4 a_3}{9}-\frac{200 a_3}{9}-\frac{80 c_4}{9}-\frac{85 \alpha _0}{216}-\frac{11 \alpha _1}{36}-\frac{\alpha _2}{6}-\frac{13160}{81}\right) \delta ^5
\\[1ex]
\rule{0pt}{3.5ex} & 
+\left(\frac{2 c_4 a_3}{3}+\frac{200 a_3}{3}+\frac{200 c_3}{27}-\frac{16 c_3 c_4}{27}-\frac{58 c_4}{27}+\frac{37 \alpha _0}{18}+\frac{5 \alpha _1}{3}+\alpha _2+\frac{41620}{27}\right) \delta ^4
\\[1ex]
\rule{0pt}{3.5ex} & 
+\left(-\frac{166}{9} c_4 a_3-\frac{220 a_3}{9}-\frac{200 c_3}{9}-\frac{2 c_3 c_4}{9}+74 c_4-\frac{575 \alpha _0}{216}-\frac{85 \alpha _1}{36}-\frac{11 \alpha _2}{6}-\frac{325660}{81}\right) \delta ^3
\\[1ex]
\rule{0pt}{3.5ex} & 
+\left(-12 c_4 a_3+140 a_3+\frac{220 c_3}{27}+\frac{166 c_3 c_4}{27}+\frac{1612 c_4}{27}+\alpha _2+\frac{86080}{27}\right) \delta ^2
\\[1ex]
\rule{0pt}{3.5ex} & 
+\left(40 c_4 a_3-200 a_3-\frac{140 c_3}{3}+4 c_3 c_4-\frac{700 c_4}{3}+\alpha _1-400\right) \delta +\frac{200 c_3}{3}-\frac{40 c_3 c_4}{3}+\frac{200 c_4}{3}+\alpha _0
\\[1ex]
\hline
\rule{0pt}{3.5ex} \multirow{2}{*}{$\frac{F_2}{(\delta-1)\,(\delta -2)\,(\delta - 3)}$} &
\left(-\frac{20}{9} c_4 a_3-\frac{20 a_3}{9}+\frac{100 c_4}{9}-\frac{85 \alpha _0}{216}-\frac{11 \alpha _1}{36}-\frac{\alpha _2}{6}-\frac{21260}{81}\right) \delta ^2
\\[1ex]
\rule{0pt}{3.5ex}  & 
+\left(-\frac{20}{3} c_4 a_3+\frac{100 a_3}{3}+\frac{20 c_3}{27}+\frac{20 c_3 c_4}{27}+\frac{860 c_4}{27}-\frac{11 \alpha _0}{36}-\frac{\alpha _1}{6}\right) \delta -\frac{100 c_3}{9}+\frac{20 c_3 c_4}{9}-\frac{100 c_4}{9}-\frac{\alpha _0}{6}
\\[1ex]
 \hline
\rule{0pt}{3.5ex} \multirow{2}{*}{$\frac{F_3}{(\delta-1)\,(\delta -2)}$}  & 
\left(-\frac{85 \alpha _0}{432}-\frac{11 \alpha _1}{72}-\frac{\alpha _2}{12}-\frac{11080}{81}\right) \delta ^3+\left(\frac{80 c_4}{27}+\frac{7 \alpha _0}{16}+\frac{3 \alpha _1}{8}+\frac{\alpha _2}{4}+\frac{13430}{27}\right) \delta ^2+\left(\frac{70 c_4}{9}+\frac{3 \alpha _0}{8}+\frac{\alpha _1}{4}-\frac{2750}{9}\right) \delta
\\[1ex]
\rule{0pt}{3.5ex} & +\frac{50 c_4}{3}+\frac{\alpha _0}{4}
\\[1ex]
\hline
\rule{0pt}{3.5ex} \multirow{2}{*}{$\frac{F_4}{(\delta-1)\,(\delta -2)\,(\delta - 3)}$}  &
\left(\frac{80 a_3}{27}-\frac{400}{27}\right) \delta ^3+\left(\frac{70 a_3}{9}-\frac{80 c_3}{81}-\frac{85 \alpha _0}{432}-\frac{11 \alpha _1}{72}-\frac{\alpha _2}{12}-\frac{4690}{27}\right) \delta ^2+\left(\frac{50 a_3}{3}-\frac{70 c_3}{27}-\frac{11 \alpha _0}{72}-\frac{\alpha _1}{12}+\frac{400}{27}\right) \delta
\\[1ex]
\rule{0pt}{3.5ex} & 
-\frac{50 c_3}{9}-\frac{\alpha _0}{12}
\\[1ex]
\midrule\midrule 
  \caption{
  Explicit preliminary results for the seven cases that emerged in the analysis of Section~\ref{sec:prelim}.
  }
  \label{tab:PrelimResult}
\end{xltabular}
}

\section{Analytic continuation in string mass level}
\label{adx:qscRegge}

The quantum spectral curve (QSC)
is an integrability-based method originally developed to solve the spectral problem for 
for local operators in planar $\mathcal{N} = 4$ SYM developed in $\mathcal{N} = 4$ SYM in~\cite{Gromov:2013pga,Gromov:2014caa}.
Its modification to describe an analytic continuation in spin-label was developed in~\cite{Alfimov:2014bwa,Gromov:2015wca,Alfimov:2018cms} (see~\cite{Alfimov:2020obh} for a recent review)
(see also~\cite{Klabbers:2023zdz,Ekhammar:2024neh}).
For a thorough review of the QSC, see~\cite{Gromov:2017blm,Kazakov:2018hrh,Levkovich-Maslyuk:2019awk}. For a practical user's manual, see~\cite{Gromov:2023hzc}. Below we present an abridged review of the fundamentals of the QSC construction.

A state in $\mathcal{N} = 4$ SYM can be uniquely identified by its \texttt{State ID}, introduced in~\cite{Gromov:2023hzc}. It is of the form
\begin{gather}
\texttt{State ID}:    {}_{\Delta_0}[n_{{\bf b}_1}\ n_{{\bf b}_2} \  n_{{\bf f}_1}\ n_{{\bf f}_2},n_{{\bf f}_3}\ n_{{\bf f}_4}\  n_{{\bf a}_1}\ n_{{\bf a}_2}]_{\texttt{sol}}\;.
\end{gather}
Here $\Delta_0$ is the bare/classical/engineering dimension of the state, \textit{i.e.} its scaling dimension at 't Hooft coupling $\lambda = 0$. 
The $n_{\bf a}$, $n_{\bf b}$ and $n_{\bf f}$ are oscillator numbers~\cite{Beisert:2003jj,Gunaydin:1984fk,Bars:1982ep,Beisert:2004ry,Marboe:2017dmb}, which can be used to quantify the field content at zero 't Hooft coupling. Finally $\texttt{sol}$ is a multiplicity label, which is used to break the degeneracy amongst states which have the same $\Delta_0$ and oscillator/field content. The 
quantum numbers of a state: $[\ell_1\;\ell_2;\;q_1\;p\;q_2]$ can be recovered from the oscillator numbers as
\begin{gather}
\label{eqn:osctoDynkin}
    \ell_1 = n_{\mathbf{b}_{2}} - n_{\mathbf{b}_{1}},\  \ell_2 = n_{\mathbf{a}_{1}} - n_{\mathbf{a}_{2}},\  q_1 = n_{\mathbf{f}_{1}} - n_{\mathbf{f}_{2}},\  p = n_{\mathbf{f}_{2}} - n_{\mathbf{f}_{3}},\  q_2 = n_{\mathbf{f}_{3}} - n_{\mathbf{f}_{4}}\;.
\end{gather}
Finally, in the integrability literature, the 't Hooft coupling is often rescaled to 
\begin{align}
    g\equiv \frac{\sqrt{\lambda}}{4\,\pi}\;.
\end{align}
We use both $\lambda$ and $g$ in the passages below.

The QSC construction can be summarised as follows:
\begin{enumerate}
    \item To each state, associate $2^{\mathtt{rank}(\mathrm{PSU}(2,2|4))+1} = 256$ $Q$-functions. The $Q$-functions are functions of a complex variable $u$, called the spectral parameter. 
    \item The $\mathrm{PSU}(2,2|4)$ charges of the state: $[\Delta(\lambda);\;\ell_1\;\ell_2;\;q_1\;p\;q_2]$ are contained in the large-$u$ asymptotics of the $Q$-functions. 
    \item $Q$-functions satisfy various functional relations amongst each other, called $QQ$-relations. One can start with a set of 8 distinguished $Q$-functions, and using the $QQ$-relations, build all 256 $Q$-functions. The distinguished $Q$-functions have a special notation: $\mathbf{P}_a$ and $\mathbf{Q}_i$ for $a,i=1,\dots,4$.  
    \item The distinguished $Q$-functions, are all analytic in the upper half plane. $\mathbf{P}_a$ have two branch points, at $\pm 2\,g$, which are connected by a short branch cut, \textit{i.e.} a branch cut passing through the origin. $\mathbf{Q}_i$ have an infinite ladder of branch points located at $\pm 2\,g - \mathrm{i}\,n$, for $n \in\mathbb{Z}^*$. Each pair of branch points at the same value of $n$, are connected via a short cut.  
    \item The discontinuities of the $\mathbf{Q}_i$ about their branch cut on the real axis are imposed. Together with the asymptotic, this defines a Riemann-Hilbert problem for the distinguished $Q$-functions. This Riemann-Hilbert problem can be solved in a variety of analytical~\cite{Marboe:2014gma,Alfimov:2014bwa,Marboe:2017dmb,Marboe:2018ugv} and numerical~\cite{Gromov:2015wca,Hegedus:2016eop,Gromov:2023hzc} ways. Currently QSC descriptions exist for local operators in $\mathcal{N} = 4$ SYM, ABJM~\cite{Cavaglia:2014exa,Bombardelli:2017vhk,Bombardelli:2018bqz}, AdS${}_3$~\cite{Cavaglia:2021eqr,Ekhammar:2021pys,Cavaglia:2022xld} and 1D dCFT~\cite{Grabner:2020nis,Julius:2021uka}.
\end{enumerate}

\paragraph{Basics of quantum spectral curve.} 
The large-$u$ asymptotic of the distinguished $Q$-functions are given by
\begin{align}
    \label{eqn:PQasymDef}
    {\bf P}_a \simeq \mathbb{A}_a\, u^{\mathtt{powP}_a} \;,\qquad {\bf Q}_i \simeq \mathbb{B}_i\, u^{\mathtt{powQ}_i}\;,
\end{align}
where $\mathbb{A}_a$ and $\mathbb{B}_i$ are constants, defined below in equation~\eqref{Aafix} and~\eqref{eqn:Bdown}. The asymptotic powers are
\begin{align}
    \mathtt{powP} &= \{
     -n_{{\bf f}_1} - 2 - \Lambda, -n_{{\bf f}_2} - 1 - \Lambda, -n_{{\bf f}_3} - \Lambda, -n_{{\bf f}_4} + 1 - \Lambda
    \}\;,\\
    \mathtt{powQ} &= \bigg\{
    L + \frac{\gamma}{2} + n_{{\bf b}_1}  + \Lambda, L + \frac{\gamma}{2}  + n_{{\bf b}_2} + 1 + \Lambda, - \frac{\gamma}{2}  - n_{{\bf a}_1} - 2 + \Lambda, - \frac{\gamma}{2}  - n_{{\bf a}_2} -1  + \Lambda
    \bigg\}\;,
    \label{eqn:powQ}
\end{align}
where $\Lambda$ is the ambiguity related to a translational symmetry of the power of the $Q$-functions, the anomalous dimension $\gamma$ is defined as
\begin{align}
    \gamma \equiv \Delta - \Delta_0\;,
\end{align}
and $\Delta_0$ and $L$ can be written in terms of the oscillator numbers as
\begin{align}
\label{Delta0inn}
    \Delta_ 0 &= \sum_{i = 1}^4 \frac{n_{{\bf f}_i}}{2} +\sum_{i = 1}^2 n_{{\bf a}_i}\;, \\ \nonumber
    \label{Linn}
    L &= \sum_{i = 1}^4 \frac{n_{{\bf f}_i}}{2} +
    \sum_{i = 1}^2 \left(\frac{ n_{{\bf a}_i}}{2} - \frac{n_{{\bf b}_i}}{2} \right)\;.
\end{align}
Finally the asymptotic coeffieicnts are given by
\begin{equation} \label{Aafix}
\mathbb{A}_a =  \, \frac{(\mathtt{powP}_a+\mathtt{powQ}_1 + 1)(\mathtt{powP}_a+\mathtt{powQ}_2+1)}{ \prod\limits_{b > a} i (\mathtt{powP}_a-\mathtt{powP}_b)}\;,
\end{equation}
\begin{equation}\label{eqn:Bdown}
  \mathbb{B}_j =\begin{cases}
    \dfrac{1}{ \prod_{k > j} i  (-\mathtt{powQ}_j + \mathtt{powQ}_k)}\;, & \text{for }j=1,2\;,\\[15pt]
    \dfrac{\prod_{k} (\mathtt{powP}_k + \mathtt{powQ}_j + 1)}{ \prod_{k > j}  i(\mathtt{powQ}_j - \mathtt{powQ}_k)}\;, & \text{for }j=3,4\;,
  \end{cases}
\end{equation}
\paragraph{Description of the numerical method.}
Due to the one-cut nature of the $\mathbf{P}_a$, they are parameterised in terms of the Zhukovsky variable in a convergent expansion.
\begin{align}\label{eqn:ZhukDef}
    x(u) \equiv \frac{u + \sqrt{u-2g}\sqrt{u+2g}}{2g}\;.
\end{align}
as
\begin{align}
    \mathbf{P}_a = \mathbb{A}_a\,(g\,x)^{\mathtt{powP}_a}\bigg[1 + \sum_{n=1}^\mathtt{cutP} \frac{c_{a,n}}{x^n} \bigg]
    \;.
\end{align}
Here, $\mathtt{cutP}$ is a cutoff introduced in numerical implementations of the QSC, which gives us a handle on the precision of the obtained numerical data. 
The parameters of our numerical problem, denoted as $\mathtt{params}$, which will be fixed by solving the Riemann-Hilbert problem are
\begin{align}\label{eqn:paramsdef}
    \mathtt{params}
    =
    \bigg\{\Delta\,,\;c_{1,1}\,,\;\dots\,,\;c_{1,\mathtt{cutP}}\,,\;c_{2,1}\,,\;\dots\,,\;c_{2,\mathtt{cutP}}\,,\;c_{3,1}\,,\;\dots\,,\;c_{3,\mathtt{cutP}}\,,\;c_{4,1}\,,\;\dots\,,\;c_{4,\mathtt{cutP}}\,,\;\bigg\}
    \;.
\end{align}
It is easier to impose the discontinuity relations on the $\mathbf{Q}_i$, so one needs to build $Q_i$ from the $\mathbf{P}_a$. 
This is done by using the $QQ$-relations as mentioned earlier. 
Specifically we need
\begin{align}
    \label{eqn:QfromP}
    \mathbf{Q}_i &= -\mathbf{P}^b Q_{b|i}\;,\\
    \label{eqm:FinDiff}
    Q_{a|i}^+ - Q_{a|i}^- &= \mathbf{P}_a \mathbf{Q}_i\;.
\end{align}
Here $Q_{a|i}$ are intermediate $Q$-functions that we need to construct, and the indices are raised and lowered by the fully anti-symmetric matrix\footnote{Note that this is not the case in general, but indeed is the case for all the states considered in this paper. For a through discussion in full generality, see~\cite{Gromov:2023hzc}.} 
\begin{equation}
  \chi_{ab} = -\chi^{ab}  =  \begin{pmatrix}
0 & 0 & 0 & 1 \\
0 & 0 & -1 & 0\\
0 & 1 & 0 & 0 \\
-1 & 0 & 0 & 0\\
\end{pmatrix}.
\end{equation}
In the numerical solution, we begin with initial numerical values for the parameters $\mathtt{params}$~\eqref{eqn:paramsdef} at some value of $\lambda$. Then, we build $\mathbf{P}_a$, and combine equations~\eqref{eqn:QfromP} and~\eqref{eqm:FinDiff} to construct $Q_{a|i}$ in large-$u$ regime, by solving this finite difference equation:
\begin{align}\label{eqn:PPQai}
    Q_{a|i}^+ - Q_{a|i}^- = -\mathbf{P}_a\mathbf{P}^b Q_{b|i}^+
    \;.
\end{align}
The result obtained for $Q_{a|i}$ as asymptotic series in $1/u$ and is as such, valid only in the large-$u$ regime.
In order to impose the discontinuity relation, we need to build $\mathbf{Q}_i$ above the branch cut on the real axis. To do so, we rewrite equation~\eqref{eqn:PPQai} as 
\begin{align}
    Q_{a|i}(u - i/2) = \bigg[\mathbf{P}_a(u)\mathbf{P}^b(u) - \delta_a^b\bigg] Q_{b|i}(u + i/2)\;.
\end{align}
Using the above equation recursively, we can write, for some point $u_0$, on the real axis, that
\begin{align}\label{eqn:QQrecurse}
    Q_{a|i}(u_0 + i/2) = 
    Q_{b|i}(u_0 + \mathtt{QaiShift}\,i/2)
    \prod_{n = 1}^{\mathtt{QaiShift}}\bigg[\mathbf{P}_a(u_0 + n\,\mathrm{i}/2)\mathbf{P}^b(u_0 + n\,\mathrm{i}/2) - \delta_a^b\bigg]
    \;.
\end{align}
Here, $\mathtt{QaiShift}$ is a large enough integer, so that the asymptotic expansion of $Q_{a|i}$, at $(u_0 + \mathtt{QaiShift}\,i/2)$ has the desired precision.
Notice that the block $[\mathbf{P}_a\mathbf{P}^b - \delta_a^b]$ is made out of $\mathbf{P}_a$ and $\mathbf{P}^a$, and as such, can be constructed anywhere on the $u$-plane, not just in the large-$u$ region. This is the critical point that allows equation~\eqref{eqn:QQrecurse} to be used to reduce $\mathrm{Im}(u)$, until we approach the real axis. 

We can now construct $\mathbf{Q}_i$ on the main Riemann sheet, using equation~\eqref{eqm:FinDiff}, on various points, above the branch cut on the real axis. 
Using an important property of the Zhukovsky variable, that the transformation $x\rightarrow 1/x$ takes us to the second Riemann sheet, we get that $\tilde{\mathbf{P}}_a = \mathbf{P}_a(1/x)$, and therefore, that
\begin{align}
    \tilde{\mathbf{Q}}_i = -\tilde{\mathbf{P}}^bQ_{b|i}^+\;.
\end{align}
Now that we have the $\mathbf{Q}_i$ on the main sheet, and $\tilde{\mathbf{Q}}_i$, on the second sheet, we can impose the discontinuity relations. They are most easily written in the form
\begin{align}\label{Qgluing}
    \tilde{\mathbf{Q}}_{i} = G_{i}^{\; j}\, \bar{\mathbf{Q}}_{j}\;,
\end{align}
where $G_{i}^{\; j}(u)$ is called the gluing matrix, and the discontinuity relations are also known as gluing conditions. 
The exact form of the gluing matrix depends on the class of states being considered. We will specify it in more detail ahead. 
One can define an optimisation problem on the gluing condition and solve it using the numerical methods developed in~\cite{Gromov:2015wca,Gromov:2023hzc}.
The solution to this optimisation problem fixes $\mathtt{params}$.
Below we present the gluing matrix for classes of states that we consider in this paper. 

\paragraph{Form of gluing matrices.} The first case we will consider is that of local operators. In this case, the gluing matrix takes the form \cite{Gromov:2014caa}
\begin{align}
    G_{i}^{\; j} =
    \left(\begin{array}{c c c c}
        0 & 0 & \alpha & 0 \\
        0 & 0 & 0 & -\bar{\alpha} \\
        \frac{1}{\bar{\alpha}} & 0 & 0 & 0 \\
        0 & -\frac{1}{\alpha} & 0 & 0
    \end{array}\right)_{i\,j}
    \;,
    \label{GluingMatrice}
\end{align}
where $\alpha$ is some constant. The second case that we consider is for states with non-integer values of the spin-label $\ell$. In general, the gluing matrix for such states, has the form~\cite{Gromov:2017blm}
\begin{align}
    G_{i}^{\; j} =
    \left(\begin{array}{c c c c}
        0 & 0 & \alpha & 0 \\
        \beta & 0 & \gamma & -\bar{\alpha} \\
        \frac{1}{\bar{\alpha}} & 0 & 0 & 0 \\
        \frac{\gamma}{\alpha\,\bar{\alpha}} & -\frac{1}{\alpha} & \bar{\beta} & 0
    \end{array}\right)_{i\,j}
    \;.
    \label{GluingMatriceAnaSpin}
\end{align}
Here $\beta$ and $\gamma$ can be non-trivial $\mathrm{i}$-periodic functions of $u$.
For the special case of states that are of Type I, which are non-degenarate due to a parity symmetry of the spectral parameter $u$, \textit{i.e.} invariance under $u \rightarrow -u$ (see~\cite{Gromov:2023hzc} for details) , the form of the gluing matrix 
at non-integer values of the spin-label $\ell$ is~\cite{Alfimov:2018cms}
\begin{align}\label{eqn:typeIanaspin}
\begin{split}
    G_{i}^{\; j} &=  \chi_{i\,k}L^{k\,j}\;\\ L^{k\,j} &= 
    \left(\begin{array}{c c c c}
        L^{(1)}_{1\,1} & L^{(1)}_{1\,2} & L^{(1)}_{1\,3} & 0 \\
        L^{(1)}_{1\,2} & 0 & 0 & 0 \\
        L^{(1)}_{1\,3} & 0 & L^{(1)}_{3\,3} & L^{(1)}_{3\,4} \\
        0 & 0 & L^{(1)}_{3\,4} & 0
    \end{array}\right)_{k\,j}
    + 
    e^{2\,\pi\,u}
    \left(\begin{array}{c c c c}
        0 & 0 & L^{(2)}_{1\,3} & 0 \\
        0 & 0 & 0 & 0 \\
        L^{(2)}_{3\,1} & 0 & 0 & 0 \\
        0 & 0 & 0 & 0
    \end{array}\right)_{k\,j}
    + 
    e^{-2\,\pi\,u}
    \left(\begin{array}{c c c c}
        0 & 0 & L^{(2)}_{3\,1} & 0 \\
        0 & 0 & 0 & 0 \\
        L^{(2)}_{1\,3} & 0 & 0 & 0 \\
        0 & 0 & 0 & 0
    \end{array}\right)_{k\,j}
    \;.
\end{split}
\end{align}
Here the $L^{(n)}_{i\,j}$ are constants. 
For the case of states that are of Type II, which come in exactly degenerate doublets due to the absence of parity symmetry of the spectral parameter (see~\cite{Gromov:2023hzc} for details), the gluing matrix is of the form 
\begin{align}\label{eqn:typeIIanaspin}
\begin{split}
    G_{i}^{\; j} &=  \chi_{i\,k}L^{k\,j}\;\\ L^{k\,j} &= 
    \left(\begin{array}{c c c c}
        L^{(1)}_{1\,1} & L^{(1)}_{1\,2} & L^{(1)}_{1\,3} & 0 \\
        \bar{L}^{(1)}_{1\,2} & 0 & 0 & 0 \\
        \bar{L}^{(1)}_{1\,3} & 0 & L^{(1)}_{3\,3} & L^{(1)}_{3\,4} \\
        0 & 0 & \bar{L}^{(1)}_{3\,4} & 0
    \end{array}\right)_{k\,j}
    + 
    e^{2\,\pi\,u}
    \left(\begin{array}{c c c c}
        0 & 0 & L^{(2)}_{1\,3} & 0 \\
        0 & 0 & 0 & 0 \\
        \bar{L}^{(2)}_{1\,3} & 0 & 0 & 0 \\
        0 & 0 & 0 & 0
    \end{array}\right)_{k\,j}
    + 
    e^{-2\,\pi\,u}
    \left(\begin{array}{c c c c}
        0 & 0 & L^{(3)}_{1\,3} & 0 \\
        0 & 0 & 0 & 0 \\
        \bar{L}^{(3)}_{1\,3} & 0 & 0 & 0 \\
        0 & 0 & 0 & 0
    \end{array}\right)_{k\,j}
    \;.
\end{split}
\end{align}
The above form of the gluing matrix is a was obtained by us in this work. To obtain it, we used a remarkable property of the QSC, that not all the gluing conditions are independent~\cite{Gromov:2017blm}.
Thus, it was enough for us to impose just the $\tilde{\mathbf{Q}}_1 = \alpha\,\bar{\mathbf{Q}}_3$ gluing condition, to make the numerical solution converge.
As this particular relation is the same for both the integer and non-integer spin-label cases, we used it to obtain some non-integer data points, before doing a numerical fit to infer the form displayed in~\eqref{eqn:typeIIanaspin}. 

\paragraph{Extracting spectral data on sub-leading Regge trajectories.}
Consider any state from Table~\ref{tab:EvenReggeStates}; such a state is associated with a value of $\delta$ and $\ell$. In addition, $\ell = 2(\delta - n)$ for some value of $n$, \textit{i.e.} the state is associated with Regge trajectory number $n$. This can be rewritten as 
\begin{align}\label{eqn:delfromell}
    \delta = \frac{\ell}{2} + n
    \;.
\end{align}
Our procedure is detailed below in the following steps:
\begin{enumerate}
    \item We start at a particular value of the 't Hooft coupling $\lambda$; in practice we use $\lambda = 16/25\,\pi^2$. We can obtain the spectral data at this point from the database of/using the method developed in~\cite{Gromov:2023hzc}.\footnote{The original numerical method to solve the QSC was developed in~\cite{Gromov:2015wca} and the first \texttt{C++} implementation of this method was developed in~\cite{Hegedus:2016eop}.}
    \item Next, we want to perturb this point slightly in the spin-label $\ell$, say by a small amount $d\ell$. Remember that for these states $\ell_1 = \ell_2 = \ell$. In terms of Dynkin labels, this means, \textit{cf.} equation~\eqref{eqn:osctoDynkin}, that $\ell = n_{\mathbf{b}_{2}} - n_{\mathbf{b}_{1}} = n_{\mathbf{a}_{1}} - n_{\mathbf{a}_{2}}$. Therefore we get
    \begin{align}
        \ell \rightarrow \ell + d\ell \quad\Rightarrow\quad n_{\mathbf{b}_{2}} \rightarrow n_{\mathbf{b}_{2}} + d\ell\quad\text{and}\quad n_{\mathbf{a}_{1}} \rightarrow n_{\mathbf{a}_{1}} + d\ell\;.
    \end{align}
    This changes the asymptotic powers~\eqref{eqn:powQ} and the asymptotic coefficients~\eqref{Aafix} and~\eqref{eqn:Bdown} correspondingly. 
    \item Let $\mathtt{params}(\lambda,\ell)$ be the converged values of the parameters of this state for some given precision. We make a starting point for the numerical solution, $\mathtt{params}_0(\lambda,\ell+d\ell)$, by using an interpolating polynomial, like in the implementation of~\cite{Alfimov:2018cms}.
    \item Then, we follow the steps of methods developed in the literature to develop a numerical solution. 
    We use a different gluing condition depending on the Type of the state: 
    for Type I states, we use the gluing matrix~\eqref{eqn:typeIanaspin}, whilst for Type II states, we use the gluing matrix~\eqref{eqn:typeIIanaspin}.\footnote{It is also of technical importance to point out that since the value of the spin-label $\ell$ is no longer integer, therefore, there is no longer a null vector obtained while solving~\eqref{eqn:PPQai} in the large-$u$ expansion of $Q_{a|i}$, and as such, there are no extra constraints on the $c_{a,n}$, as one obtains in the case of local operators for Type II (and indeed Type IV) states, \textit{cf.}~\cite{Gromov:2023hzc}.} 
    \item Initially, we need very small values of $d\ell$ to ensure that the numerical method converges. In practice, we start with $d\ell = 1/10/2^{28} = 1/2684354560$. Then, as we generate more points, we gradually increase the $d\ell$. 
    \item We begin with the sates on the sub-leading Regge trajectory with $\ell = 0$. These have \texttt{State ID} ${}_{4}[0\;0\;2\;2\;2\;2\;0\;0]_{1}$ and $[0\;0\;2\;2\;2\;2\;0\;0]_{2}$ 
    For both states, as mentioned earlier, we start at $\lambda = 16/25\,\pi^2$. By slowly varying $\ell$, we produce data points for various values of $\ell$, until $\ell = 4$.
    At $\ell = 2$ and $\ell = 4$ we see that the analytically continued scaling dimension of the state with \texttt{State ID} ${}_{4}[0\;0\;2\;2\;2\;2\;0\;0]_{1}$ matches the state with  \texttt{State ID} ${}_{6}[0\;2\;2\;2\;2\;2\;2\;0]_{2}$
    and ${}_{8}[0\;4\;2\;2\;2\;2\;4\;0]_{1}$ respectively.
    We also see that the at $\ell = 2$ and $\ell = 4$, the analytically continued scaling dimension of the state with \texttt{State ID} $[0\;0\;2\;2\;2\;2\;0\;0]_{2}$ matches the state with  \texttt{State ID} ${}_{6}[0\;2\;2\;2\;2\;2\;2\;0]_{3}$
    and ${}_{8}[0\;4\;2\;2\;2\;2\;4\;0]_{4}$ respectively.
    Thus, we identify that the states with \texttt{State ID} 
    ${}_{4}[0\;0\;2\;2\;2\;2\;0\;0]_{2}$, ${}_{6}[0\;2\;2\;2\;2\;2\;2\;0]_{3}$ and ${}_{8}[0\;4\;2\;2\;2\;2\;4\;0]_{4}$
    to be on one of the sub-leading Regge trajectories, associated with the sub-leading quadratic Casimir eigenvalue $j_{1;1}$, and 
    ${}_{4}[0\;0\;2\;2\;2\;2\;0\;0]_{1}$, ${}_{6}[0\;2\;2\;2\;2\;2\;2\;0]_{2}$
    and ${}_{8}[0\;4\;2\;2\;2\;2\;4\;0]_{1}$
    as being on another sub-leading Regge trajectory, with associated sub-leading quadratic Casimir eigenvalue $j_{1;2}$.
    Then we move on to other states with $\ell = 2$. We have the exactly degenerate Type II states ${}_{6}[0\;2\;2\;2\;2\;2\;2\;0]_{4/5}$. 
    For these states, we see that at $\ell = 4$, the scaling dimension matches that of 
    the states with \texttt{State ID} ${}_{8}[0\;4\;2\;2\;2\;2\;4\;0]_{13/14}$,
    thus identifying the third/fourth sub-leading Regge trajectory.
    Finally, we move on to $\ell = 4$, and construct the fifth/sixth Regge trajectory, associated with states with \texttt{State ID} ${}_{8}[0\;4\;2\;2\;2\;2\;4\;0]_{15/16}$.
    \item Now that we have identified the six sub-leading Regge trajectories, we can produce more spectral data on a given trajectory. On each of the sub-leading trajectories, we take the states with $\ell = 0,2,4$ for the Type I states, and $\ell = 2,4,6$ for the Type II states ($\ell = 2$ is not available on the fifth/sixth sub-leading Regge trajectories). 
    For each such state, we begin with the point that we have produced, at $\lambda = 16/25\,\pi^2$, and systematically increase the value of $\lambda$ until we reach the strong coupling regime.
    \item At strong coupling, the string mass level $\delta$ of these states is given by equation~\eqref{eqn:delfromell}, with $n = 2$, since all the states are on sub-leading Regge trajectories, and therefore have Regge trajectory number 2. 
    Using the fitting procedure described in~\cite{Gromov:2023hzc}, we obtain strong coupling predictions for the scaling dimensions of each of these states; 
    we obtain a fit for the first sub-leading coefficient $j_k$ in~\eqref{eqn:StrongExpCasimir}. 
    \item Once we have obtained a prediction for $j_1$ for various states on the a particular sub-leading Regge trajectory, we are able to see how this number changes along this trajectory and fit it to the ansatz~\eqref{eqn:j1Anz}. 
\end{enumerate}
Applying the above procedure on the first two sub-leading Regge trajectories yields the formulas for $j_{1;1}$ and $j_{1;2}$ displayed in equations~\eqref{eqn:j11pred} and~\eqref{eqn:j12pred} respectively. 

For the third/fourth and fifth/sixth Regge trajectories, we do not have enough numerical precision on individual data points to perform a fit to the ansatz~\eqref{eqn:j1Anz}. Nevertheless, we display below the value of $j_1$ for individual states on the third/fourth and fifth/sixth Regge trajectories, to the best precision that we have:
{
\begin{xltabular}[c]{\textwidth}{C|C|C}
\delta &
 j_{1;3} & j_{1;4} 
 \\
 \midrule\midrule
 \rule{0pt}{3.5ex}4 & 56.4551 & 47.2115 \\[1ex]
\hline
 \rule{0pt}{3.5ex}5 & 94.99 & 83.331 \\[1ex]
  \caption{
  Predictions for the sub-leading quadratic Casimir eigenvalue for states on the third/fourth and fifth/sixth sub-leading Regge trajectories, at specific values of $\delta$.
  }
  \label{tab:j13j14fitdata}
\end{xltabular}
}

\bibliographystyle{JHEP.bst}
\bibliography{references}

\providecommand{\href}[2]{#2}\begingroup\raggedright\begin{thebibliography}{10}

\bibitem{Alday:2022uxp}
L.~F. Alday, T.~Hansen, and J.~A. Silva, {\it {AdS Virasoro-Shapiro from dispersive sum rules}},  {\em JHEP} {\bf 10} (2022) 036, [\href{http://arxiv.org/abs/2204.07542}{{\tt arXiv:2204.07542}}].

\bibitem{Alday:2022xwz}
L.~F. Alday, T.~Hansen, and J.~A. Silva, {\it {AdS Virasoro-Shapiro from single-valued periods}},  {\em JHEP} {\bf 12} (2022) 010, [\href{http://arxiv.org/abs/2209.06223}{{\tt arXiv:2209.06223}}].

\bibitem{Alday:2023mvu}
L.~F. Alday and T.~Hansen, {\it {The AdS Virasoro-Shapiro amplitude}},  {\em JHEP} {\bf 10} (2023) 023, [\href{http://arxiv.org/abs/2306.12786}{{\tt arXiv:2306.12786}}].

\bibitem{Caron-Huot:2017vep}
S.~Caron-Huot, {\it {Analyticity in Spin in Conformal Theories}},  {\em JHEP} {\bf 09} (2017) 078, [\href{http://arxiv.org/abs/1703.00278}{{\tt arXiv:1703.00278}}].

\bibitem{Alday:2017vkk}
L.~F. Alday and S.~Caron-Huot, {\it {Gravitational S-matrix from CFT dispersion relations}},  {\em JHEP} {\bf 12} (2018) 017, [\href{http://arxiv.org/abs/1711.02031}{{\tt arXiv:1711.02031}}].

\bibitem{Homrich:2022mmd}
A.~Homrich, D.~Simmons-Duffin, and P.~Vieira, {\it {Complex Spin: The Missing Zeroes and Newton's Dark Magic}},  \href{http://arxiv.org/abs/2211.13754}{{\tt arXiv:2211.13754}}.

\bibitem{Henriksson:2023cnh}
J.~Henriksson, P.~Kravchuk, and B.~Oertel, {\it {Missing local operators, zeros, and twist-4 trajectories}},  {\em JHEP} {\bf 07} (2024) 248, [\href{http://arxiv.org/abs/2312.09283}{{\tt arXiv:2312.09283}}].

\bibitem{Alday:2023flc}
L.~F. Alday, T.~Hansen, and J.~A. Silva, {\it {On the spectrum and structure constants of short operators in N=4 SYM at strong coupling}},  {\em JHEP} {\bf 08} (2023) 214, [\href{http://arxiv.org/abs/2303.08834}{{\tt arXiv:2303.08834}}].

\bibitem{Julius:2023hre}
J.~Julius and N.~Sokolova, {\it {Conformal field theory-data analysis for $\mathcal{N}$ = 4 Super-Yang-Mills at strong coupling}},  {\em JHEP} {\bf 03} (2024) 090, [\href{http://arxiv.org/abs/2310.06041}{{\tt arXiv:2310.06041}}].

\bibitem{Alday:2023jdk}
L.~F. Alday, T.~Hansen, and J.~A. Silva, {\it {Emergent Worldsheet for the AdS Virasoro-Shapiro Amplitude}},  {\em Phys. Rev. Lett.} {\bf 131} (2023), no.~16 161603, [\href{http://arxiv.org/abs/2305.03593}{{\tt arXiv:2305.03593}}].

\bibitem{Alday:2024yax}
L.~F. Alday, S.~M. Chester, T.~Hansen, and D.-l. Zhong, {\it {The AdS Veneziano amplitude at small curvature}},  {\em JHEP} {\bf 05} (2024) 322, [\href{http://arxiv.org/abs/2403.13877}{{\tt arXiv:2403.13877}}].

\bibitem{Alday:2024ksp}
L.~F. Alday and T.~Hansen, {\it {Single-valuedness of the AdS Veneziano amplitude}},  {\em JHEP} {\bf 08} (2024) 108, [\href{http://arxiv.org/abs/2404.16084}{{\tt arXiv:2404.16084}}].

\bibitem{Maldacena:1997re}
J.~M. Maldacena, {\it {The Large N limit of superconformal field theories and supergravity}},  {\em Adv. Theor. Math. Phys.} {\bf 2} (1998) 231--252, [\href{http://arxiv.org/abs/hep-th/9711200}{{\tt hep-th/9711200}}].

\bibitem{Gubser:1998bc}
S.~S. Gubser, I.~R. Klebanov, and A.~M. Polyakov, {\it {Gauge theory correlators from noncritical string theory}},  {\em Phys. Lett.} {\bf B428} (1998) 105--114, [\href{http://arxiv.org/abs/hep-th/9802109}{{\tt hep-th/9802109}}].

\bibitem{Gromov:2023hzc}
N.~Gromov, A.~Hegedus, J.~Julius, and N.~Sokolova, {\it {Fast QSC solver: tool for systematic study of $ \mathcal{N} $ = 4 Super-Yang-Mills spectrum}},  {\em JHEP} {\bf 05} (2024) 185, [\href{http://arxiv.org/abs/2306.12379}{{\tt arXiv:2306.12379}}].

\bibitem{Gromov:2009tv}
N.~Gromov, V.~Kazakov, and P.~Vieira, {\it {Exact Spectrum of Anomalous Dimensions of Planar N=4 Supersymmetric Yang-Mills Theory}},  {\em Phys. Rev. Lett.} {\bf 103} (2009) 131601, [\href{http://arxiv.org/abs/0901.3753}{{\tt arXiv:0901.3753}}].

\bibitem{Gromov:2009zb}
N.~Gromov, V.~Kazakov, and P.~Vieira, {\it {Exact Spectrum of Planar ${\cal N}=4$ Supersymmetric Yang-Mills Theory: Konishi Dimension at Any Coupling}},  {\em Phys. Rev. Lett.} {\bf 104} (2010) 211601, [\href{http://arxiv.org/abs/0906.4240}{{\tt arXiv:0906.4240}}].

\bibitem{Roiban:2009aa}
R.~Roiban and A.~A. Tseytlin, {\it {Quantum strings in AdS(5) x S**5: Strong-coupling corrections to dimension of Konishi operator}},  {\em JHEP} {\bf 11} (2009) 013, [\href{http://arxiv.org/abs/0906.4294}{{\tt arXiv:0906.4294}}].

\bibitem{Tseytlin:2009fw}
A.~A. Tseytlin, {\it {Quantum strings in AdS(5) x S**5 and AdS/CFT duality}},  {\em Int. J. Mod. Phys. A} {\bf 25} (2010) 319--331, [\href{http://arxiv.org/abs/0907.3238}{{\tt arXiv:0907.3238}}].

\bibitem{Frolov:2010wt}
S.~Frolov, {\it {Konishi operator at intermediate coupling}},  {\em J. Phys. A} {\bf 44} (2011) 065401, [\href{http://arxiv.org/abs/1006.5032}{{\tt arXiv:1006.5032}}].

\bibitem{Passerini:2010xc}
F.~Passerini, J.~Plefka, G.~W. Semenoff, and D.~Young, {\it {On the Spectrum of the $AdS_{5}$ x $S^{5}$ String at large $\lambda$}},  {\em JHEP} {\bf 03} (2011) 046, [\href{http://arxiv.org/abs/1012.4471}{{\tt arXiv:1012.4471}}].

\bibitem{Gromov:2011de}
N.~Gromov, D.~Serban, I.~Shenderovich, and D.~Volin, {\it {Quantum folded string and integrability: From finite size effects to Konishi dimension}},  {\em JHEP} {\bf 08} (2011) 046, [\href{http://arxiv.org/abs/1102.1040}{{\tt arXiv:1102.1040}}].

\bibitem{Roiban:2011fe}
R.~Roiban and A.~A. Tseytlin, {\it {Semiclassical string computation of strong-coupling corrections to dimensions of operators in Konishi multiplet}},  {\em Nucl. Phys. B} {\bf 848} (2011) 251--267, [\href{http://arxiv.org/abs/1102.1209}{{\tt arXiv:1102.1209}}].

\bibitem{Vallilo:2011fj}
B.~C. Vallilo and L.~Mazzucato, {\it {The Konishi multiplet at strong coupling}},  {\em JHEP} {\bf 12} (2011) 029, [\href{http://arxiv.org/abs/1102.1219}{{\tt arXiv:1102.1219}}].

\bibitem{Basso:2011rs}
B.~Basso, {\it {An exact slope for AdS/CFT}},  \href{http://arxiv.org/abs/1109.3154}{{\tt arXiv:1109.3154}}.

\bibitem{Gromov:2011bz}
N.~Gromov and S.~Valatka, {\it {Deeper Look into Short Strings}},  {\em JHEP} {\bf 03} (2012) 058, [\href{http://arxiv.org/abs/1109.6305}{{\tt arXiv:1109.6305}}].

\bibitem{Frolov:2012zv}
S.~Frolov, {\it {Scaling dimensions from the mirror TBA}},  {\em J. Phys. A} {\bf 45} (2012) 305402, [\href{http://arxiv.org/abs/1201.2317}{{\tt arXiv:1201.2317}}].

\bibitem{Beccaria:2012xm}
M.~Beccaria, S.~Giombi, G.~Macorini, R.~Roiban, and A.~A. Tseytlin, {\it {'Short' spinning strings and structure of quantum $AdS_5 \times S^5$ spectrum}},  {\em Phys. Rev. D} {\bf 86} (2012) 066006, [\href{http://arxiv.org/abs/1203.5710}{{\tt arXiv:1203.5710}}].

\bibitem{Gromov:2014bva}
N.~Gromov, F.~Levkovich-Maslyuk, G.~Sizov, and S.~Valatka, {\it {Quantum spectral curve at work: from small spin to strong coupling in $ \mathcal{N} $ = 4 SYM}},  {\em JHEP} {\bf 07} (2014) 156, [\href{http://arxiv.org/abs/1402.0871}{{\tt arXiv:1402.0871}}].

\bibitem{Gromov:2015wca}
N.~Gromov, F.~Levkovich-Maslyuk, and G.~Sizov, {\it {Quantum Spectral Curve and the Numerical Solution of the Spectral Problem in AdS5/CFT4}},  {\em JHEP} {\bf 06} (2016) 036, [\href{http://arxiv.org/abs/1504.06640}{{\tt arXiv:1504.06640}}].

\bibitem{Hegedus:2016eop}
A.~Heged\H{u}s and J.~Konczer, {\it {Strong coupling results in the AdS$_{5}$ /CFT$_{4}$ correspondence from the numerical solution of the quantum spectral curve}},  {\em JHEP} {\bf 08} (2016) 061, [\href{http://arxiv.org/abs/1604.02346}{{\tt arXiv:1604.02346}}].

\bibitem{Costa:2012cb}
M.~S. Costa, V.~Goncalves, and J.~Penedones, {\it {Conformal Regge theory}},  {\em JHEP} {\bf 12} (2012) 091, [\href{http://arxiv.org/abs/1209.4355}{{\tt arXiv:1209.4355}}].

\bibitem{Gromov:2014caa}
N.~Gromov, V.~Kazakov, S.~Leurent, and D.~Volin, {\it {Quantum spectral curve for arbitrary state/operator in AdS$_{5}$/CFT$_{4}$}},  {\em JHEP} {\bf 09} (2015) 187, [\href{http://arxiv.org/abs/1405.4857}{{\tt arXiv:1405.4857}}].

\bibitem{Alfimov:2018cms}
M.~Alfimov, N.~Gromov, and G.~Sizov, {\it {BFKL spectrum of $ \mathcal{N} $ = 4: non-zero conformal spin}},  {\em JHEP} {\bf 07} (2018) 181, [\href{http://arxiv.org/abs/1802.06908}{{\tt arXiv:1802.06908}}].

\bibitem{Gromov:2017blm}
N.~Gromov, {\it {Introduction to the Spectrum of $N=4$ SYM and the Quantum Spectral Curve}},  \href{http://arxiv.org/abs/1708.03648}{{\tt arXiv:1708.03648}}.

\bibitem{Brizio:2024nso}
N.~Brizio, A.~Cavagli\`a, R.~Tateo, and V.~Tripodi, {\it {Regge trajectories and bridges between them in integrable AdS/CFT}},  \href{http://arxiv.org/abs/2410.08927}{{\tt arXiv:2410.08927}}.

\bibitem{Ekhammar:2024rfj}
S.~Ekhammar, N.~Gromov, and P.~Ryan, {\it {New Approach to Strongly Coupled N = 4 SYM via Integrability}},  \href{http://arxiv.org/abs/2406.02698}{{\tt arXiv:2406.02698}}.

\bibitem{Caron-Huot:2022sdy}
S.~Caron-Huot, F.~Coronado, A.-K. Trinh, and Z.~Zahraee, {\it {Bootstrapping $ \mathcal{N} $ = 4 sYM correlators using integrability}},  {\em JHEP} {\bf 02} (2023) 083, [\href{http://arxiv.org/abs/2207.01615}{{\tt arXiv:2207.01615}}].

\bibitem{Aprile:2018efk}
F.~Aprile, J.~Drummond, P.~Heslop, and H.~Paul, {\it {Double-trace spectrum of $N=4$ supersymmetric Yang-Mills theory at strong coupling}},  {\em Phys. Rev. D} {\bf 98} (2018), no.~12 126008, [\href{http://arxiv.org/abs/1802.06889}{{\tt arXiv:1802.06889}}].

\bibitem{Caron-Huot:2018kta}
S.~Caron-Huot and A.-K. Trinh, {\it {All tree-level correlators in AdS$_{5}$\texttimes{}S$_{5}$ supergravity: hidden ten-dimensional conformal symmetry}},  {\em JHEP} {\bf 01} (2019) 196, [\href{http://arxiv.org/abs/1809.09173}{{\tt arXiv:1809.09173}}].

\bibitem{Aprile:2020mus}
F.~Aprile, J.~M. Drummond, H.~Paul, and M.~Santagata, {\it {The Virasoro-Shapiro amplitude in AdS$_{5}$ \texttimes{} S$^{5}$ and level splitting of 10d conformal symmetry}},  {\em JHEP} {\bf 11} (2021) 109, [\href{http://arxiv.org/abs/2012.12092}{{\tt arXiv:2012.12092}}].

\bibitem{Caron-Huot:2021usw}
S.~Caron-Huot and F.~Coronado, {\it {Ten dimensional symmetry of $ \mathcal{N} $ = 4 SYM correlators}},  {\em JHEP} {\bf 03} (2022) 151, [\href{http://arxiv.org/abs/2106.03892}{{\tt arXiv:2106.03892}}].

\bibitem{Alday:2023pzu}
L.~F. Alday, T.~Hansen, and M.~Nocchi, {\it {High Energy String Scattering in AdS}},  {\em JHEP} {\bf 02} (2024) 089, [\href{http://arxiv.org/abs/2312.02261}{{\tt arXiv:2312.02261}}].

\bibitem{Alday:2024xpq}
L.~F. Alday, M.~Nocchi, C.~Virally, and X.~Zhou, {\it {On the Regge behaviour of the AdS Virasoro-Shapiro Amplitude}},  \href{http://arxiv.org/abs/2409.03695}{{\tt arXiv:2409.03695}}.

\bibitem{Saha:2024qpt}
A.~P. Saha and A.~Sinha, {\it {Field Theory Expansions of String Theory Amplitudes}},  {\em Phys. Rev. Lett.} {\bf 132} (2024), no.~22 221601, [\href{http://arxiv.org/abs/2401.05733}{{\tt arXiv:2401.05733}}].

\bibitem{Fardelli:2023fyq}
G.~Fardelli, T.~Hansen, and J.~A. Silva, {\it {AdS Virasoro-Shapiro amplitude with KK modes}},  {\em JHEP} {\bf 11} (2023) 064, [\href{http://arxiv.org/abs/2308.03683}{{\tt arXiv:2308.03683}}].

\bibitem{Gromov:2013pga}
N.~Gromov, V.~Kazakov, S.~Leurent, and D.~Volin, {\it {Quantum Spectral Curve for Planar $\mathcal{N} = 4$ Super-Yang-Mills Theory}},  {\em Phys. Rev. Lett.} {\bf 112} (2014), no.~1 011602, [\href{http://arxiv.org/abs/1305.1939}{{\tt arXiv:1305.1939}}].

\bibitem{Alfimov:2014bwa}
M.~Alfimov, N.~Gromov, and V.~Kazakov, {\it {QCD Pomeron from AdS/CFT Quantum Spectral Curve}},  {\em JHEP} {\bf 07} (2015) 164, [\href{http://arxiv.org/abs/1408.2530}{{\tt arXiv:1408.2530}}].

\bibitem{Alfimov:2020obh}
M.~Alfimov, N.~Gromov, and V.~Kazakov, {\it {Chapter 13: {\cal N} = 4 SYM Quantum Spectral Curve in BFKL Regime}},  \href{http://arxiv.org/abs/2003.03536}{{\tt arXiv:2003.03536}}.

\bibitem{Klabbers:2023zdz}
R.~Klabbers, M.~Preti, and I.~M. Sz\'ecs\'enyi, {\it {Regge Spectroscopy of Higher-Twist States in N=4 Supersymmetric Yang-Mills Theory}},  {\em Phys. Rev. Lett.} {\bf 132} (2024), no.~19 191601, [\href{http://arxiv.org/abs/2307.15107}{{\tt arXiv:2307.15107}}].

\bibitem{Ekhammar:2024neh}
S.~Ekhammar, N.~Gromov, and M.~Preti, {\it {Long Range Asymptotic Baxter-Bethe Ansatz for N=4 BFKL}},  \href{http://arxiv.org/abs/2406.18639}{{\tt arXiv:2406.18639}}.

\bibitem{Kazakov:2018hrh}
V.~Kazakov, {\it {Quantum Spectral Curve of $\gamma$-twisted ${\cal N}=4$ SYM theory and fishnet CFT}},  \href{http://arxiv.org/abs/1802.02160}{{\tt arXiv:1802.02160}}. [Rev. Math. Phys.30,no.07,1840010(2018)].

\bibitem{Levkovich-Maslyuk:2019awk}
F.~Levkovich-Maslyuk, {\it {A review of the AdS/CFT Quantum Spectral Curve}},  {\em J. Phys. A} {\bf 53} (2020), no.~28 283004, [\href{http://arxiv.org/abs/1911.13065}{{\tt arXiv:1911.13065}}].

\bibitem{Beisert:2003jj}
N.~Beisert, {\it {The complete one loop dilatation operator of N=4 superYang-Mills theory}},  {\em Nucl. Phys. B} {\bf 676} (2004) 3--42, [\href{http://arxiv.org/abs/hep-th/0307015}{{\tt hep-th/0307015}}].

\bibitem{Gunaydin:1984fk}
M.~Gunaydin and N.~Marcus, {\it {The Spectrum of the s**5 Compactification of the Chiral N=2, D=10 Supergravity and the Unitary Supermultiplets of U(2, 2/4)}},  {\em Class. Quant. Grav.} {\bf 2} (1985) L11.

\bibitem{Bars:1982ep}
I.~Bars and M.~Gunaydin, {\it {Unitary Representations of Noncompact Supergroups}},  {\em Commun. Math. Phys.} {\bf 91} (1983) 31.

\bibitem{Beisert:2004ry}
N.~Beisert, {\it {The Dilatation operator of N=4 super Yang-Mills theory and integrability}},  {\em Phys. Rept.} {\bf 405} (2004) 1--202, [\href{http://arxiv.org/abs/hep-th/0407277}{{\tt hep-th/0407277}}].

\bibitem{Marboe:2017dmb}
C.~Marboe and D.~Volin, {\it {The full spectrum of AdS5/CFT4 I: Representation theory and one-loop Q-system}},  {\em J. Phys. A} {\bf 51} (2018), no.~16 165401, [\href{http://arxiv.org/abs/1701.03704}{{\tt arXiv:1701.03704}}].

\bibitem{Marboe:2014gma}
C.~Marboe and D.~Volin, {\it {Quantum spectral curve as a tool for a perturbative quantum field theory}},  {\em Nucl. Phys.} {\bf B899} (2015) 810--847, [\href{http://arxiv.org/abs/1411.4758}{{\tt arXiv:1411.4758}}].

\bibitem{Marboe:2018ugv}
C.~Marboe and D.~Volin, {\it {The full spectrum of AdS$_5$/CFT$_4$ II: Weak coupling expansion via the quantum spectral curve}},  {\em J. Phys. A} {\bf 54} (2021), no.~5 055201, [\href{http://arxiv.org/abs/1812.09238}{{\tt arXiv:1812.09238}}].

\bibitem{Cavaglia:2014exa}
A.~Cavagli\`a, D.~Fioravanti, N.~Gromov, and R.~Tateo, {\it {Quantum Spectral Curve of the $\mathcal N=$ 6 Supersymmetric Chern-Simons Theory}},  {\em Phys. Rev. Lett.} {\bf 113} (2014), no.~2 021601, [\href{http://arxiv.org/abs/1403.1859}{{\tt arXiv:1403.1859}}].

\bibitem{Bombardelli:2017vhk}
D.~Bombardelli, A.~Cavagli\`a, D.~Fioravanti, N.~Gromov, and R.~Tateo, {\it {The full Quantum Spectral Curve for $AdS_4/CFT_3$}},  {\em JHEP} {\bf 09} (2017) 140, [\href{http://arxiv.org/abs/1701.00473}{{\tt arXiv:1701.00473}}].

\bibitem{Bombardelli:2018bqz}
D.~Bombardelli, A.~Cavagli\`a, R.~Conti, and R.~Tateo, {\it {Exploring the spectrum of planar AdS$_{4}$/CFT$_{3}$ at finite coupling}},  {\em JHEP} {\bf 04} (2018) 117, [\href{http://arxiv.org/abs/1803.04748}{{\tt arXiv:1803.04748}}].

\bibitem{Cavaglia:2021eqr}
A.~Cavagli\`a, N.~Gromov, B.~Stefa\'nski, Jr., and A.~Torrielli, {\it {Quantum Spectral Curve for AdS$_{3}$/CFT$_{2}$: a proposal}},  {\em JHEP} {\bf 12} (2021) 048, [\href{http://arxiv.org/abs/2109.05500}{{\tt arXiv:2109.05500}}].

\bibitem{Ekhammar:2021pys}
S.~Ekhammar and D.~Volin, {\it {Monodromy bootstrap for SU(2|2) quantum spectral curves: from Hubbard model to AdS$_{3}$/CFT$_{2}$}},  {\em JHEP} {\bf 03} (2022) 192, [\href{http://arxiv.org/abs/2109.06164}{{\tt arXiv:2109.06164}}].

\bibitem{Cavaglia:2022xld}
A.~Cavagli\`a, S.~Ekhammar, N.~Gromov, and P.~Ryan, {\it {Exploring the Quantum Spectral Curve for AdS$_{3}$/CFT$_{2}$}},  {\em JHEP} {\bf 12} (2023) 089, [\href{http://arxiv.org/abs/2211.07810}{{\tt arXiv:2211.07810}}].

\bibitem{Grabner:2020nis}
D.~Grabner, N.~Gromov, and J.~Julius, {\it {Excited States of One-Dimensional Defect CFTs from the Quantum Spectral Curve}},  {\em JHEP} {\bf 07} (2020) 042, [\href{http://arxiv.org/abs/2001.11039}{{\tt arXiv:2001.11039}}].

\bibitem{Julius:2021uka}
J.~Julius, {\em {Modern techniques for solvable models}}.
\newblock PhD thesis, King's Coll. London, 2021.

\end{thebibliography}\endgroup

\end{document}